\documentclass{aa}  
\usepackage{graphicx}
\usepackage{txfonts}
\usepackage{subfigure}
\usepackage{longtable}
\usepackage{color}
\usepackage{icomma}
\usepackage{bm}
\usepackage{marginnote}
\usepackage{textcomp}
\usepackage{float}
\usepackage[table,xcdraw]{xcolor}
\usepackage{amstext}
\usepackage[normalem]{ulem}

\def\mum{$\mu$m}
\def\deg{$^\circ$}

\def\hei{${\rm \ion{He}{I}}$}
\def\hei{${\rm \ion{He}{I}}$ (10830 \AA)}

\def\brg{${\rm Br}\gamma$}
\def\pab{${\rm Pa}\beta$}

\begin{document}

   \title{New insights on the near-infrared veiling of young stars using CFHT/SPIRou data\thanks{Based on observations obtained at the Canada-France-Hawaii Tele-scope  (CFHT)  which  is  operated  by  the  National  Research  Council (NRC) of Canada, the Institut National des Sciences de l’Univers of the Centre National de la Recherche Scientifique (CNRS) of France, and the University of Hawaii. Based on observations obtained with SPIRou, an international project led by Institut de Recherche en Astrophysique et Planétologie, Toulouse, France.}}

   \subtitle{}

   \author{A. P. Sousa\inst{1} 
           \and
          J. Bouvier\inst{1}
           \and
          S. H. P. Alencar\inst{2}          
           \and
          J.-F. Donati\inst{3}
           \and
          C. Dougados\inst{1}
           \and
          E. Alecian\inst{1}
           \and
          A. Carmona\inst{1}
           \and
          L. Rebull\inst{4}
           \and
          N. Cook\inst{5}
           \and
          E. Artigau\inst{5}
           \and
          P. Fouqu\'e\inst{3}
           \and
          R. Doyon\inst{5}
           \and
          the SLS consortium
}                
\institute{Univ. Grenoble Alpes, CNRS, IPAG, 38000 Grenoble, France\\
               \email{alana.sousa@univ-grenoble-alpes.fr}
      \and
      Departamento de F\'isica-Icex-UFMG Ant\^onio Carlos, 6627, 31270-901. Belo Horizonte, MG, Brazil
      \and
      Univ. de Toulouse, CNRS, IRAP, 14 avenue Belin, 31400 Toulouse, France
      \and 
      Infrared Science Archive (IRSA), IPAC, California Institute of Technology, 1200 E. California Blvd., Pasadena, CA, 91125, USA
      \and
      Universit\'e de Montr\'eal, D\'epartement de Physique, IREX, Montr\'eal, QC H3C 3J7, Canada
}
 
   \date{Received September 15, 1996; accepted March 16, 1997}

 
  \abstract
   {Veiling is ubiquitous at different wavelength ranges in classical T Tauri stars. However, the origin of the veiling in the infrared (IR) domain is not well understood at present. The accretion spot alone is not enough to explain the shallow photospheric IR lines in accreting systems, suggesting that another source is contributing  to the veiling in the near-infrared (NIR). The inner disk is often quoted as the additional emitting source meant to explain the IR veiling. }
   {In this work, we aim to measure and discuss the NIR veiling to understand its origins and variability timescale.}
   {We used a sample of 14 accreting stars observed with the CFHT/SPIRou spectrograph,  within the framework of the SPIRou Legacy Survey, to measure the NIR veiling along the $YJHK$ bands. We compared the veiling measurements with accretion and inner disk diagnostics. We also analyzed circumstellar emission lines and photometric observations from the literature. }
   {The measured veiling grows from the $Y$ to the $K$ band for most of the targets in our sample. The IR veiling agrees with NIR emission excess obtained using photometric data. However, we also find a linear correlation between the veiling and the accretion properties of the system,  showing that accretion contributes to the inner disk heating and, consequently, to the inner disk emission excess. We also show a connection between the NIR veiling and the system's inclination with respect to our line of sight. This is probably due to the reduction of the visible part of the inner disk edge, where the NIR emission excess is expected to arise, as the inclination of the system increases. Our search for periods on the veiling variability showed that the IR veiling is not clearly periodic in the typical timescale of stellar rotation -- which, again, is broadly consistent with the idea that the veiling comes from the inner disk region. 
   The NIR veiling appears variable on a timescale of a day, showing the night-by-night dynamics of the optical veiling variability.  In the long term, the mean NIR veiling seems to be stable for most of the targets on timescales of a month to a few years. However, during occasional episodes of high accretion in classical T Tauri stars, which affect the system's dynamic, the veiling also seems to be much more prominent at such times, as we found in the case of the target RU Lup.
 }
  {We provide further evidence that  for most targets in our sample, the veiling that mainly occurs  in the $JHK$ bands arises from dust in the inner disk.}


   \maketitle
\section{Introduction}

The photospheric lines of young low-mass accreting systems, commonly referred to as Classical T Tauri stars (CTTS), are shallower and present smaller equivalent widths than those of non-accreting stars with a similar spectral type. This phenomenon is known as the veiling of the photospheric lines \citep[e.g.,][]{1991ApJ...382..617H,1993AJ....106.2024V,1999A&A...352..517F,2011ApJ...730...73F}. The presence of veiling suggests an additional emitting source, beyond the stellar photosphere contributing to the spectra of the targets, that is responsible for filling in the photospheric lines 
  \citep[e.g.,][]{1990ApJ...349..190H,1998ApJ...492..323G,1998ApJ...509..802C,2001ApJ...561.1060J}. 
  
The veiling in the optical and in the IR wavelengths has been studied using different approaches in the aim to understand its origins and variability \citep[e.g.,][]{1990ApJ...363..654B,2006ApJ...646..319E,2011ApJ...730...73F,2017A&A...606A..48A,2017ApJ...836..200G,2013ApJ...767..112I,2013ApJ...769...73M,2021ApJ...922...27K}. The veiling variability along the stellar spectra depends on wavelength \citep[e.g.,][]{2011ApJ...730...73F,2012PASP..124.1137F,2013ApJ...769...73M,2018A&A...610A..40R}, and optical veiling is often associated with the accretion process, while the accretion shock is thought to be at the origin of an additional continuum emitting source to the stellar spectrum. \citep[e.g.,][]{1998ApJ...492..323G}. Usually, the accretion spot presents an emission contribution maximum around the ultraviolet domain, which decreases as the wavelength increases \citep[e.g.,][]{1998ApJ...509..802C}. 
Nevertheless, we do not expect a significant contribution of the accretion spot continuum emission in the IR wavelength; therefore, the accretion spot alone cannot explain the veiling in the IR domain. 

In the IR region, the veiling increases with wavelength and in some cases, it becomes greater than the veiling in the optical domain \citep[e.g.,][]{2013ApJ...769...73M}. 
The central star illuminates the inner disk and this region absorbs photons from the star,  the accretion spot, and even the accretion funnel, then re-emitting them in the infrared (IR) as the system rotates \citep[e.g.,][]{1997ApJ...490..368C,1999ApJ...519..279C}. Therefore, the inner disk is suggested as the origin of the additional continuum emission that is essential to explaining the near-infrared (NIR) veiling, although the measured veiling is often too great to be explained as coming merely from the disk emission, based on model predictions \citep[e.g.,][]{1999A&A...352..517F,2001ApJ...561.1060J}.  
Many authors have used different techniques to connect the observed veiling with inner disk emission, such as measuring the temperature of the region where the veiling comes from and using a black body fit to the veiling. For most of these systems, they found temperatures compatible with the dust temperature in the inner disk   \citep[e.g.,][]{2011ApJ...730...73F,2017A&A...606A..48A,2021A&A...652A..72A}.
For a few targets, the blackbody temperature measured using veiling is too high for dust to survive in the inner disk; this would indicate that the veiling should arise from the gas in the inner disk inside the star-disk co-rotation radius \citep[e.g.,][]{2017A&A...606A..48A,2021A&A...652A..72A}. However, \cite{2013ApJ...769...73M} found no evidence of hot gas inside the inner disk, which would be responsible for the NIR veiling. Instead, they explained the IR veiling as the combined emission from the accretion shock on the stellar surface and dust around the sublimation rim of the inner disk. 

The veiling around $1$\mum\ and the veiling in the $K$ band and beyond can also have different origins. While the accretion spot emission contribution is not very substantial around $1$\mum, we do not expect a significant contribution from the inner disk either. The origin of the veiling in this spectral domain is poorly understood. However, significant veiling was measured around $1$\mum, primarily for high accretion rate systems \citep[e.g.,][]{2006ApJ...646..319E,2011ApJ...730...73F,2013ApJ...767..112I}. In the literature, there are only a few plausible explanations given for that case of veiling, such as a contribution from emission lines, filling in of the photospheric lines, or origination from the accretion shock \citep[e.g.,][]{2015A&A...580A..82S,2013AstL...39..389D}. Even if we do not  detect these extra emission lines directly, they can contribute to making  the photospheric lines of the stellar spectra shallower, which increases the veiling measurement.

We cannot exclude other possible explanations to the IR veiling, such as an envelope around a star that can also be a source of  additional emission to the photospheric continuum, with emission compatible with NIR veiling \citep[e.g.,][]{1997ApJ...481..912C}. However, CTTSs are usually Class II stars and we would not expect a significant contribution from a dusty envelope. Furthermore, the veiling in the $K$ band is higher than the dust envelope emission can explain \citep{1999A&A...352..517F}.

In this work, we study the veiling and its variability in the NIR, using a sample of young accreting stars observed with the  Canada-France-Hawaii Telescope  SPectropolarimètre
InfraROUge (CFHT/SPIRou). Our sample comprises stars with different properties, such as the mass accretion rate, spectral type, and inclination with respect to our line of sight. We computed the veiling for the $YJHK$ bands and compared the results with accretion and inner disk diagnostics.

We organized the paper as follows. In Sect. \ref{sec:data}, we present the sample of stars that we used in this work, and we describe the data used. In Sect. \ref{sec:method} we show the procedures to measure the veiling. We describe the results obtained from the veiling measurements in Sect. \ref{sec:result}. In Sect. \ref{sec:discussion}, we discuss the possible origin of the veiling, and we compare our results with those of previous works. In Sect. \ref{sec:concl}, we present our conclusions.

\section{Observations and targets selection}\label{sec:data}

\begin{table*}
 \centering
 \tiny
 \addtolength{\tabcolsep}{-2pt}  
 \caption{Sample of stars and number of observations per observational period}
 \label{tab:obs}
 \begin{tabular}{llllllllllllll}
  \hline
  \hline
  Star & SpT & v$\sin{i}$ & Av & i$^*$\tablefootmark{a} & 2018 & 2019a & 2019b & 2020a & 2020b & 2021a & 2021b &  2022a & References\tablefootmark{b} \\
   &  & (km/s) & (mag) & (\deg) &  &  &  &  &  &  &  & &  \\
  \hline
\multicolumn{13}{c}{Accreting systems} \\
  \hline
    CI Tau           & K4    &9.5$\pm$0.5  & 0.65     & 55$^{+35}_{-10}$ & 2  &6  &26 &5  &39 & - & - & - &  (1),(2),(2),(2)\\ 
    DoAr 44          & K2-K3 &17.0$\pm$1.1 & 2.0$\pm$0.2 &30$\pm$5  & -  &8  &-  &-  &-  & - & - & - & (3),(3),(3),(3)\\

    GQ Lup           & K7  &5$\pm$1      &0.7       & $\sim$30     & -  &8  &-  &18 &6  &-  & - & - &  (4),(5),(4),(5)\\   
    TW Hya           & K6  &6.0$\pm$1.2  &0.0       & 18$\pm$10     & -  &12 &-  & 14 &-  &  27  & - & - &  (1),(1),(6),(7)\\  
    V2129 Oph        & K5  &14.5$\pm$0.3 &0.6        & 60     & 9 &-  &-  &17 &8 & -  & - & - &  (1),(1),(8),(9)\\    
    BP Tau           & K5  &9.0$\pm$0.5  & 0.45     & $\sim$45    & -  &-  &21 &-  & - & -  &  34 & - & (10),(11),(6),(11)\\    
    V347 Aur         & M2-M3 &11.7$^{+0.16}_{-0.24}$        & 3.4 &40  & -  &-  &18 &-  &13& 12  &  22 & - & (12),(13),(12),(14) \\    
    DG Tau           & K6    &24.7$\pm$0.7  &1.60$\pm$0.15  & 38$\pm$2  & -  &-  &-  &2  &29& -  & - & - & (10),(10),(6),(15)\\
    
    RU Lup           & K7    &8.5$\pm$4.8   &0.0  & 24   & -  &-  &-  &9  & - &  17  &  13 &  1 &  (16),(17),(14),(18)\\ 
    V2247 Oph        & M0    &20.5$\pm$0.5    & 0.98$\pm$0.02 & 45$\pm$10   &-   &-  &-  &9 &7 & -  & - & - &  (19),(20),(19),(20)\\
 
    DO Tau           & M0    &14.3$\pm$0.5     &0.75  & 37.0$\pm$3.7 & -  &-  &-  &3  &8 & -  & - & - &  (6),(21),(6),(15) \\
  
    J1604$^*$            & K3    &17.3$\pm$0.4  & 1.0  & >61  & -  &-  &-  &12 &- & -  & - & - &  (22),(22),(24),(23)\\
    
    PDS 70           & K7    &16.0$\pm$0.5  &0.01$\pm$0.07 & 50$\pm$8 &-  &-  &-  & 4 & - & - & - &  6 &  (19),(25),(19),(25)\\
    GM Aur           & K4-K5 &14.9$\pm$0.3  &0.3$\pm$0.3 &  $\geq$63 &-  &-  &-  & - & 2 & - & 34 & - &  (26),(26),(26),(26)\\
   \hline
   \multicolumn{13}{c}{Non-accreting systems} \\
   \hline
    V819 Tau         & K4    & 9.5        &   &   & -  &-  &-  &1  &-  & -  & - & - &  (27),(27)\\
    TWA 25           & M0.5   &12.9$\pm$1.2  &   &  & -  &-  &25 &14 & - & -  & - &- &  (6),(1)\\  
    TWA 9A           & K6     &7$\pm$3       &   &  & -  &-  &-  &1  & - & -  & - & - &  (6),(1)\\
 \hline
 \end{tabular} 
 \tablefoot{
 \tablefoottext{*}{ RX J1604.3-2130A, elsewhere in the text, we used J1604 for brevity.}
 \tablefoottext{a}{
 The inclination (see references) between the stellar rotation axis and our line of sight obtained from v$\sin{i}$, period, and radius of each target, except for the inclinations of DO Tau and DG Tau that were obtained from the outer disk parameters.}
 \tablefoottext{b}{ References for the SpT, $v\sin{i}$, Av and inclination ($i$), respectively.}
\tablebib{ (1)~\citet{2006...460..695T};
(2) \citet{2020MNRAS.491.5660D}; (3) \citet{2020...643A..99B}; (4) \citet{2017...600A..20A}; (5) \citet{2012MNRAS.425.2948D}; (6) \citet{2014ApJ...786...97H}; (7) \citet{2002ApJ...571..378A}; (8) \citet{2007MNRAS.380.1297D}; (9) \citet{2012AA...541A.116A}; (10)  \citet{2012ApJ...745..119N}; (11) \citet{2008MNRAS.386.1234D}; (12) \cite{2010AJ....140.1214C}; (13)
\citet{2019ApJ...882...75F}; (14) 
Alecian et al. (in prep.); 
(15) \citet{2016ApJ...831..169S}; (16) \cite{2014AA...561A...2A}; (17) \citet{2017AA...602A..33F};
(18) \citet{2007A&A...461..253S}; (19)  \citet{2016MNRAS.461..794P}; (20) \citet{2010MNRAS.402.1426D}; (21) \citet{2019AJ....157..196K}; (22) \citet{2012ApJ...745...56D}; (23) \citet{2019MNRAS.484.1926D}; (24) \citet{1999AJ....117.2381P}; (25) \citet{2020ApJ...892...81T}; (26) Bouvier et al. (in prep.); (27) \citet{2015MNRAS.453.3706D}.

}
}
 
\end{table*}

The sample of stars used in this work is composed of well-known young stars that are part of the SPIRou Legacy Survey-SLS science program: "Magnetic PMS star/planet survey" of some 50 class I, II, and III stars. The SPectropolarim\`etre InfraROUge (SPIRou) is a high-resolution velocimeter and polarimeter  spectrograph ($R\sim75\,000$) covering the NIR wavelength  range $\sim0.98-2.35\,\mu$m, corresponding to the spectral domain of the $YJHK$ bands \citep{2018haex.bookE.107D}. The main science goals of CFHT/SPIRou Legacy Survey are the search for and characterization of planets around low-mass stars, and investigating the impact of the stellar magnetic field on planet and star formation in young systems \citep{2020MNRAS.498.5684D}.

We aim to investigate the veiling of accreting young stars. Therefore, we selected, among the sample of stars observed by the SLS, 13 stars reported as accreting systems in the literature and for which we had a reasonable number of observations in time in comparison to the stellar rotation period. Most of these targets are classified as Class II and CTTS systems, and only V347 Aur is a Class I target. In addition, we added the T Tauri star J1604 (RX J1604.3-2130A) to the sample, which is not part of the SLS program, however, its  CFHT/SPIRou observations are available. 

In Table \ref{tab:obs}, we show the list of young stars analyzed in this work and the number of observations that we have for each target and each observational period. We also list three non-accreting T Tauri stars that cover our sample's spectral types and are also slow rotators ($v\sin{i}<15\,$km/s). We used these stars as templates to compute veiling and the residual profiles, as described in the following sections.

Each CFHT/SPIRou observation consists of four sub-exposures to measure Stokes V, taken at different orientations of the polarimeter and used to compute the non-polarized and the circularly polarized profiles. As the focus of this work is to analyze only the non-polarized component of the spectrum, we averaged the four sub-exposures to increase the signal-to-noise ratio (S/N) of the spectra obtained at each night. For the non-accreting systems, we  averaged all the observations obtaining a mean spectra that were used as templates. The CFHT/SPIRou data were reduced, while the telluric-corrected  spectra were obtained using the data reduction system APERO,  versions 0.6.131, 0.6.132, and 0.7.232 \citep{2022arXiv221101358C}. The spectra were corrected for the barycentric velocity and locally normalized to the continuum level, using a polynomial function to fit the continuum.

\section{Procedures to measure the veiling} \label{sec:method}

\begin{figure*}
    \centering
    \includegraphics[width=9.1cm]{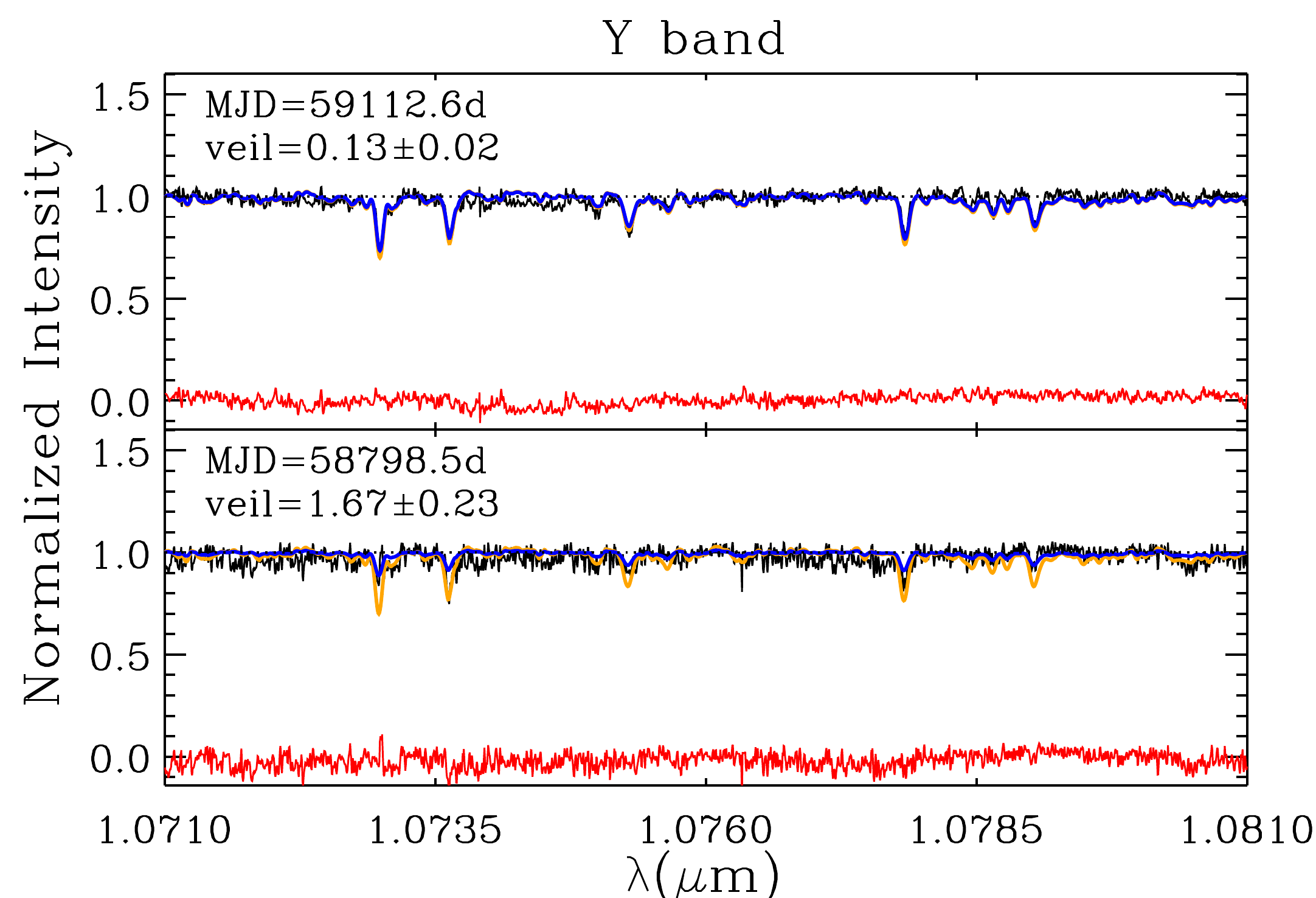}
    \includegraphics[width=9.1cm]{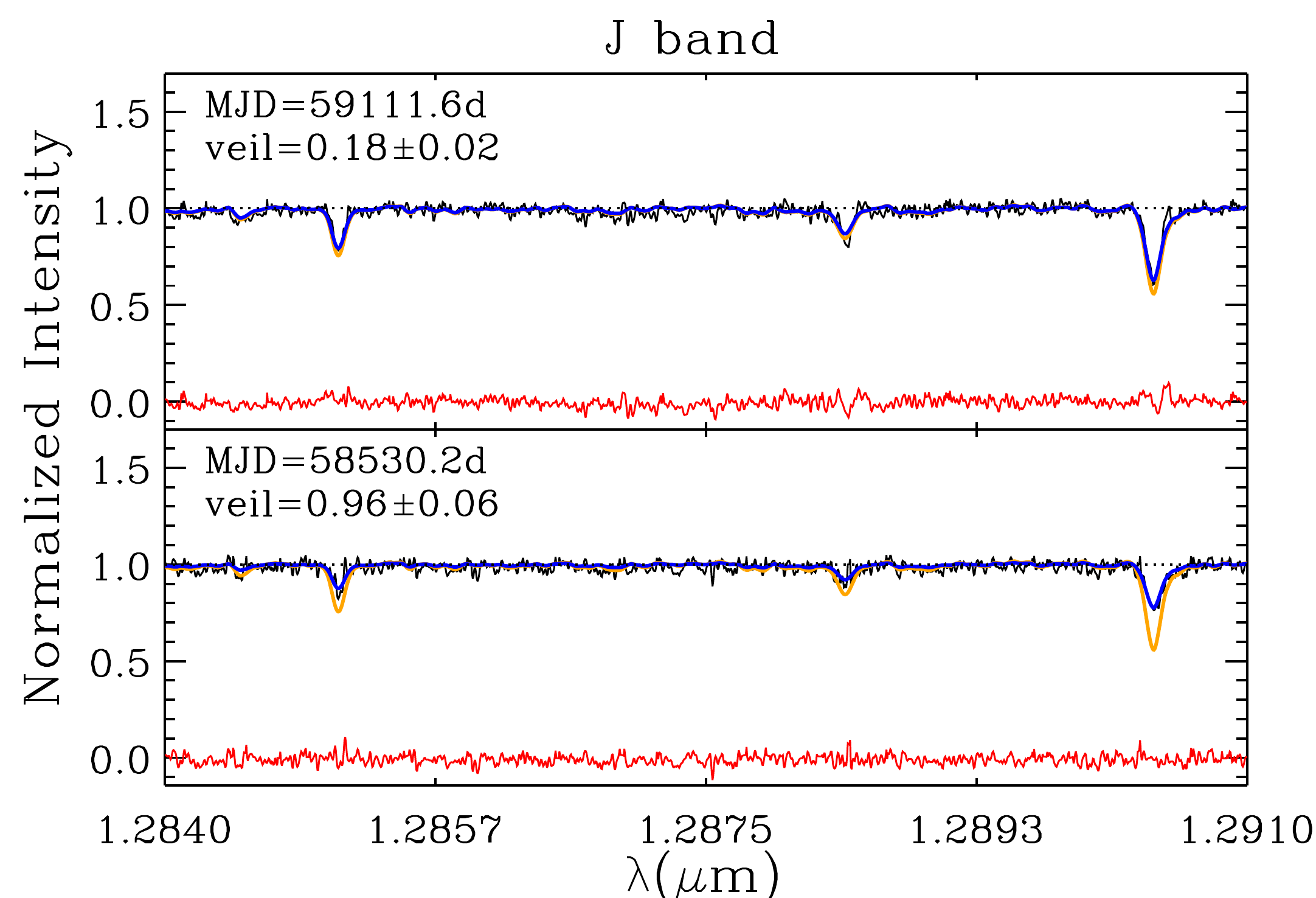}\\
    \includegraphics[width=9.1cm]{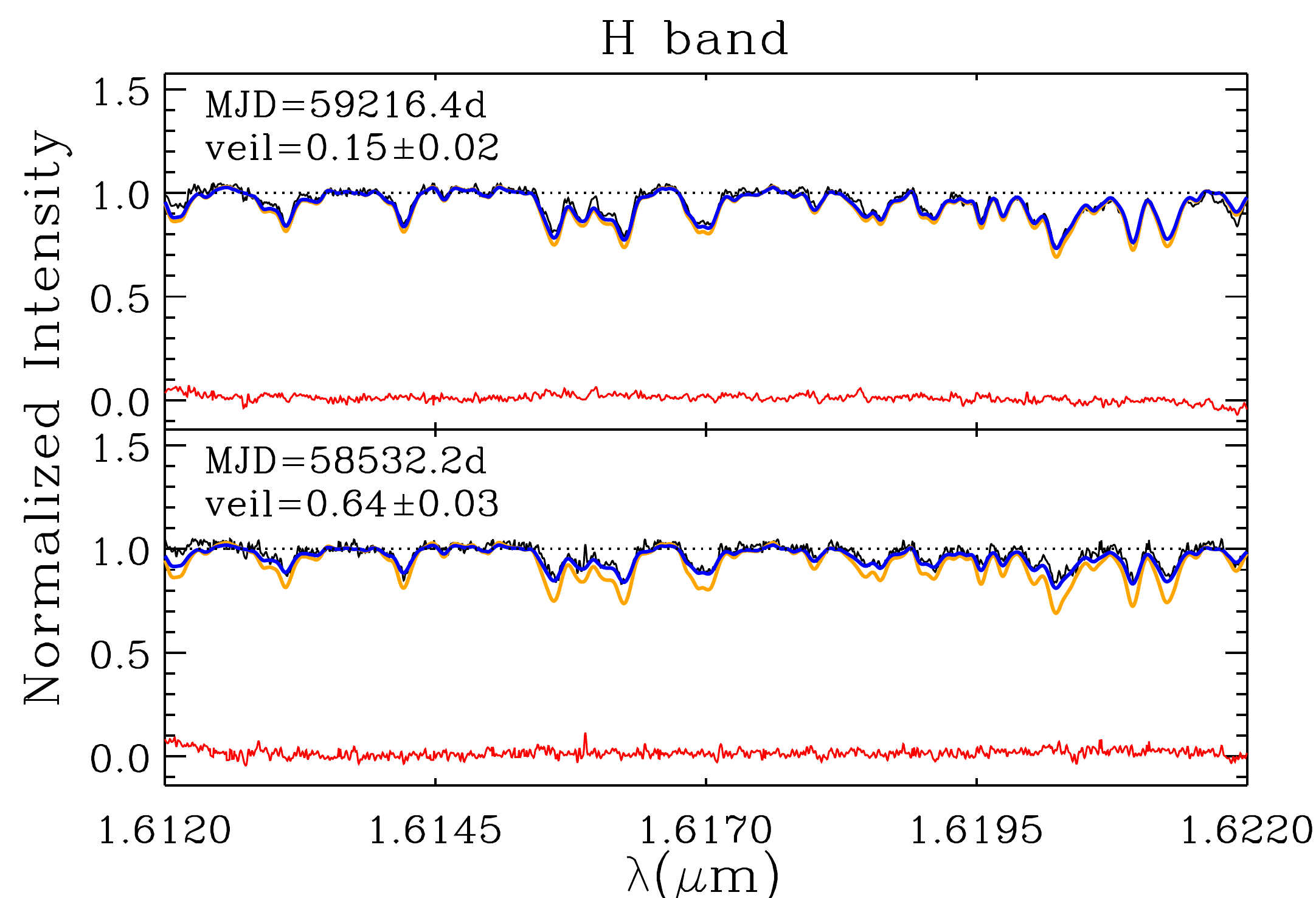}
    \includegraphics[width=9.1cm]{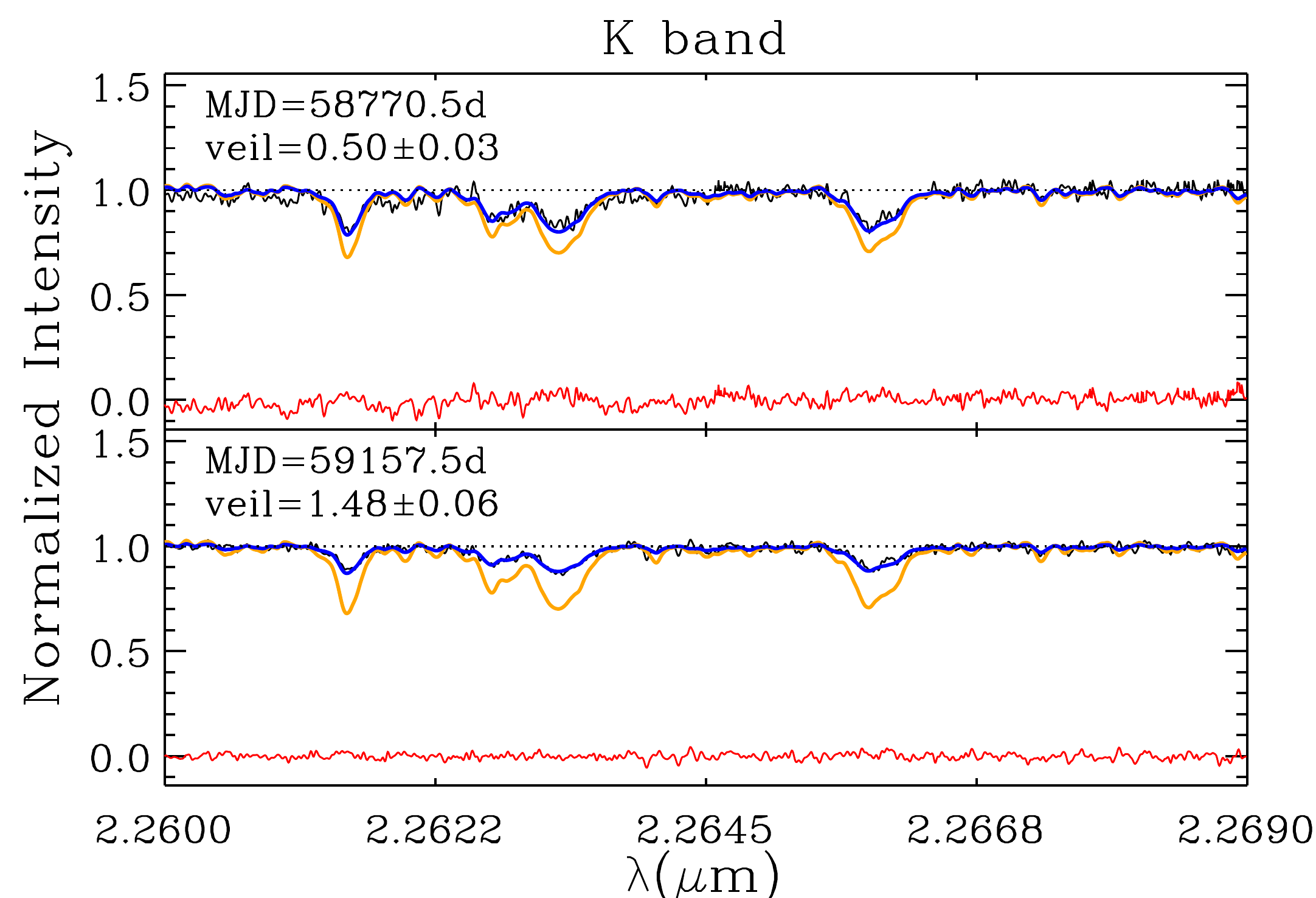}
    \caption{Examples of the four spectral regions used to measure the veiling of CI Tau. In each panel,  we show two different nights, representing a small and a high veiling estimated for this target. We show the CI Tau spectrum in black and the residual profile (in red) obtained after subtracting the veiled template. The V819 Tau template spectra are also displayed before (orange) and after (blue) applying the veiling correction. The photospheric lines were removed in the residual profiles, showing that the veiling was accurately determined for most nights.}
    \label{fig:phot_regions}
\end{figure*}
We computed the IR veiling of the targets following the method described by \cite{1989ApJS...70..899H}, where we compare the spectra of the target with the spectrum of a non-accreting T Tauri star of a similar spectral type.  The Zeeman broadening of the photospheric lines can affect the veiling measurements. Therefore, we used WTTSs as templates that should present a similar magnetic activity as the CTTS \citep[e.g.,][]{2000ApJ...539..815J}. The WTTSs also present comparable physical properties to the CTTSs, such as chromospheric activity and surface gravity, which make WTTSs stars suitable to measure the veiling in accreting systems. We list the templates applied to each star in Table \ref{tab:param}.

Before comparing the target and template spectra, we shifted and broadened the template spectra to match the target spectra, using the radial velocities of the targets.\footnote{ For most targets, the radial velocities were computed using the CCF profiles generated by the SPIRou pipeline, implementing a numerical mask corresponding to the target spectral type. For TW Hya, we computed the radial velocity cross-correlating the target spectra with a WTTS with a similar spectral type.} and the literature $v\sin{i}$ values of the targets. Due to the IR veiling wavelength dependence \citep[e.g.,][]{2021A&A...652A..72A}, we measured the veiling in four different spectral regions, 10710\AA-10810\AA, 12840\AA-12910\AA, 16120\AA-16220\AA, and 22600\AA-22690\AA, which we call $r_Y$, $r_J$, $r_H$, and $r_K$, representing the $YJHK$ veilings, respectively. The stars RU Lup, DO Tau and, DG Tau present many emission lines along the spectra, probably originating from the accretion shock, which is a characteristic of high-mass accretion rate systems, which prevented us from using the same $Y$ and $J$ spectral regions for the veiling calculations. Then, for these targets, we used the spectral regions 10760\AA-10785\AA\ and 12400\AA-12550\AA\ to measure the $r_Y$ and $r_J$ veilings, respectively. On some nights, the 10710\AA-10810\AA\ region of the TW Hya and V2247 Oph spectra presented features that  made veiling measurements impossible and we had to use the 10864\AA-10920\AA\ region to measure $r_Y$ veiling instead.  

We determined the best veiling value for each target spectrum through a $\chi^2$ minimization routine. The veiling was defined as the ratio of the continuum excess flux to the stellar photospheric flux ($r_\lambda=F_{\lambda_{Excess}}/F_{\lambda_{Phot}}$). Then, zero veiling means the system has no additional excess to the stellar photosphere. 
In the spectral range of our sample (K2 to M3), the choice of template does not interfere much in the computed veiling, as shown by \cite{1999A&A...352..517F}, since the difference between templates of different spectral types is small and produces an almost null relative veiling, within the uncertainties.  We also measured the systematic veiling between our templates, which corresponds to the average veiling resulting from the computed veiling comparing each template with the other. We found it to be $r_Y=0.098\pm0.068$, $r_J=0.02\pm0.02$, $r_H=0.05\pm0.02$ and, $r_K=0.04\pm0.02$, which should only affect the veiling determination of stars with very small or no veiling.

In Fig. \ref{fig:phot_regions},   we present the results for the four photospheric regions used to measure the veiling for CI Tau from two nights, representing spectra with smaller and higher veiling values. We show the target spectrum, and the unveiled and veiled template spectra. We also computed the residual profile, which is the target's spectrum subtracted from the veiled template. Most of the residual profiles show almost no features at the location of photospheric lines, indicating that the veiling measurements were correctly determined for most nights, and that the photospheric lines were correctly removed. However, some of the spectra were quite noisy and this affected the veiling determinations. The veiling error obtained for each night, written in the plot, comes from a Chi-square minimization process, where we compare the target with the template spectra. Besides taking into account the noise of the target and template spectra\footnote{ We used as the spectral noise, the standard deviation measured in the continuum of each night, in the spectral region used to measure the veiling.} to compute the veiling, there are other errors, such as those associated with the normalization processes that we did not consider in estimating the veiling error.

\section{Results}\label{sec:result}

\begin{figure}
\centering
{\includegraphics[width=4.4cm]{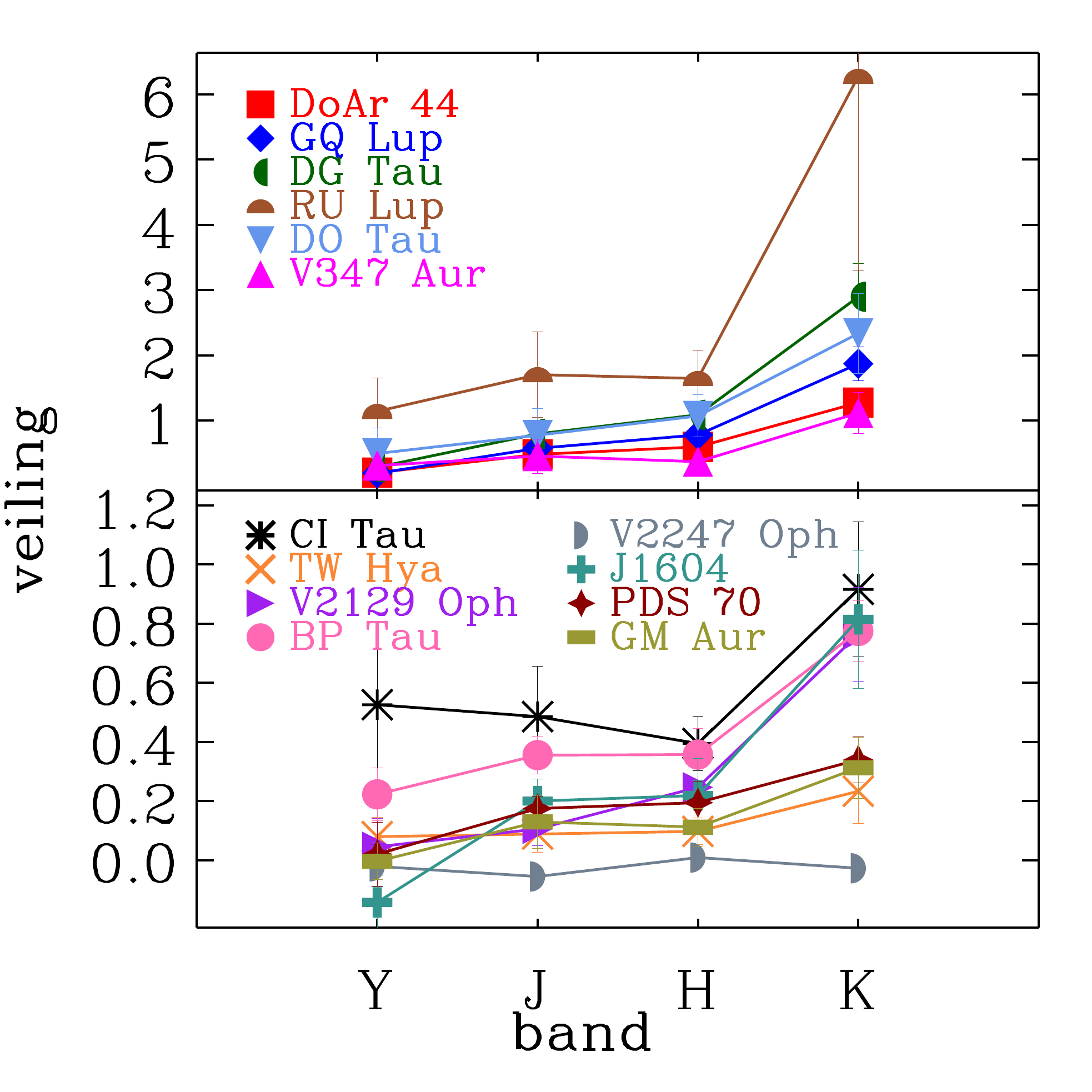}}
{\includegraphics[width=4.4cm]{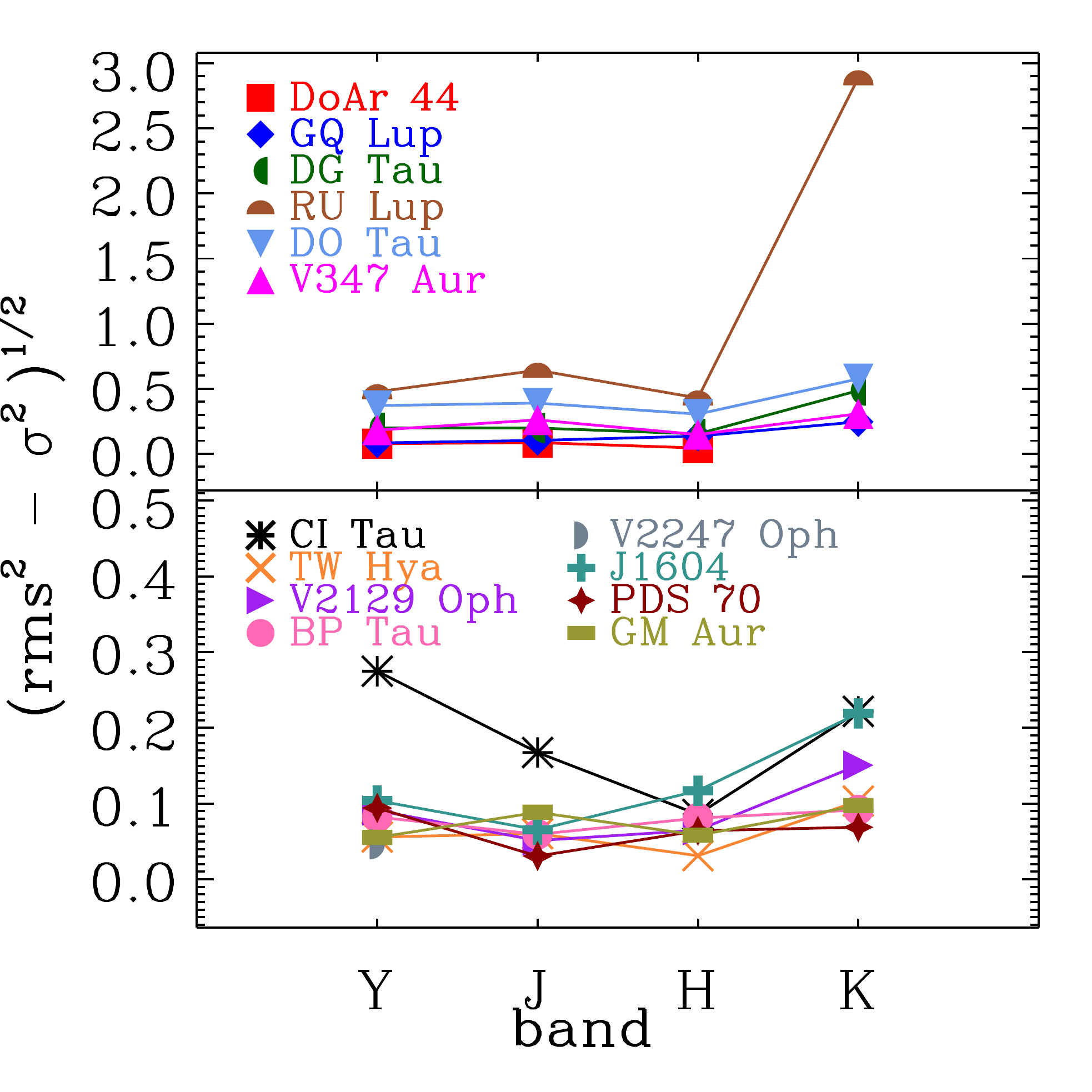}}
\caption{\label{fig:veilband} Average NIR veiling (\textit{left}) and the veiling variability diagnostic (\textit{right}) measured in different wavelength regions. The top and bottom panels show the targets with $r_K$ higher and lower than 1, respectively.  The veiling increases from the $Y$ to the $K$ band for most of the targets. The error bars in the left panel represent the standard deviation of the average veiling over all the observation nights.
In this figure and the following figures, the color and symbol codes in the panels identify each target.} 
\end{figure}

We present in Fig. \ref{fig:veilband} the NIR average veiling over the observation nights, measured for all the four regions, referred to as $r_Y$, $r_J$, $r_H$, and $r_K$.  For  readability, we split the targets into two groups; systems with $r_K$ higher and smaller than 1. We also show the averaged veiling obtained for each target in Table \ref{tab:param}. The veiling values increase from the $Y$ to the $K$ band for most of the targets, similar to the results found in previous works \citep[e.g.,][]{2013ApJ...769...73M,2021A&A...649A..68S,2021A&A...652A..72A}. Besides this general result, the average veiling for some individual targets remains the same from the $Y$ to $K$ band. For example,  the veiling measured for V2247 Oph. However, this target does not present significant veiling in any band, probably due to the fact that it is a more evolved system, where the accretion and the dust in the inner disk are too faint to be detected. Furthermore, this M-type star makes detecting excess in the IR difficult due to the low contrast between the stellar photosphere and the inner disk emission \citep[e.g.,][]{2009MNRAS.394L.141E}. Another example is CI Tau, where the average veiling decreases from the $Y$ to the $H$ band, followed by an increase to the $K$ band. In that case, each band's veiling variability is high and the difference in the $Y$ to $H$ average veiling is smaller than the standard deviation.

\begin{figure}
\centering
{\includegraphics[width=4.4cm]{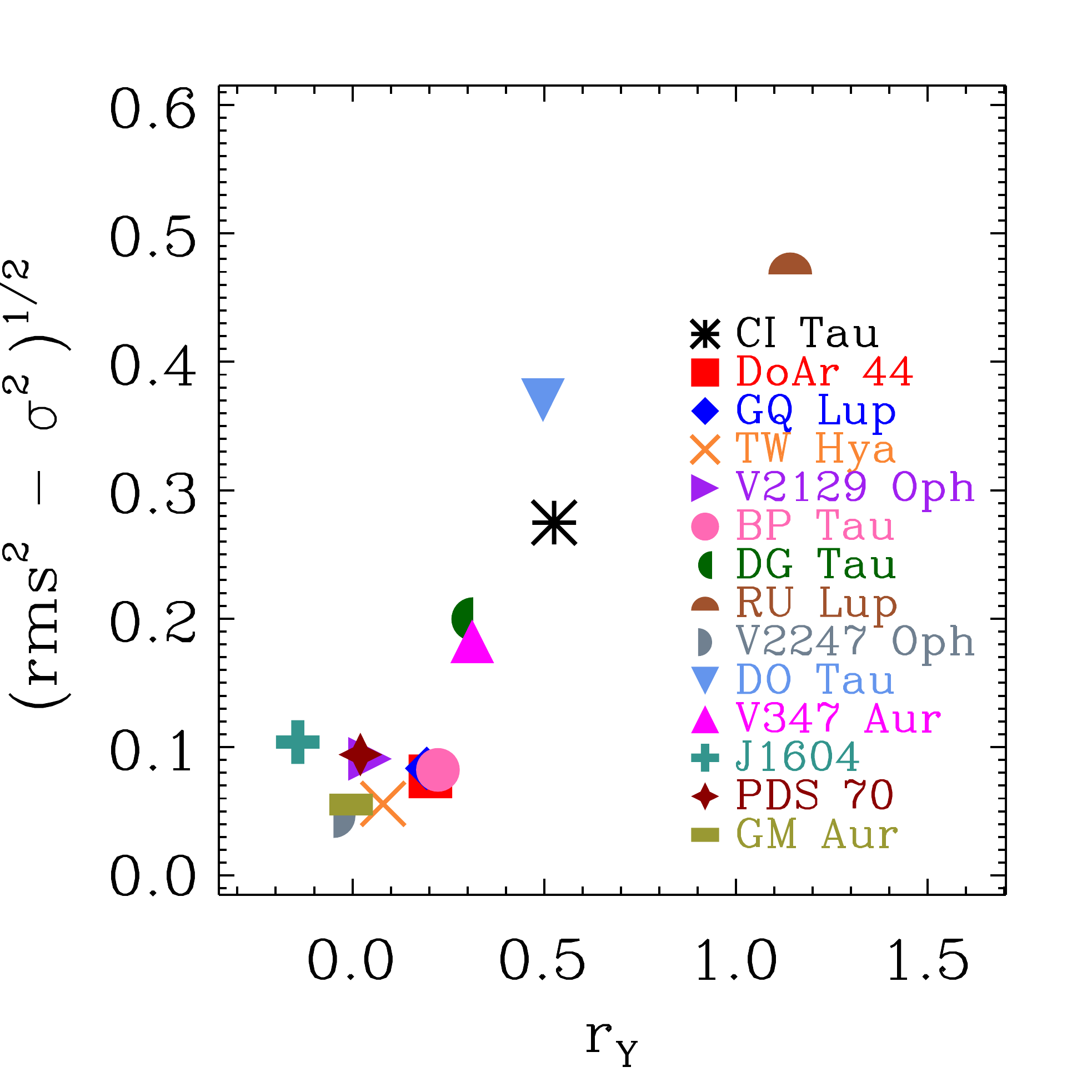}}
{\includegraphics[width=4.4cm]{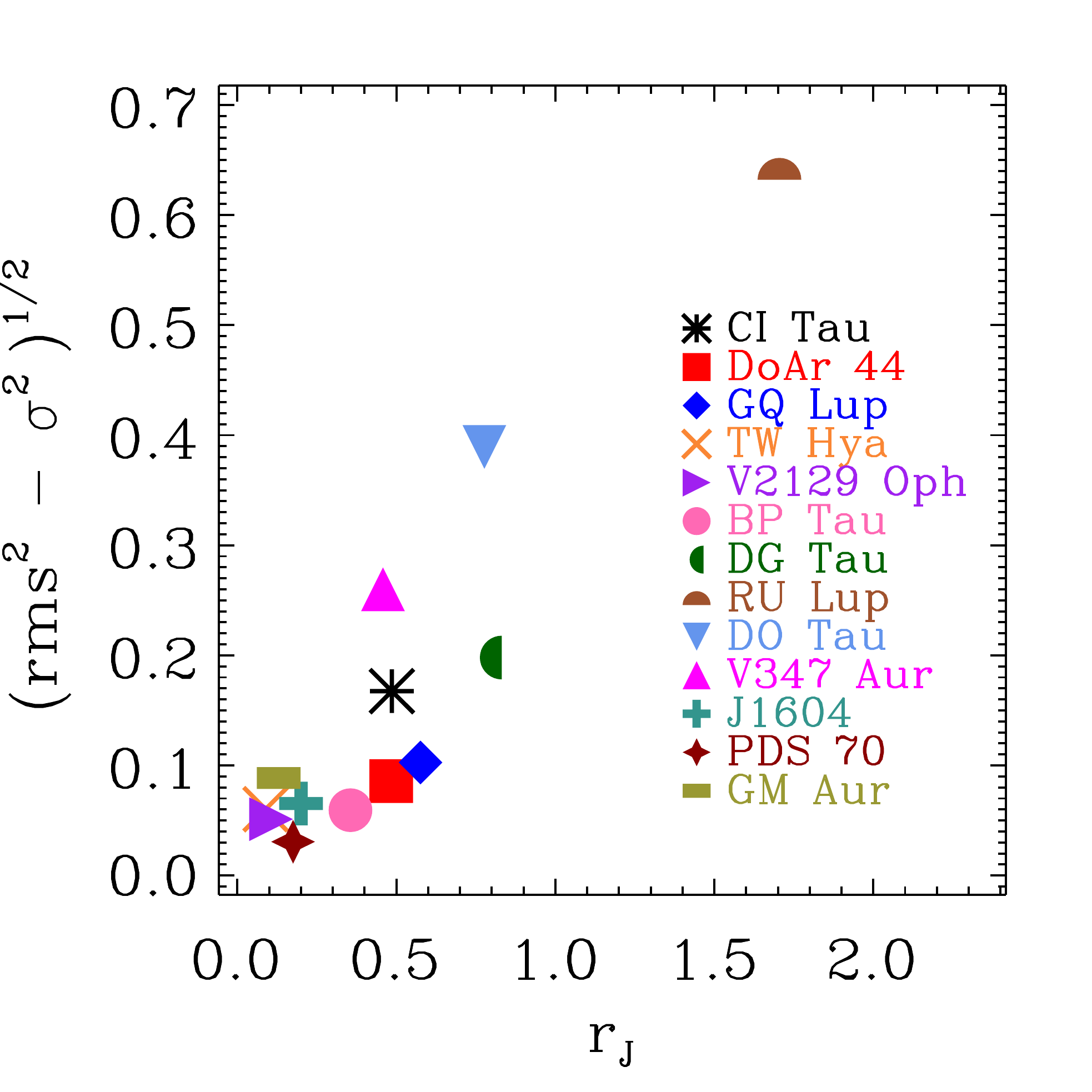}}\\
{\includegraphics[width=4.4cm]{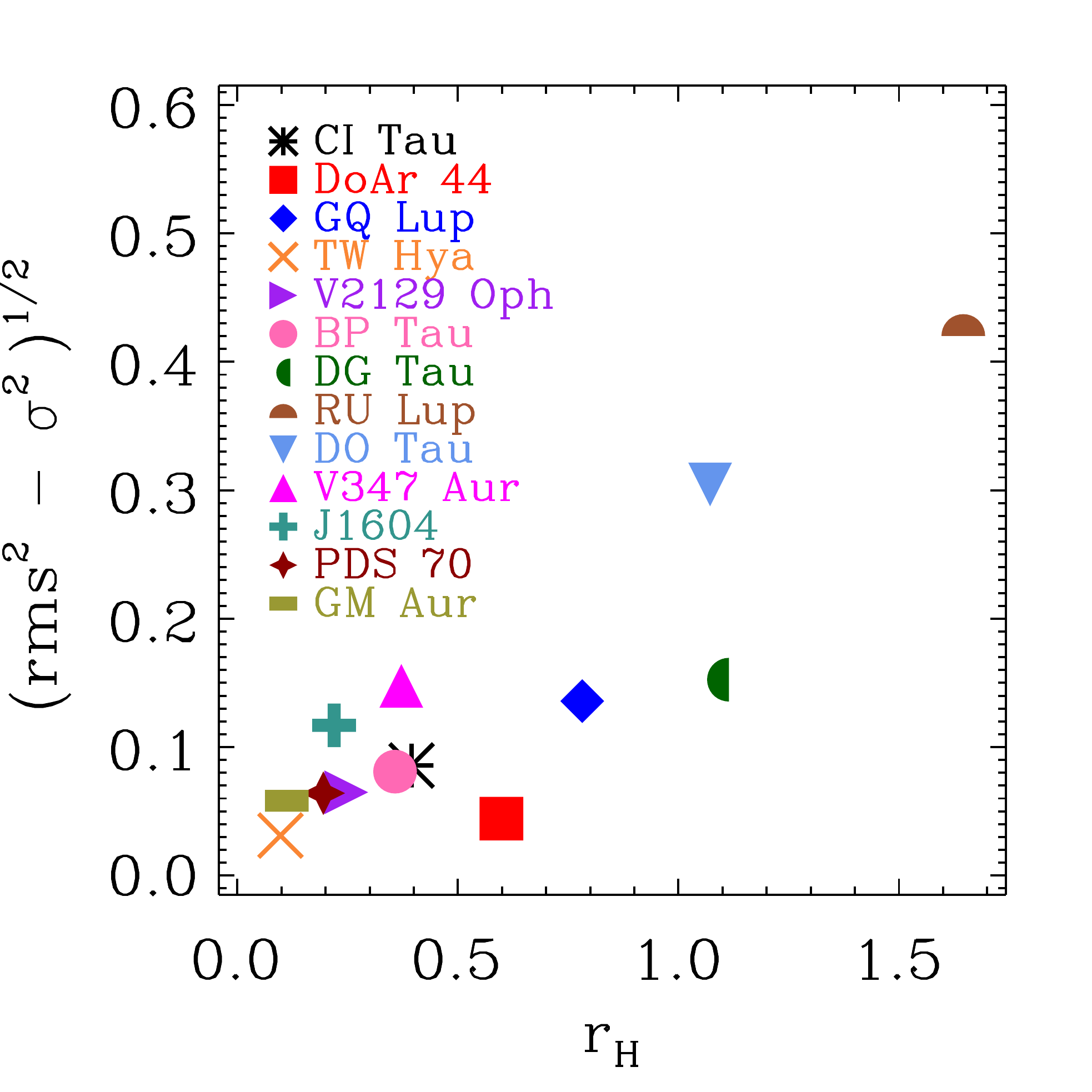}}
{\includegraphics[width=4.4cm]{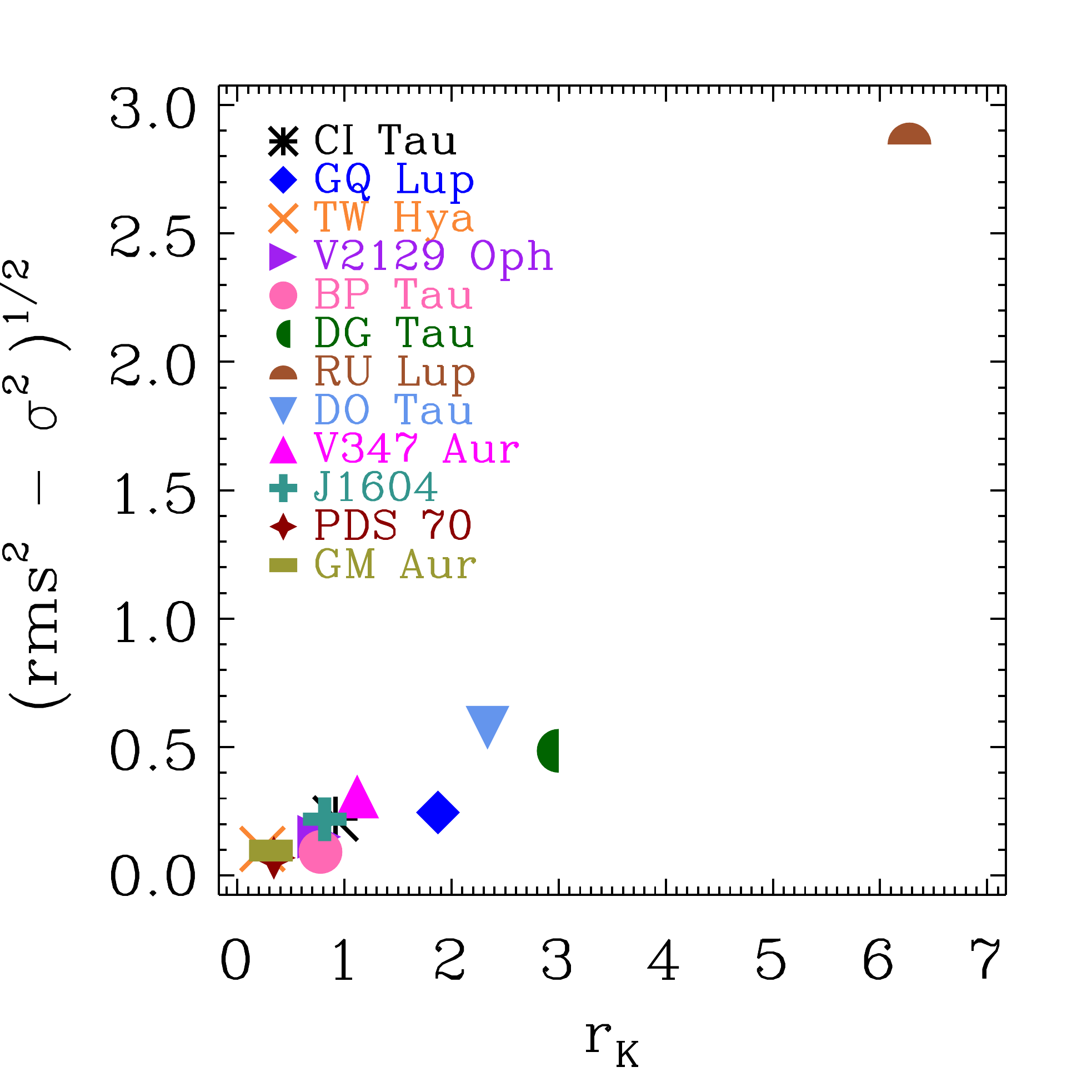}}
\caption{\label{fig:rmsvsveil} Veiling variability diagnostic as a function of the average veiling. The rms value refers to the root-mean-square of the veiling variability and $\sigma$ is the average error on the veiling measurements. Each panel shows the veiling $r_\lambda$ measured in a different band ($YJHK$). 
 Systems with higher veiling also present higher veiling variability. } 
\end{figure}

Due to the small number of photospheric lines and the lower  S/N values in the $Y$ region of the spectra, the veiling in this region was not as well determined, compared to the other bands. For some nights or targets, the veiling is even slightly negative, which has no physical meaning -- in such cases, we can assume that the system in this region has zero veiling. 

We also checked the variability of the veiling measured in each band, computing for each target the RMS of the $YJHK$ veilings, using all the observed nights. To characterize the variability over the noise, we subtracted the average error of the veiling ($\sigma$) from the RMS. Then, we computed $S=\sqrt{rms^2-\sigma^2}$, as a veiling variability diagnostic, following \cite{2014AJ....147...82C}.  We show the results in Figs. \ref{fig:veilband} and \ref{fig:rmsvsveil} as a function of the band and veiling measured, respectively. Most of the targets present variability above the error level. The veiling in the $H$ band appears to be less variable than the other bands, while the $K$ band presents the highest variability, driven by the systems with the highest veilings, which are also the most variable ones (see Fig. \ref{fig:rmsvsveil}). 

 The average veiling measured in RU Lup is relatively high compared to the other targets, mainly in the $K$ band. These high average veilings are primarily due to the veiling measured in the 2021a period of observations (see Sect. \ref{sec:variability}). In that period, RU Lup also presented an increase in the photometric AAVSO $V$ band brightness (about $0.3$\,mag smaller than the next observational period, where the veiling starts to go down). Therefore, the average veiling of RU Lup probably is overestimated and does not represent a quiescent value.
 
\subsection{Veiling compared to inner disk diagnostics}
\begin{figure}
\centering
{\includegraphics[width=4.4cm]{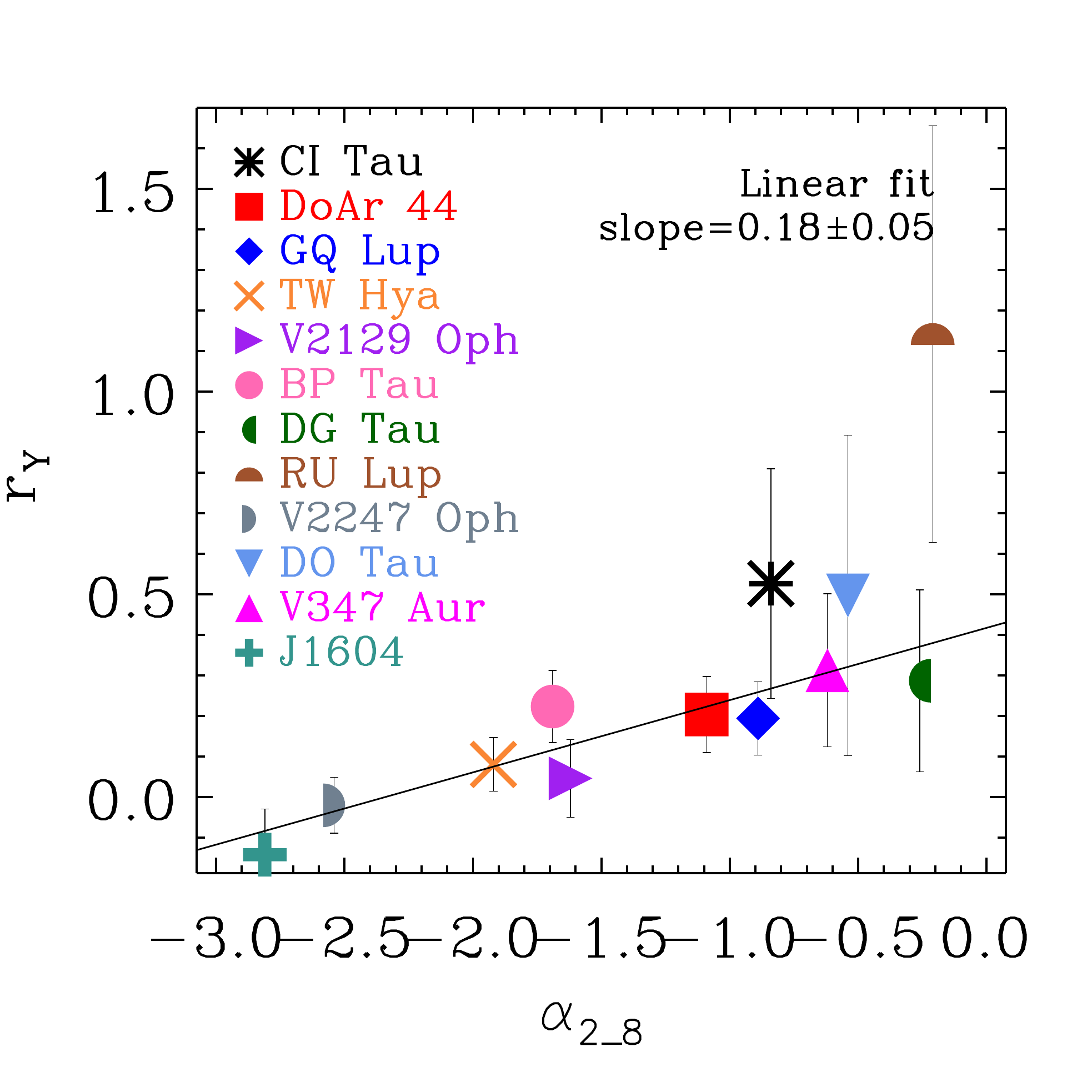}}
{\includegraphics[width=4.4cm]{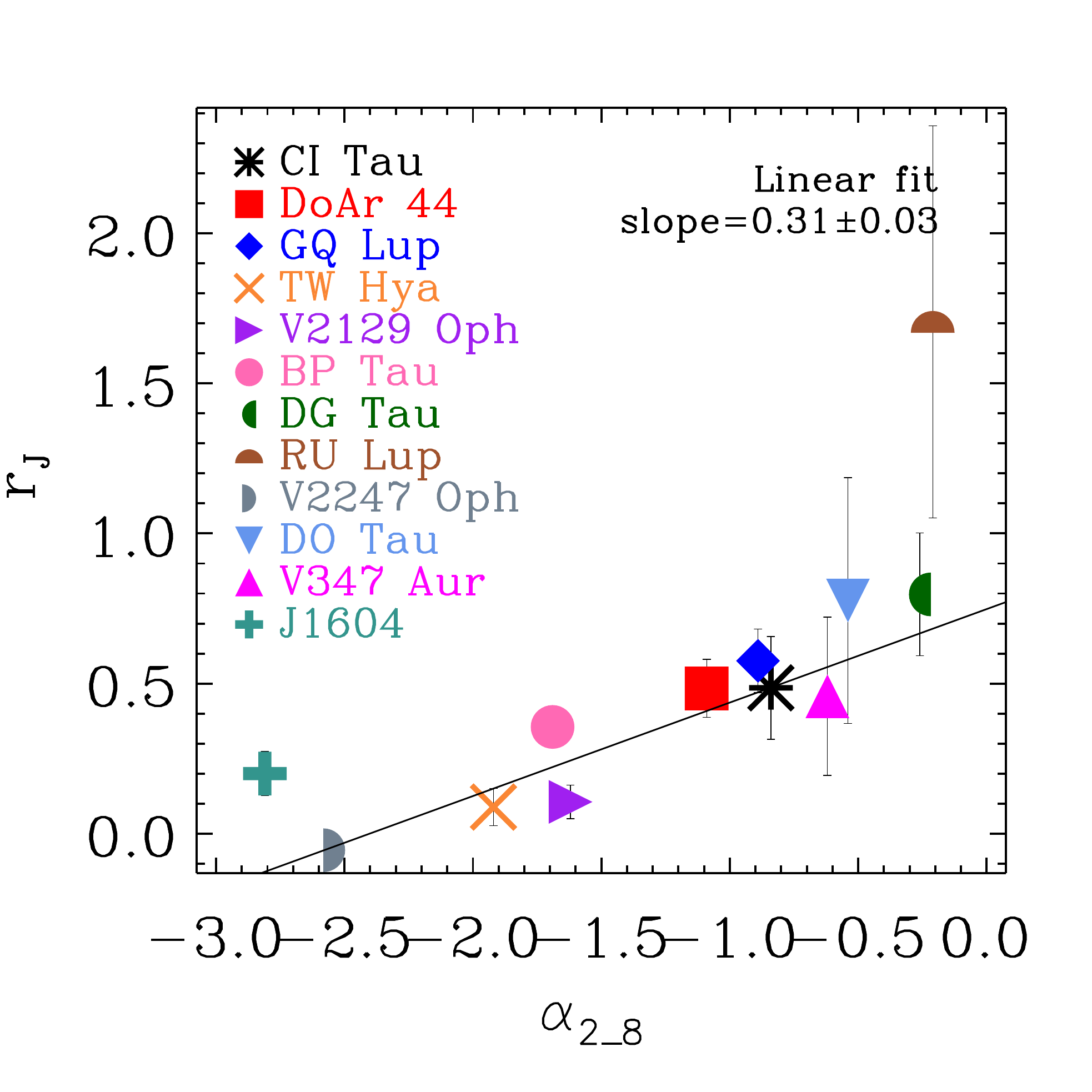}}\\
{\includegraphics[width=4.4cm]{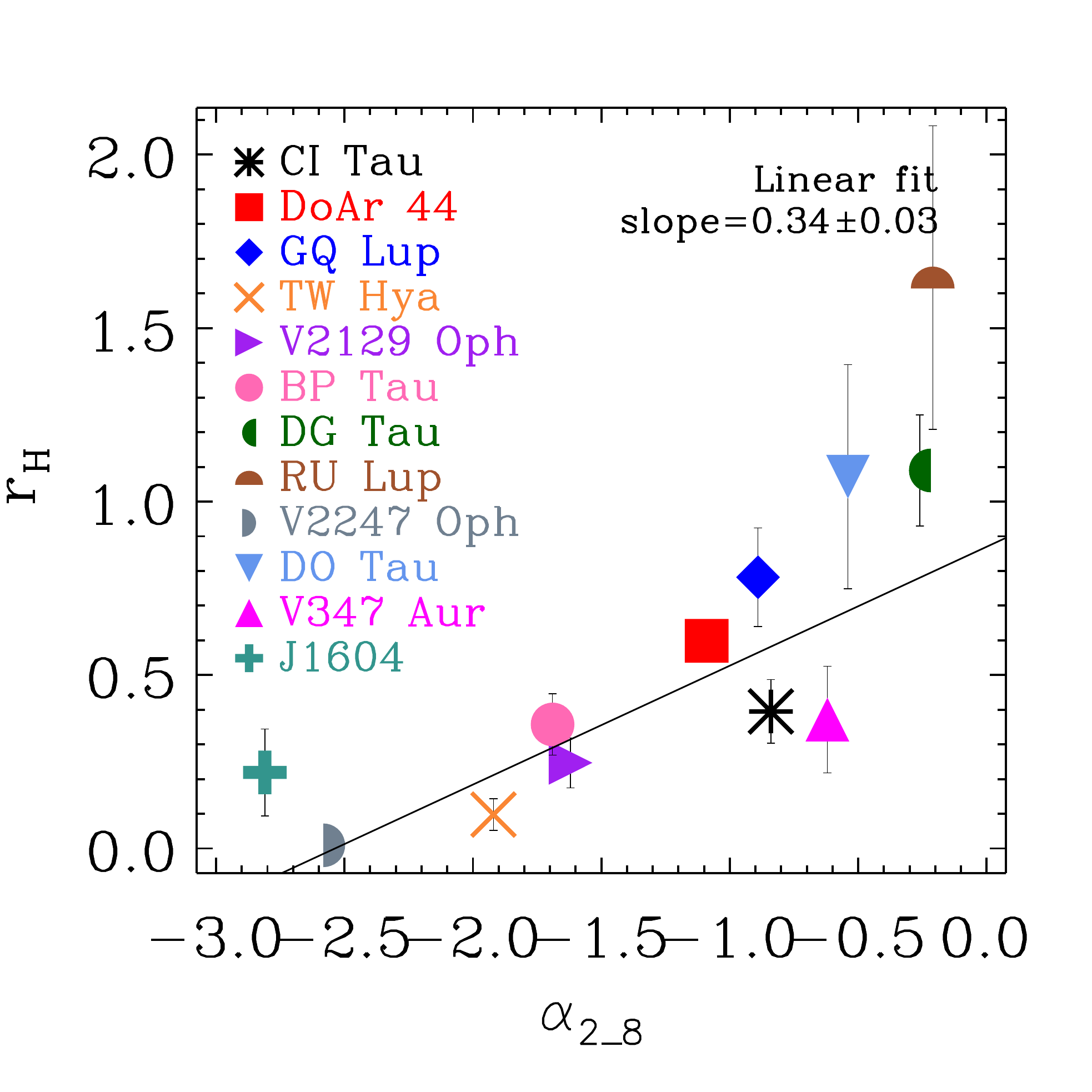}}
{\includegraphics[width=4.4cm]{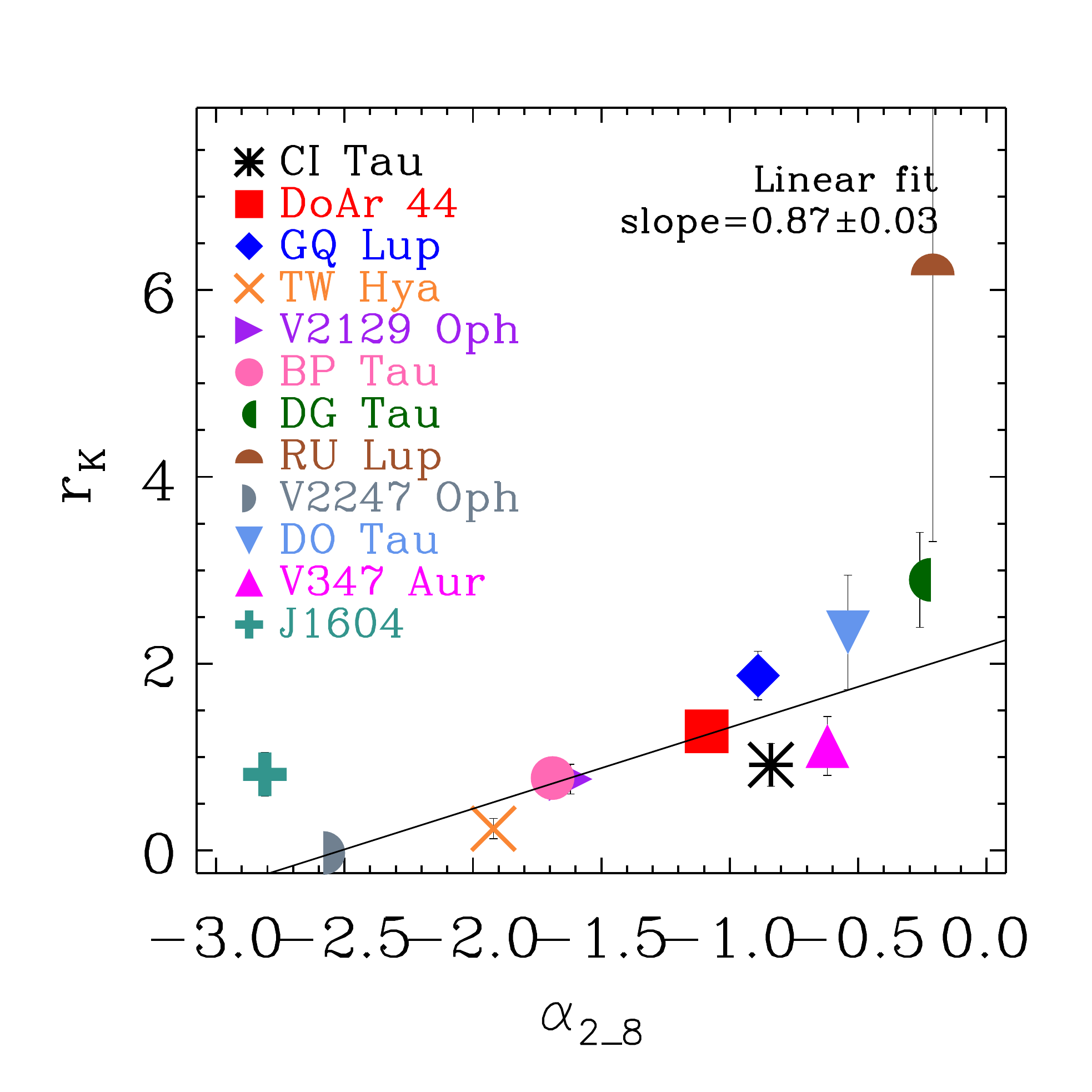}}
\caption{\label{fig:veilvsphot} Average NIR veiling  ($r_\lambda$) as a function of  spectral energy distribution slope between 2 and 8\,\mum\  ($\alpha_{2\_8}$), which we used as the NIR disk emission diagnostic. The solid line is the linear fit to the data, and the correspondent slope is written in each panel. The NIR veiling seems to scale with the SED slope.} 
\end{figure}

The emission from the inner disk is claimed to contribute to the veiling. In such a case, we would expect a correlation between the veiling and other inner disk emission diagnostics from the photometric data. 

The slope of the spectral energy distribution (SED) is often used as a disk emission diagnostic \citep[e.g.,][]{Lada2006,2010ApJ...708.1107M,2012A&A...540A..83T}. Using photometric data from different surveys, such as SDSS \citep{1998AJ....116.3040G}, Gaia DR2 \citep{2018A&A...616A...1G}, 2MASS, WISE \citep{2010AJ....140.1868W}, Spitzer \citep{2004ApJS..154...10F,2004ApJS..154...25R}, Herschel (PACS), Akari (IRC and FIS), we constructed the SED of the targets and measured the slope of the SED between 2 and 8\,\mum\  ($\alpha_{2-8}$), which is the spectral range that indicates significant inner disk emission. The $\alpha_{2-8}$ slope is smaller (more negative) for systems with less or no inner disk emission, and is higher (and even positive) for systems that present inner disk emission excess \citep[e.g.,][]{Lada2006,2010ApJ...708.1107M}.   We list the $\alpha_{2-8}$ slope computed for the targets in Table \ref{tab:param}.

We show in Fig. \ref{fig:veilvsphot} the NIR veiling in the four spectral regions ($YJHK$), averaged over all the observation nights, as a function of   $\alpha_{2-8}$. The veiling presents a clear linear correlation with the SED slope, mainly from the $J$ to $K$ band. 
The $Y$ band veiling values are still correlated to the SED slope but less than the other bands, probably due to the contribution from the inner disk emission becoming more important at longer wavelength. The $W_1-W_2$ ([3.4]-[4.6]) color index (not shown here) comes from the WISE telescope \citep{2010AJ....140.1868W} is also correlated with the NIR veiling, presenting similar results as $\alpha_{2-8}$.

\begin{figure}
\centering
{\includegraphics[width=4.3cm]{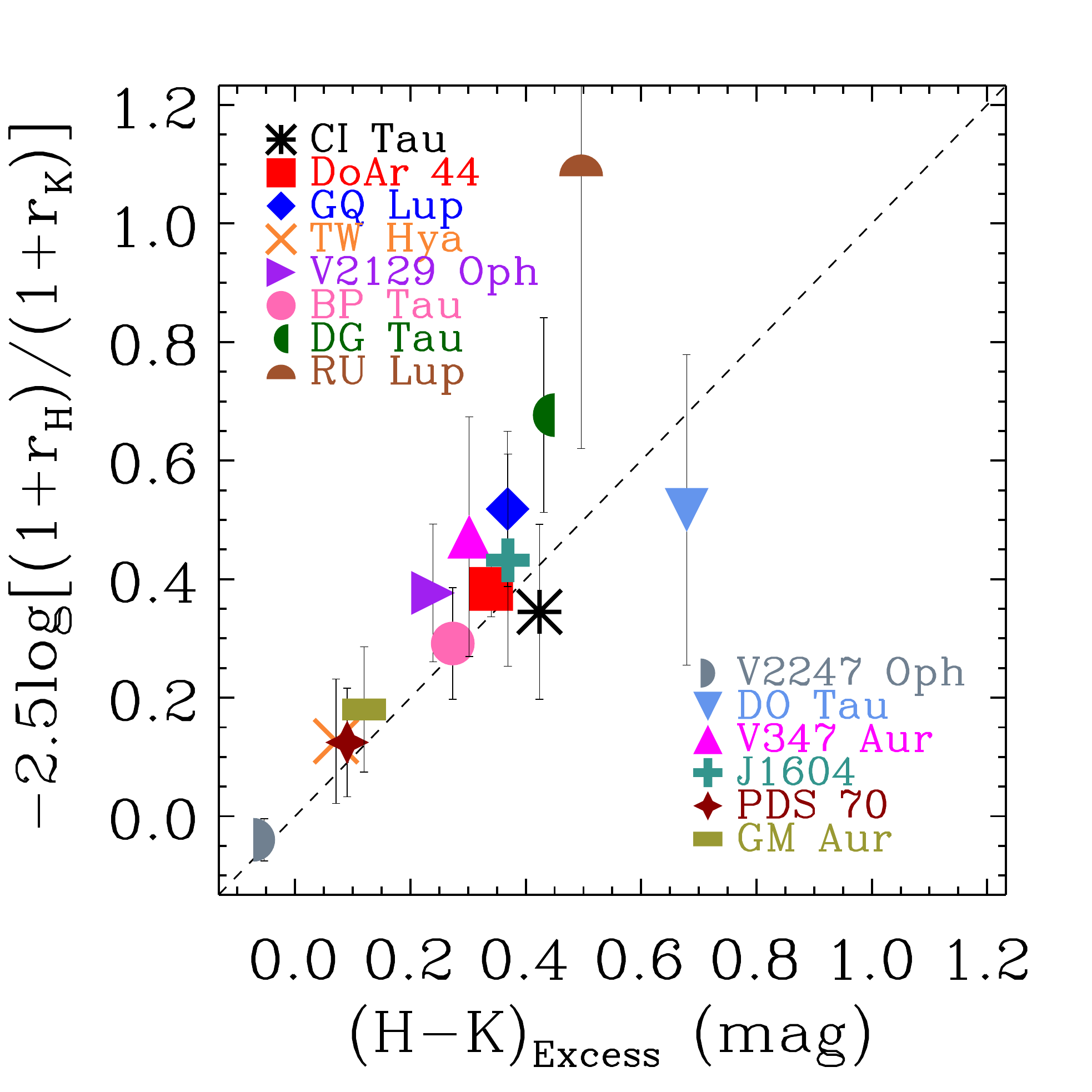}}
{\includegraphics[width=4.3cm]{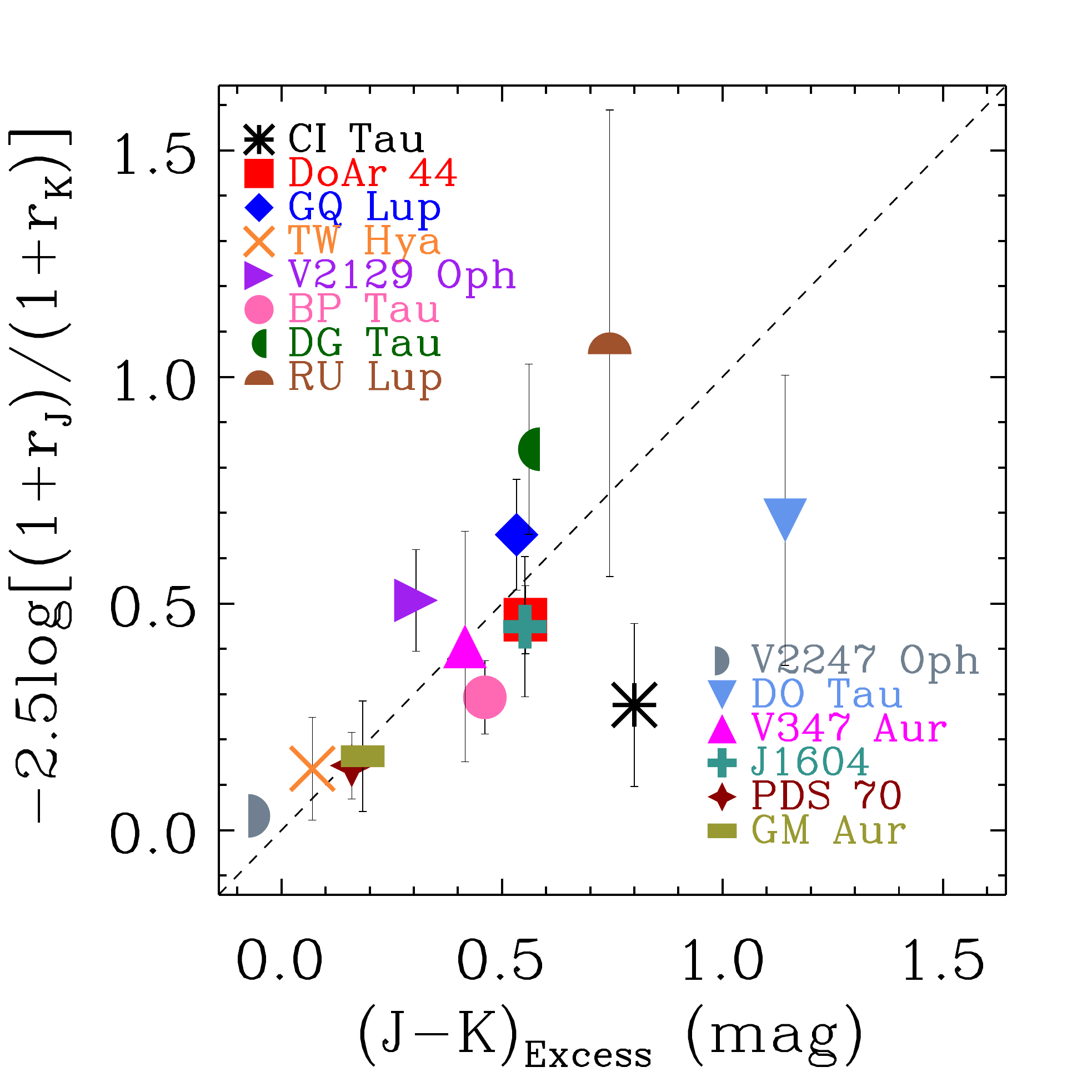}}
\caption{\label{fig:veilvsphotexc} Comparison between the color excess computed using the average NIR veiling and the 2MASS photometry. {\it left}: $(H-K_{\rm s})_{excess}$, {\it right}: $(J-K_{\rm s})_{excess}$. See the text for the color excess definition. The dashed line represents a slope equal to 1. The NIR color excesses computed using the average veiling and the 2MASS magnitudes agree for most targets.} 
\end{figure}

We could expect that the NIR excess computed using the veiling would scale with the emission excess from NIR photometric data, which is often used as an inner disk emission indicator \citep{1998AJ....116.1816H,2001AJ....121.1676R,2002AJ....123.1528R}. We computed the $(H - K_{\rm s})_{excess}$, using the observed $H-K_{\rm s}$ color from 2MASS, corrected for extinction, using the $A_V$ quoted in Table \ref{tab:obs} and the SVO Filter \citep{2012ivoa.rept.1015R,2020sea..confE.182R} $A_\lambda/A_V$ relations to obtain the $A_H$ and $A_K$ extinctions. Then, we compared this dereddened color excess to the intrinsic color $(H-K_{\rm s})_o$ expected for an object with the same spectral type \citep{2013ApJS..208....9P}. The color excess is $(H-K_{\rm s})_{excess}= (H-K_{\rm s})_{obs,dred} - (H-K_{\rm s})_o$. We also relate the color excess and the excess flux, leading to $(H-K_{\rm s})_{excess}=-2.5 \log{[(1+r_H)/(1+r_K)]}$, where we used the veiling definition as $r_\lambda=F_{\lambda_{excess}}/F_{\lambda_{Phot}}$. Then we can directly compare the color excess computed using the veiling and using the photometric measurements. We compare the two sides of this equation in Fig. \ref{fig:veilvsphotexc}, which shows a linear tendency and similar values considering the error of the measurements. 
Performing similar procedures, we computed the $(J-K_{\rm s})_{excess}$, and the results are also presented in Fig. \ref{fig:veilvsphotexc}. 
 The color excess computed using the 2MASS photometry is dependent on the $A_v$ of the systems, and the targets V347 Aur and CI Tau present discrepant $A_v$ values in the literature. CI Tau presents $A_v=0.65\,$mag \citep{2020MNRAS.491.5660D} and $A_V=1.90\,$mag \citep{2014ApJ...786...97H}, while the $(J-K_{\rm s})_{excess}$ computed using the largest $A_V$ better agrees with the color excess calculated using veiling, the $(H-K_{\rm s})_{excess}$, and the mass accretion rate computed in the next section seem to be in better agreement with $A_V=0.65\,$mag; thus, we used this extinction value in the paper. The $A_V$ range of V347 Aur is even larger, with values from $2.2$ to $7.5\,$mag  \citep[e.g.,][]{2020AJ....160..278D}. The $A_V=3.4\,$mag computed using NIR colors by \cite{2010AJ....140.1214C} seems to better represent the value obtained from the veiling calculations. 
 
The 2MASS photometric magnitudes used for this analysis were obtained years before our observations. However, we do not expect a significant change in the NIR magnitudes over the next few years, aside from the daily timescale variation \citep[e.g.,][]{2001AJ....121.3160C}. It is likely that  RU Lup will stand as an exception, as the average veiling was measured in a non-quiescent period. Then, the color excess computed using the veiling is higher than that obtained using the 2MASS magnitudes.

All these relations between the NIR veiling and the inner disk emission diagnostics show that a higher veiled system also presents higher inner disk emission, which is expected if the veiling has a contribution from the inner disk.  However,  to draw this conclusion, we assumed that these inner disk indicators, such as the slope of the spectral energy distribution, are suitable inner disk diagnostics. \cite{2021ApJ...922...27K} showed that some of the targets classified as Class III using these disk indicators still show some inner disk emission based on the $K$ band excess. They checked the emission excess of V819 Tau, which we used as a template to compute the veiling,  but the $H$ and $K$ excesses found were very small, similar to the systematic veiling we obtained in Sect. \ref{sec:method}.
The relation between veiling and inner disk emission is clear for the $K$ band and less for other bands, probably due to the influence of another additional continuum source for the other spectral regions and/or a smaller contribution from the inner disk to these shorter wavelengths.

\subsection{Veiling compared to accretion diagnostics}

\begin{figure*}
\centering
{\includegraphics[width=6.cm]{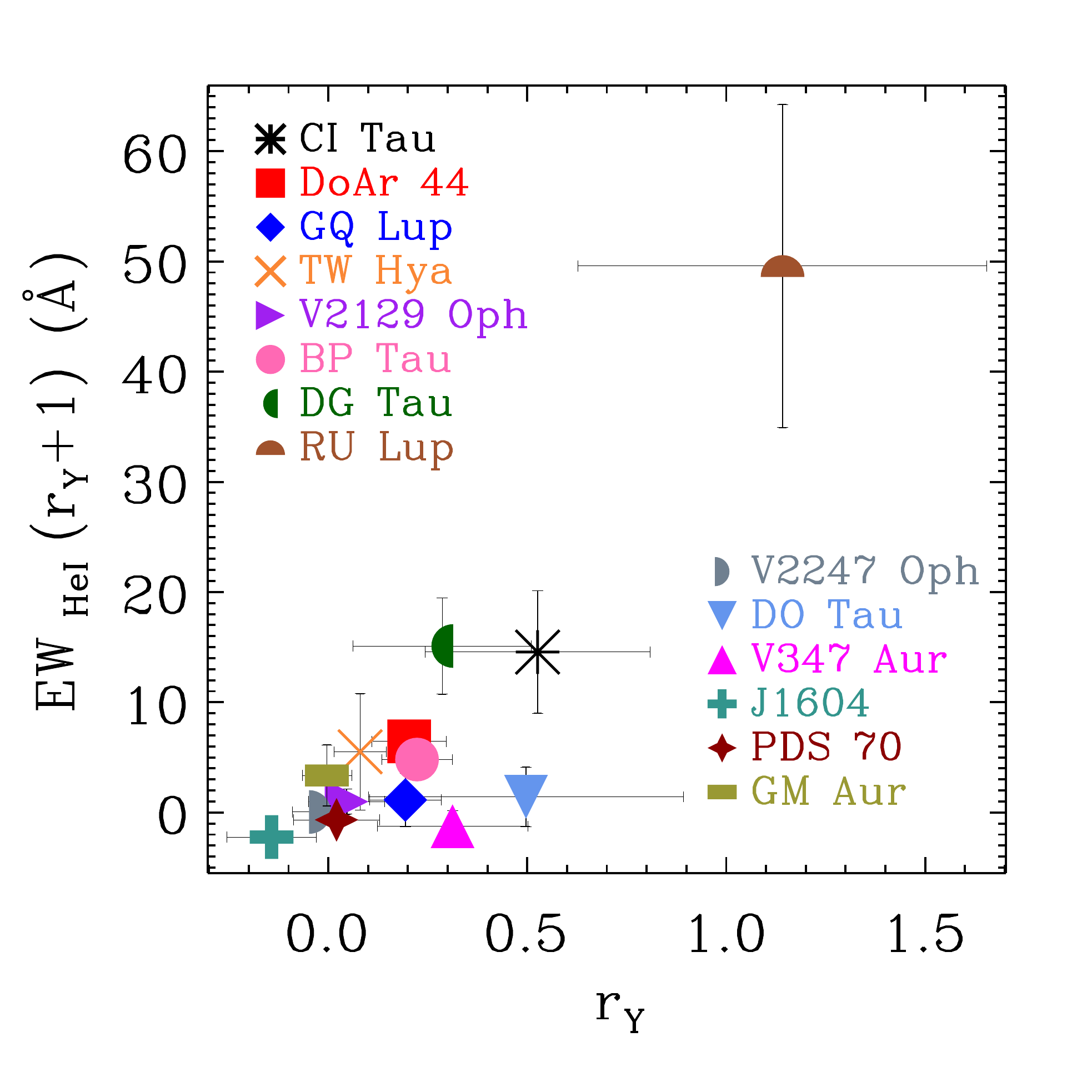}}
{\includegraphics[width=6.cm]{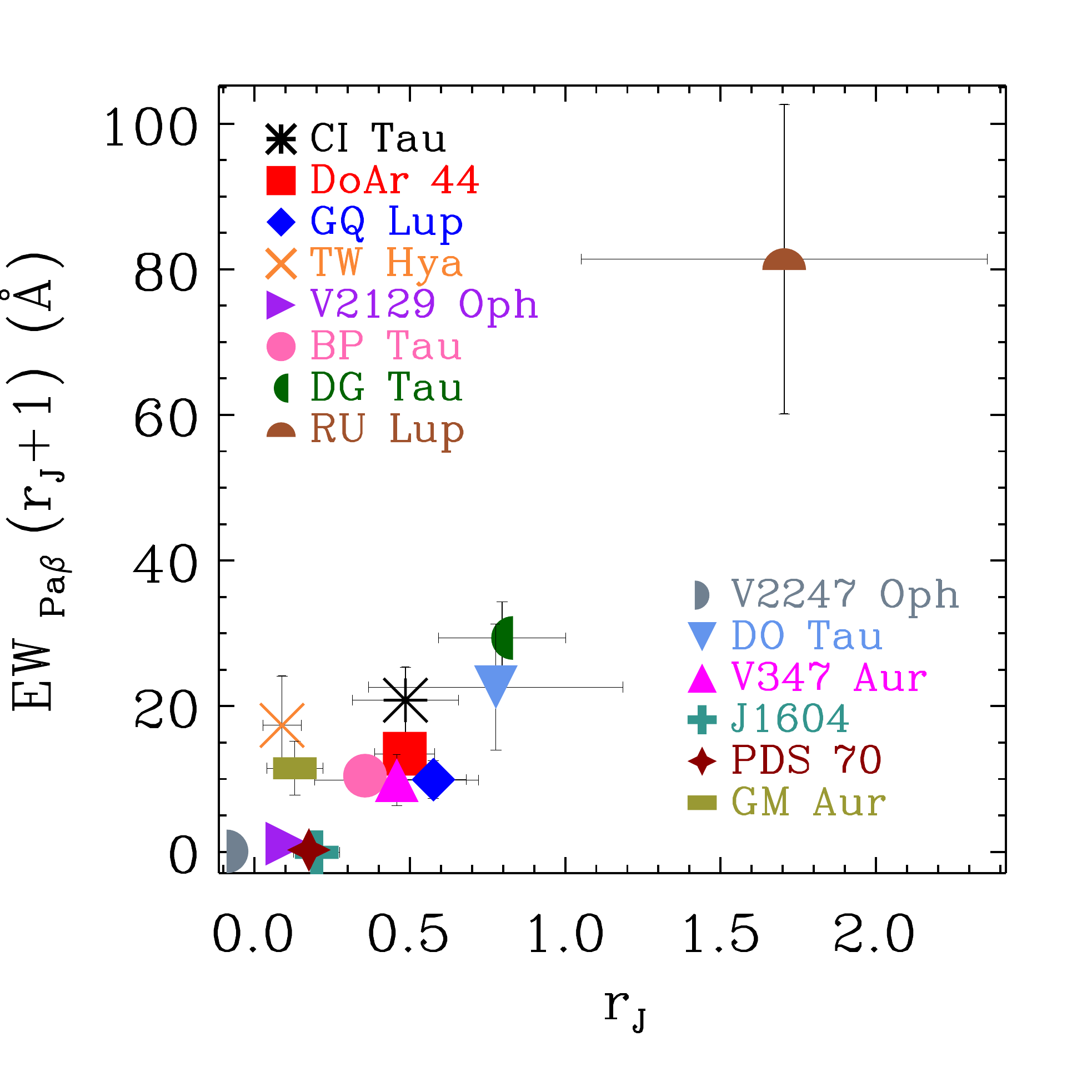}}
{\includegraphics[width=6.cm]{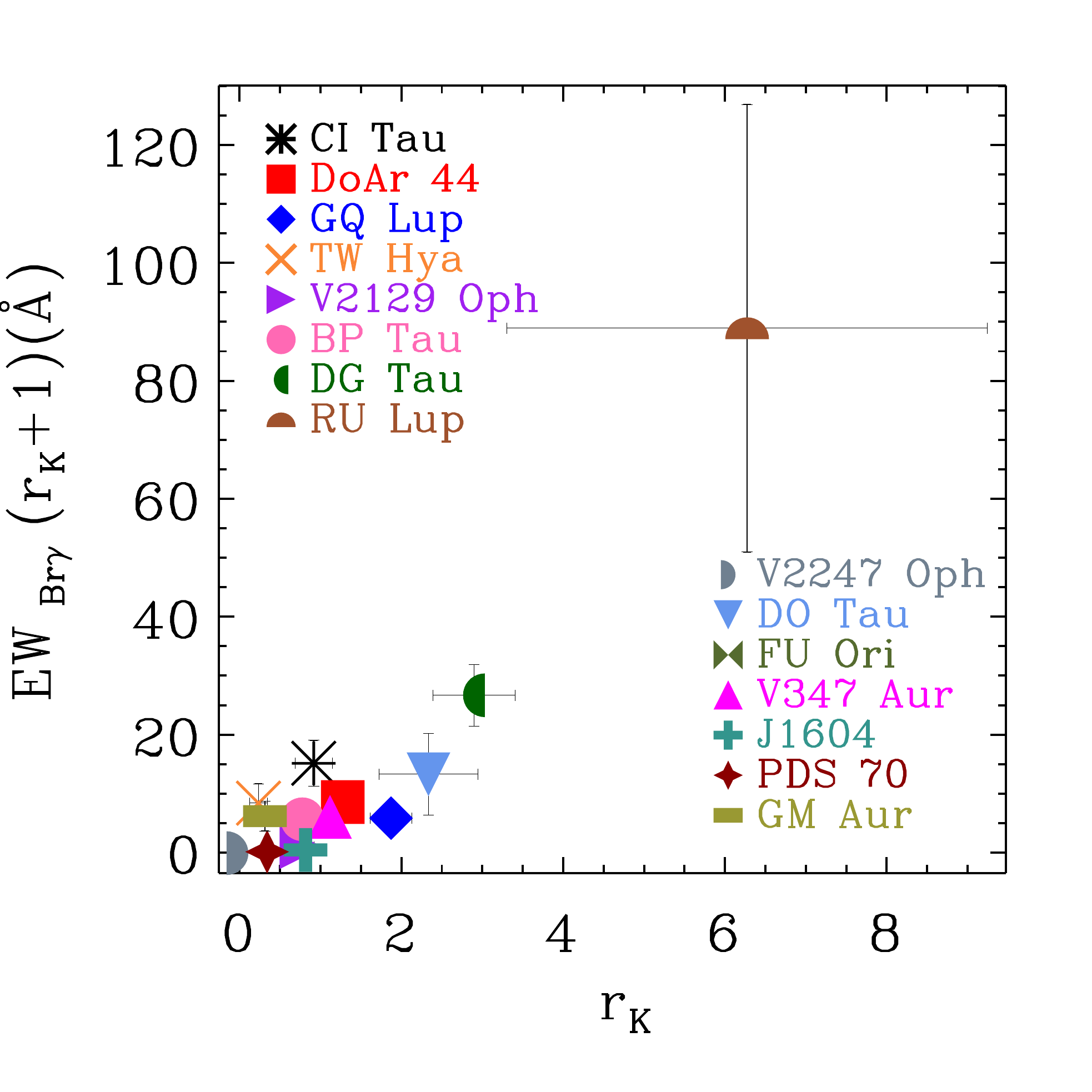}}
\caption{\label{fig:veilEW} Accretion diagnostics as a function of average NIR veiling. From left to right:\ Average equivalent width corrected from the veiling of \hei,\   \pab\ ,  and \brg \ lines. We show a connection between the system's accretion diagnostic and the NIR veiling.}
\end{figure*}

We know that CTTSs are still accreting gas from the disk and accreting systems typically present strong and variable emission lines that form in the accretion funnel or in the disk wind \citep[e.g.,][]{1998AJ....116..455M,2003ApJ...582.1109W,edwards03,2011MNRAS.411.2383K,2012AA...541A.116A}.
The CFHT/SPIRou wavelength range includes some emission lines from hydrogen and helium, such as \pab\ and \brg\ as well as the \hei\ triplet; in particular, the latter is very sensitive to accretion and ejection processes \citep[e.g.,][]{2007ApJ...657..897K,2021A&A...649A..68S}. The dynamics of the circumstellar lines for this sample of stars will be analyzed in an accompanying paper (Sousa et al. in prep.).

We measured the equivalent width of the circumstellar lines, and the average over all observing nights is listed in Table \ref{tab:param}. First, we used the equivalent width as an accretion diagnostic \citep{2017...600A..20A}, as systems that present larger equivalent widths are supposed to present higher mass accretion rates as well. 

\begin{table*}[htb!]
\tiny
\addtolength{\tabcolsep}{-3.0pt}  
\caption{\label{tab:param} Parameters of the targets derived in this work. }
\begin{center}
\begin{tabular}{lllllllllllll}
  \hline\hline 
Star & Std &$\mathrm{v_{rad}}$  & $\mathrm{r_Y}$ &  $\mathrm{r_J}$ & $\mathrm{r_H}$  & $\mathrm{r_K}$ & $\mathrm{EW}_{\mathrm{HeI}\ }$\tablefootmark{a} & $\mathrm{EW}_{\mathrm{Pa\beta}}$\tablefootmark{a} & $\mathrm{EW}_{\mathrm{Br\gamma}}$\tablefootmark{a} & $\mathrm{\dot M_{\mathrm{Pa\beta}}}$ &$\mathrm{\dot M_{\mathrm{Br\gamma}}}$ & $\alpha_{2\_8}$\tablefootmark{b}\\
& &  & & & & &  & & & $\times10^{-8}$ & $\times10^{-8}$  & \\

& &($\mathrm{km}\ \mathrm{s}^{-1})$  & & & & & $(\mathring{\mathrm{A}})$ & $(\mathring{\mathrm{A}})$ & $(\mathring{\mathrm{A}})$ & $(M_\odot\mathrm{yr}^{-1})$ &$(M_\odot\mathrm{yr}^{-1})$&  \\
\hline
CI Tau  &   V819TAU  &   16.3  $\pm$    0.5  &   0.53  $\pm$  0.28  &   0.49  $\pm$  0.17  &  0.39  $\pm$  0.09  &   0.92  $\pm$  0.23  &    9.5  $\pm$   3.2  &   14.0  $\pm$   2.6  &    7.9  $\pm$   1.8  &   2.2  &   4.2  &  -0.84  \\
DoAr 44  &   V819TAU  &   -6.1  $\pm$    0.7  &   0.20  $\pm$  0.09  &   0.48  $\pm$  0.10  &  0.60  $\pm$  0.06  &   1.28  $\pm$  0.05  &    5.4  $\pm$   1.4  &    9.1  $\pm$   0.8  &    3.8  $\pm$   0.5  &   1.7  &   1.5  &  -1.09  \\
GQ Lup  &     TWA9A  &   -3.0  $\pm$    0.3  &   0.19  $\pm$  0.09  &   0.58  $\pm$  0.11  &  0.78  $\pm$  0.14  &   1.87  $\pm$  0.26  &    0.9  $\pm$   2.0  &    6.3  $\pm$   1.6  &    2.0  $\pm$   0.7  &   2.5  &   1.9  &  -0.89  \\
TW Hya  &     TWA25  &   12.4  $\pm$    0.1  &   0.08  $\pm$  0.07  &   0.09  $\pm$  0.06  &  0.10  $\pm$  0.05  &   0.23  $\pm$  0.11  &    5.1  $\pm$   4.9  &   15.9  $\pm$   6.2  &    6.8  $\pm$   2.6  &   0.6  &   0.3  &  -1.92  \\
V2129 Oph  &   V819TAU  &   -7.1  $\pm$    0.6  &   0.05  $\pm$  0.10  &   0.11  $\pm$  0.06  &  0.25  $\pm$  0.07  &   0.76  $\pm$  0.16  &    0.9  $\pm$   1.1  &    1.0  $\pm$   0.7  &    0.5  $\pm$   0.3  &   0.2  &   0.1  &  -1.62  \\
 BP Tau  &   V819TAU  &   15.2  $\pm$    0.6  &   0.22  $\pm$  0.09  &   0.36  $\pm$  0.06  &  0.36  $\pm$  0.09  &   0.78  $\pm$  0.10  &    3.9  $\pm$   1.4  &    7.7  $\pm$   1.4  &    3.2  $\pm$   0.8  &   1.3   &   1.1  &  -1.69  \\
DG Tau  &     TWA9A  &   15.1  $\pm$    3.4  &   0.29  $\pm$  0.22  &   0.80  $\pm$  0.20  &  1.09  $\pm$  0.16  &   2.90  $\pm$  0.51  &   11.7  $\pm$   2.7  &   16.3  $\pm$   2.1  &    6.8  $\pm$   1.0  &   6.4  &   7.6  &  -0.26  \\
RU Lup  &     TWA9A  &   -1.3  $\pm$    1.4  &   1.14  $\pm$  0.51  &   1.70  $\pm$  0.65  &  1.65  $\pm$  0.44  &   6.27  $\pm$  2.97  &   23.2  $\pm$   4.0  &   30.1  $\pm$   3.0  &   12.2  $\pm$   1.5  &   7.8  &  11.9  &  -0.21  \\
V2247 Oph  &     TWA25  &   -5.8  $\pm$    0.5  &  -0.02  $\pm$  0.07  &  -0.06  $\pm$  0.01  &  0.01  $\pm$  0.03  &  -0.03  $\pm$  0.02  &    0.0  $\pm$   0.2  &    0.1  $\pm$   0.1  &   -0.1  $\pm$   0.2  &   0.009  &    -  &  -2.54  \\
DO Tau  &     TWA25  &   16.1  $\pm$    1.0  &   0.50  $\pm$  0.40  &   0.78  $\pm$  0.41  &  1.07  $\pm$  0.32  &   2.33  $\pm$  0.61  &    1.0  $\pm$   1.8  &   12.7  $\pm$   3.9  &    4.0  $\pm$   1.9  &   2.3  &   3.4  &  -0.54  \\
 V347 Aur  &     TWA25  &    8.1  $\pm$    0.6  &   0.31  $\pm$  0.19  &   0.46  $\pm$  0.26  &  0.37  $\pm$  0.15  &   1.12  $\pm$  0.31  &   -0.9  $\pm$   1.1  &    6.7  $\pm$   2.1  &    2.9  $\pm$   1.0  &   7.1  &   7.3  &  -0.62  \\
J1604  &   V819TAU  &   -5.8  $\pm$    0.8  &  -0.14  $\pm$  0.11  &   0.20  $\pm$  0.07  &  0.22  $\pm$  0.13  &   0.81  $\pm$  0.23  &   -2.6  $\pm$   1.8  &   -0.1  $\pm$   0.2  &    0.3  $\pm$   0.1  &     -  &   0.013  &  -2.81  \\
PDS 70  &     TWA9A  &    5.0  $\pm$    0.4  &   0.02  $\pm$  0.11  &   0.18  $\pm$  0.04  &  0.19  $\pm$  0.07  &   0.34  $\pm$  0.08  &   -0.7  $\pm$   0.9  &    0.2  $\pm$   0.2  &    0.1  $\pm$   0.2  &   0.007  &   0.003  &  -  \\
GM Aur  &     TWA9A  &   14.5  $\pm$    0.3  &  -0.00  $\pm$  0.06  &   0.13  $\pm$  0.09  &  0.11  $\pm$  0.06  &   0.31  $\pm$  0.10  &    3.4  $\pm$   2.8  &   10.2  $\pm$   3.2  &    4.7  $\pm$   1.9  &   1.3  &   1.0  &  -  \\
\hline
\end{tabular}
\end{center}
 \tablefoot{
  \tablefoottext{a}{We used the convention of positive equivalent width for emission lines, and negative values for absorption lines.}
  \tablefoottext{b}{ Slope of the spectral energy distribution measured between 2 and 8\,\mum. }
  }
 \end{table*}

We corrected the equivalent width for the veiling as $EW=EW_{measured}(r_\lambda+1)$, where the $r_\lambda$ represents the veiling computed close to each emission line.
In Fig. \ref{fig:veilEW}, we show the veiling as a function of the veiling corrected equivalent width of the circumstellar emission lines. We see a clear relationship between the NIR veiling and the accretion diagnostics. It means that higher mass-accretion rate systems also present a higher degree of veiling; this is a similar result to that found by \cite{1999A&A...352..517F}, demonstrating that although the veiling shows a contribution from the inner disk emission, it also suggests a connection with the accretion process.

We do not have photometric data simultaneous with our spectra to accurately compute the mass accretion rates using the equivalent width of the emission lines. However, most of our systems' NIR magnitudes are relatively long-term stable. We used the 2MASS $J$ and $K$ magnitudes to estimate the continuum flux and then calculate the mass accretion rate using the \pab\ and \brg\ lines, respectively. The star V347 Aur is known to present long-term photometric variations \citep[e.g.,][]{2020AJ....160..278D}, and we did not compute the mass accretion rate of this target. 

\begin{figure}
\centering
{\includegraphics[width=4.3cm]{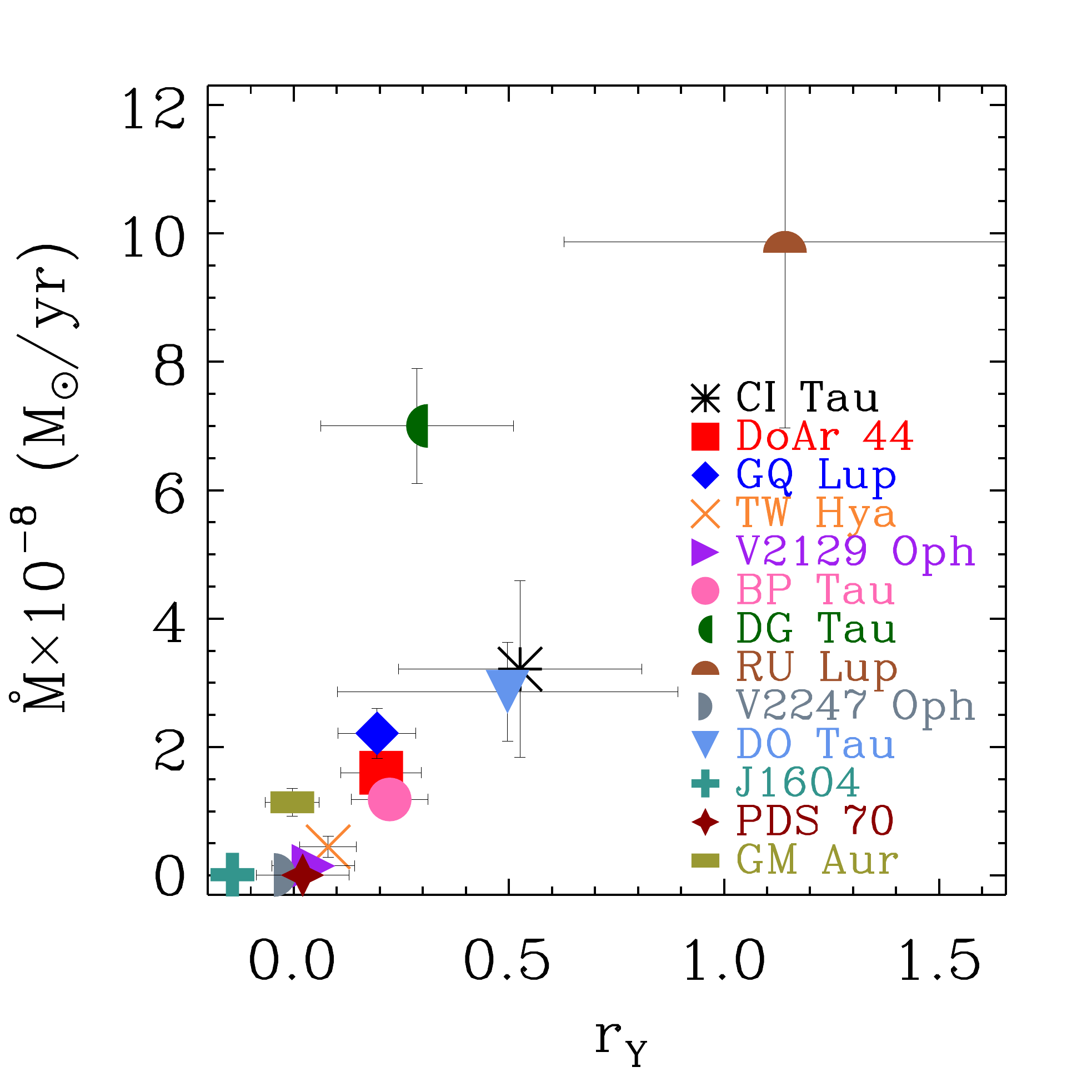}}
{\includegraphics[width=4.3cm]{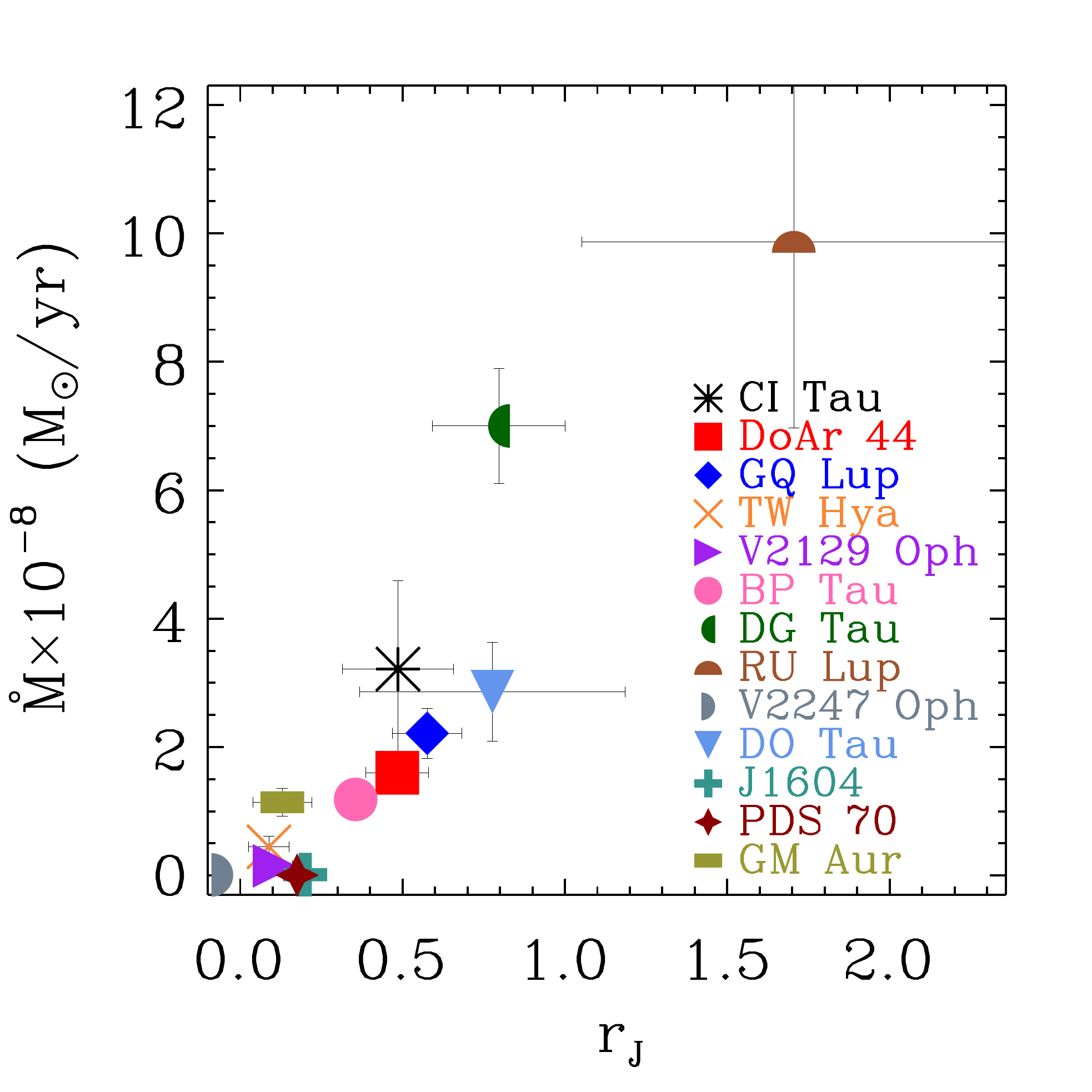}}\\
{\includegraphics[width=4.3cm]{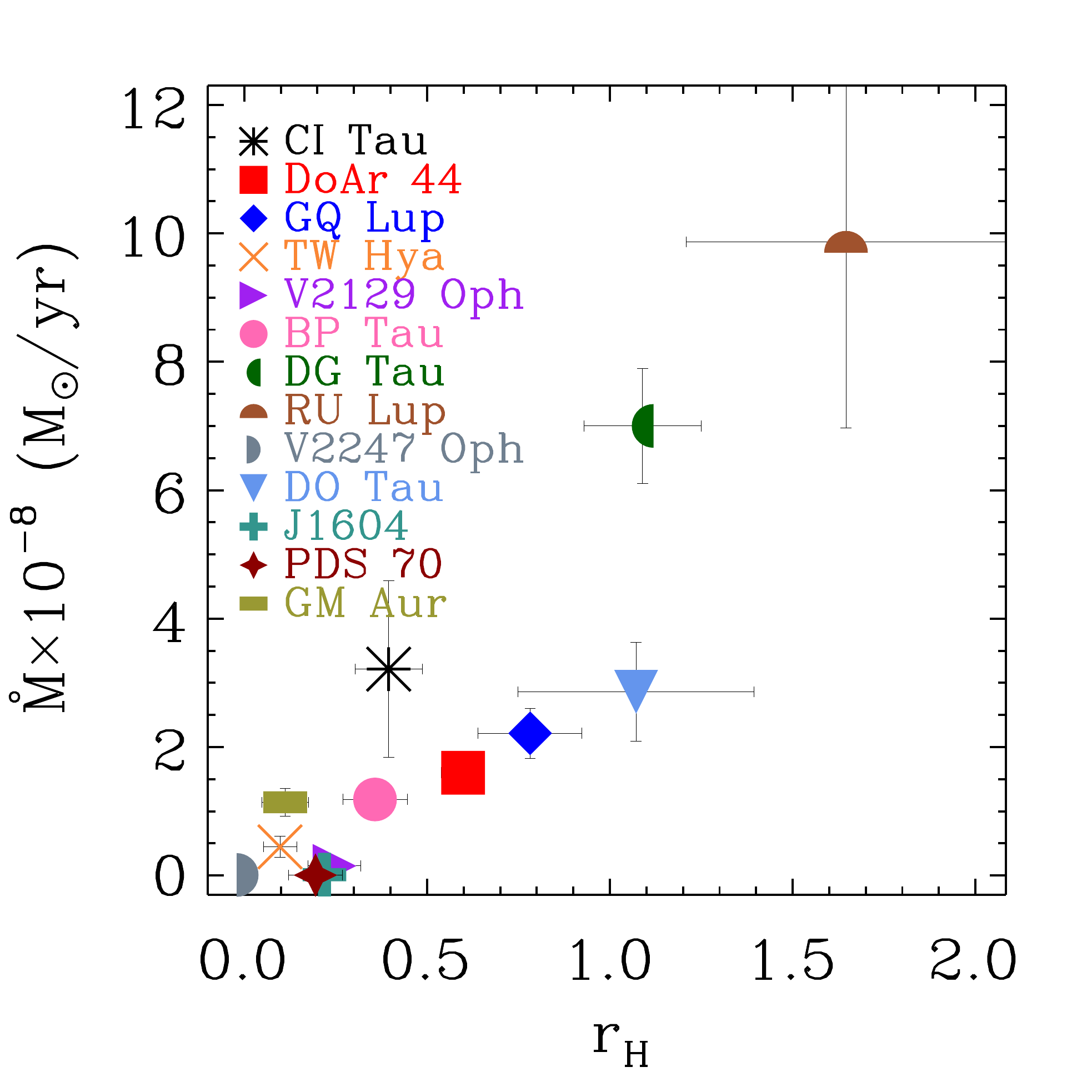}}
{\includegraphics[width=4.3cm]{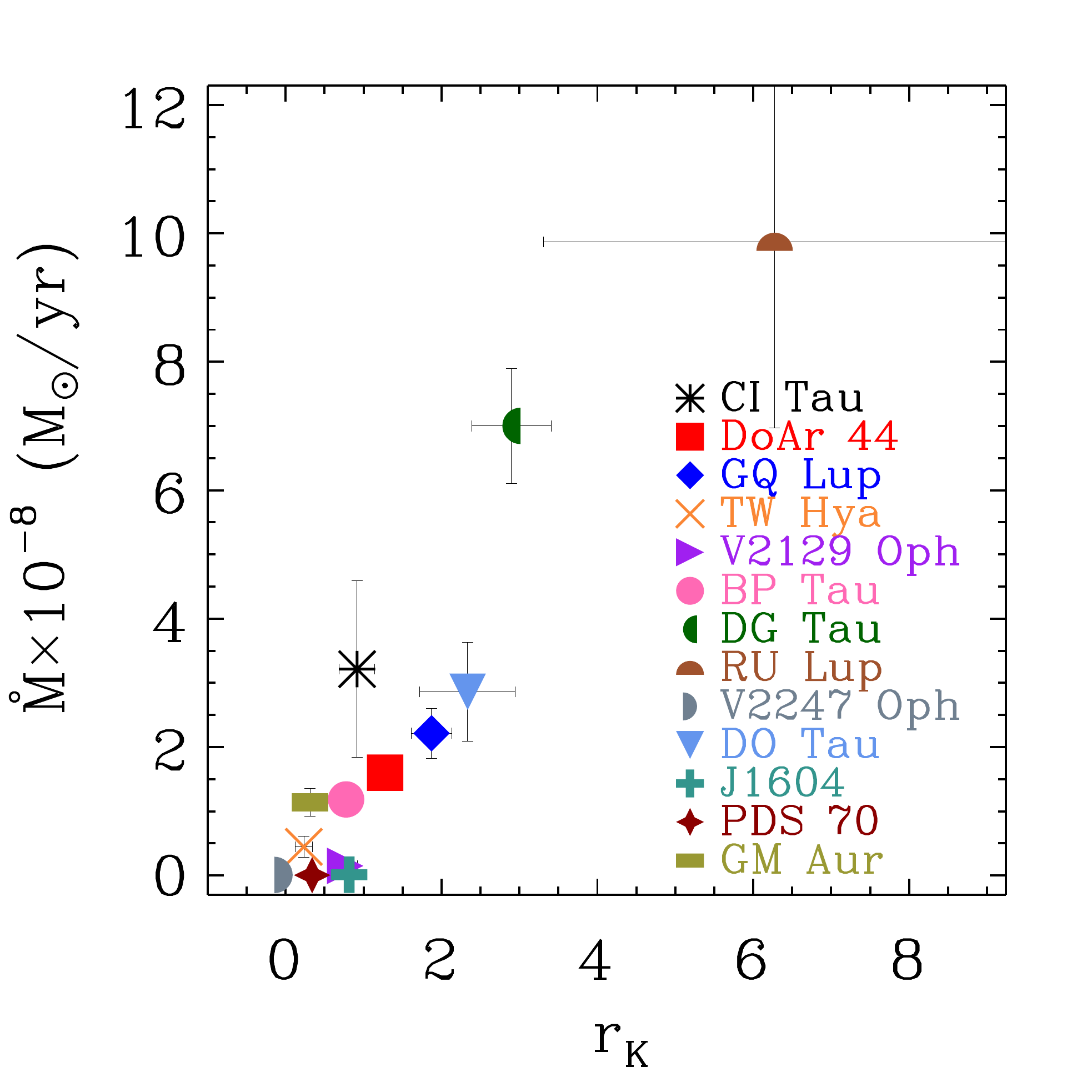}}
\caption{\label{fig:veilAccrate} Average mass accretion rate as a function of average NIR veiling. The average mass accretion was computed using the line fluxes of \pab\ and \brg. The error bar is the standard deviation between the two measurements. We see an association between the veiling and the mass accretion rate.} 
\end{figure}

We followed the procedures described by \cite{1998ApJ...492..323G} to compute the line flux and luminosity. The stellar parameters used are listed in Table \ref{tab:stellarparam}, and we used the stellar distance from the Gaia collaboration \citep{2021A&A...649A...1G}. We dereddened the 2MASS magnitudes using the same method described in the previous section. To compute the accretion luminosity, we used the fits proposed by \cite{2017...600A..20A}, which show the relation between the line and accretion luminosities. Then, we determined the accretion rate setting the system inner radius as $5R_\ast$ \citep{1998ApJ...492..323G}. 
In Table \ref{tab:param}, we show the individual mass accretion rates computed using the \pab\ and \brg\ lines. In Fig. \ref{fig:veilAccrate}, we show the average mass accretion rate as a function of the $Y$ to $K$ band veiling. Once again, we can connect the highest accreting system with the highest NIR veiling computed in the four bands.

\begin{table}[htb!]
\tiny
\caption{\label{tab:stellarparam} Stellar parameters}
\begin{center}
\begin{tabular}{llll}
  \hline\hline 
Star &M$_\ast$ &R$_\ast$ & ref   \\

         &  (M$_\sun$)    & (R$_\sun$)   & \\
\hline
CI Tau    &     0.90 &       2.0  & (1)\\
DoAr 44   &     1.20 &       2.0  & (2) \\ 
GQ Lup    &     0.86 &       2.26 & (3) \\ 
TW Hya    &     0.80 &       1.1  & (4) \\ 
V2129 Oph &     1.35 &       2.1  & (5) \\ 
BP Tau    &     0.70 &       1.99 & (6)\\ 
DG Tau    &     0.65 &       2.05 & (6)\\ 
RU Lup    &     1.15 &       2.39 & (7)\\ 
V2247 Oph &     0.35 &       2.0  & (8) \\ 
DO Tau    &     0.60 &       2.0  & (9) \\ 
J1604    &     1.24 &       1.4  & (10) \\ 
PDS 70   &     0.76 &       1.26 & (11) \\
GM Aur   &     0.95 &       1.71 & (12) \\    

\hline
\end{tabular}
\end{center}
\tablebib{(1)~ \cite{2020MNRAS.491.5660D};
(2) \cite{2020...643A..99B}; (3) \cite{2017...600A..20A}; (4) \cite{2011MNRAS.417..472D}; (5) \cite{2012AA...541A.116A}; (6) \citet{2007ApJ...664..975J}; (7) \citet{2014AA...561A...2A}; (8) \citet{2010MNRAS.402.1426D}; (9) \citet{2010A&A...512A..15R}; (10) \citet{2020AA...633A..37S}; (11) \citet{2018AA...617L...2M}; (12) Bouvier et. al. (in prep.) 
}
 
 \end{table}

\subsection{Veiling night-to-night variability} \label{sec:variability}

\begin{figure*}[htb]
  {\includegraphics[width=4.4cm]{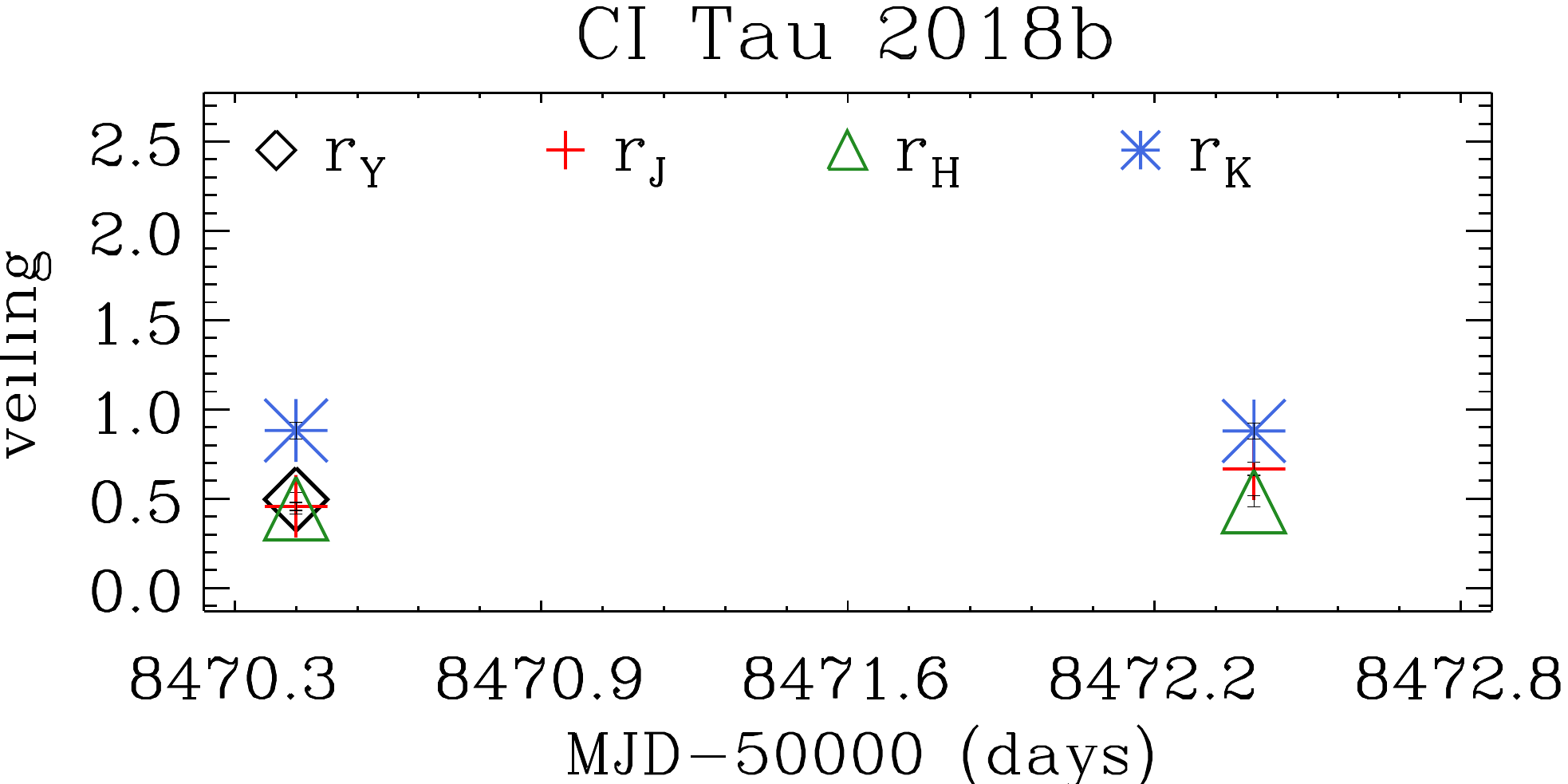}}
 {\includegraphics[width=4.4cm]{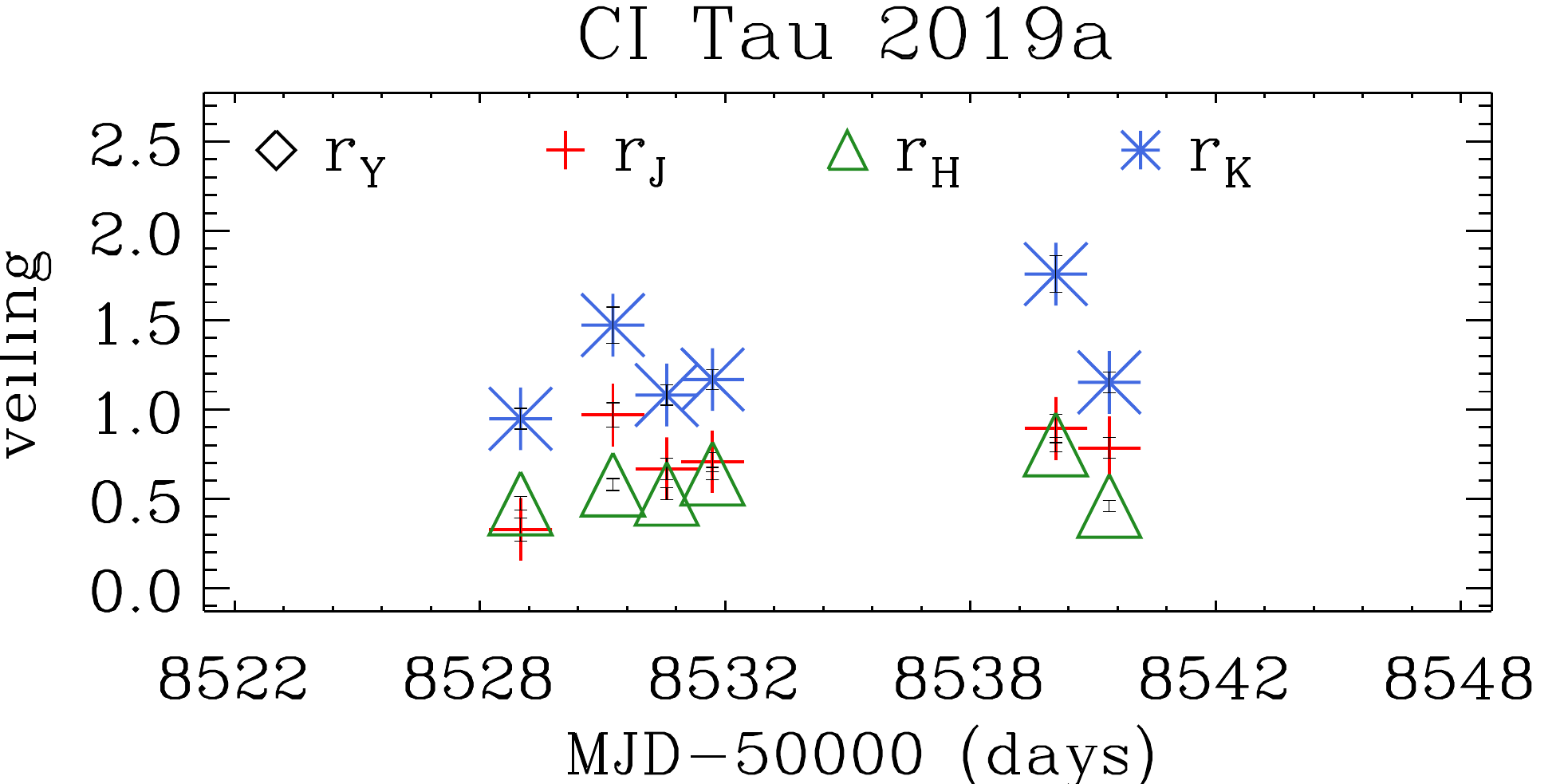}}
 {\includegraphics[width=4.4cm]{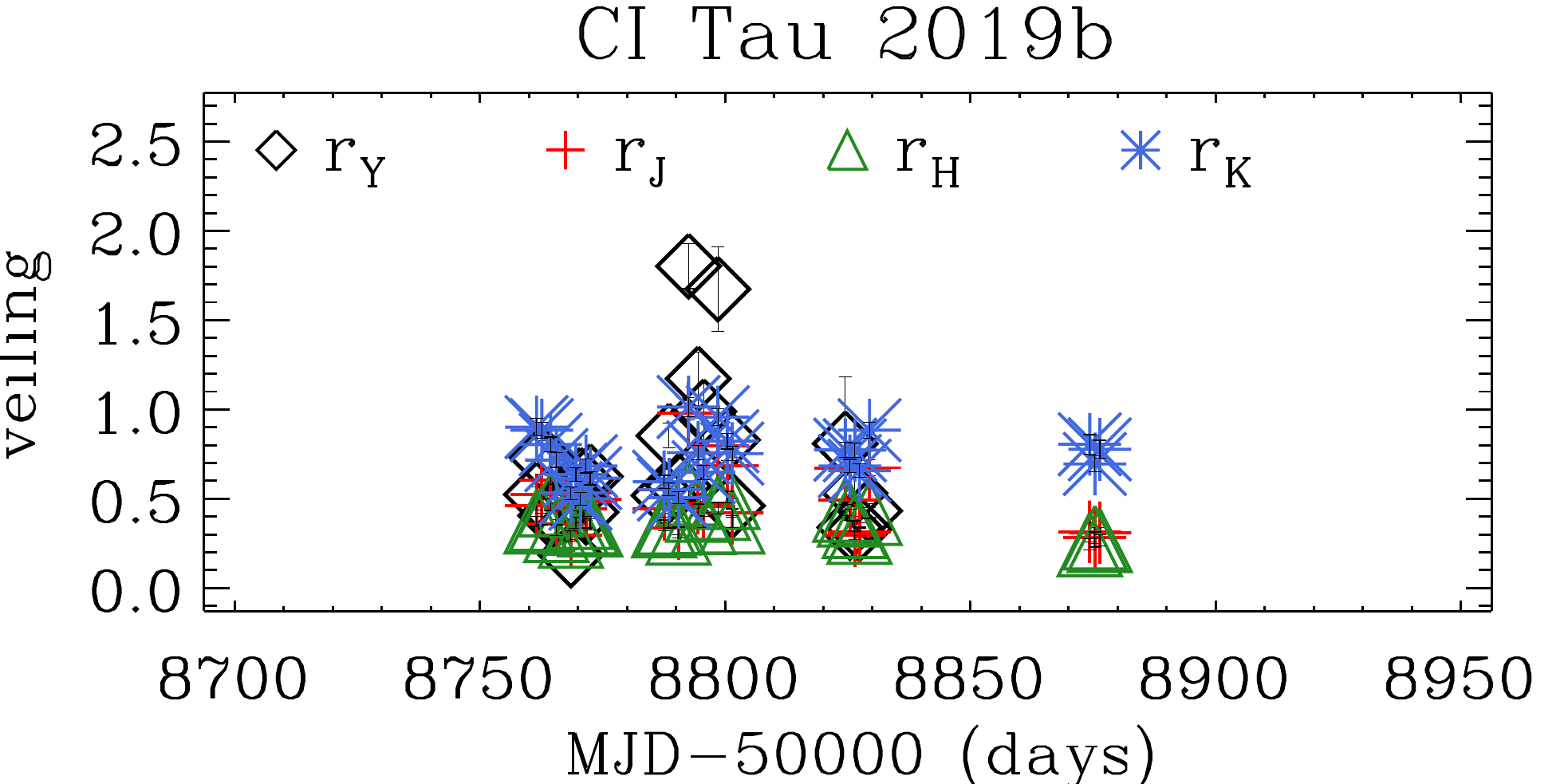}}
 {\includegraphics[width=4.4cm]{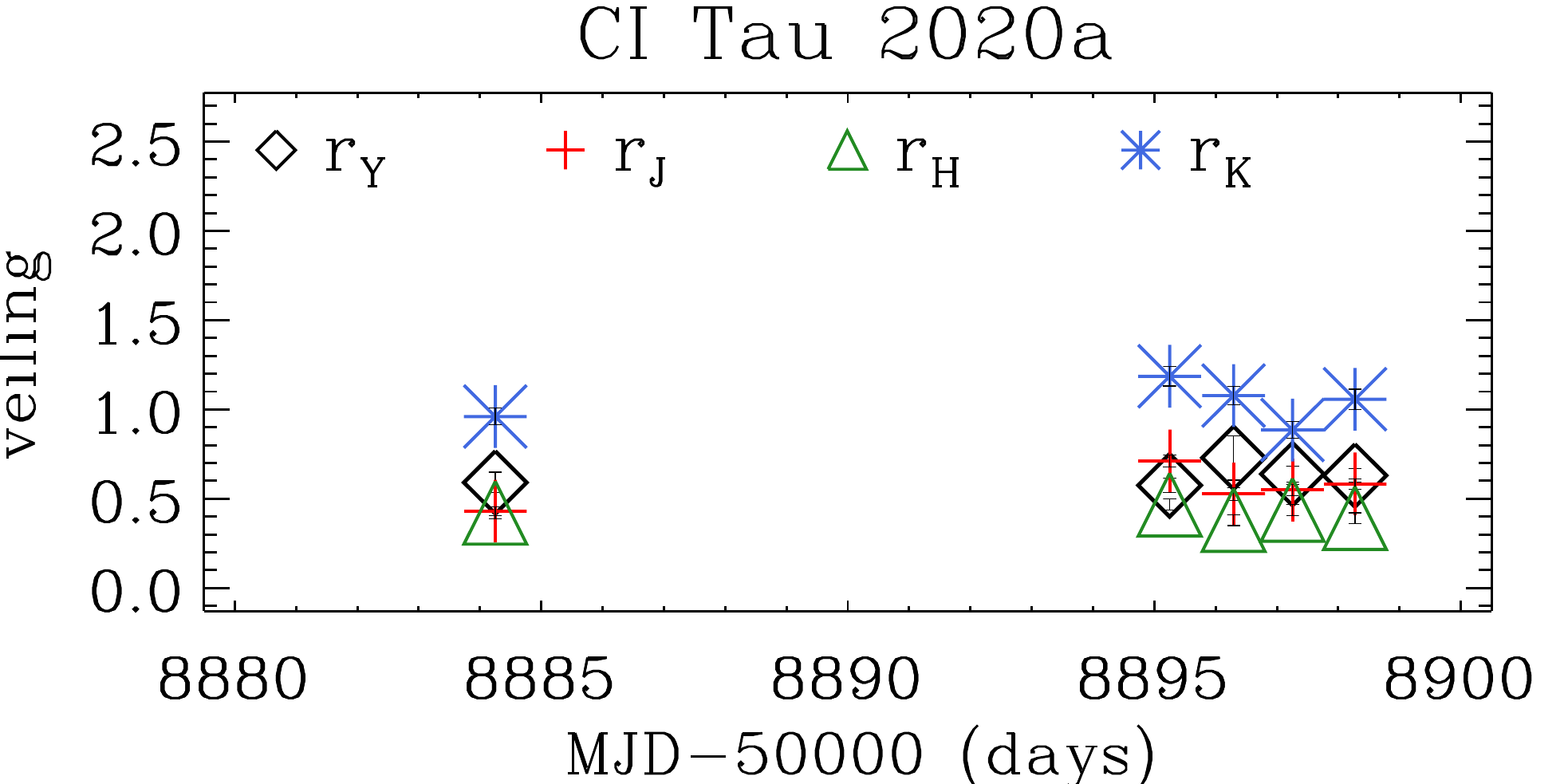}}
 {\includegraphics[width=4.4cm]{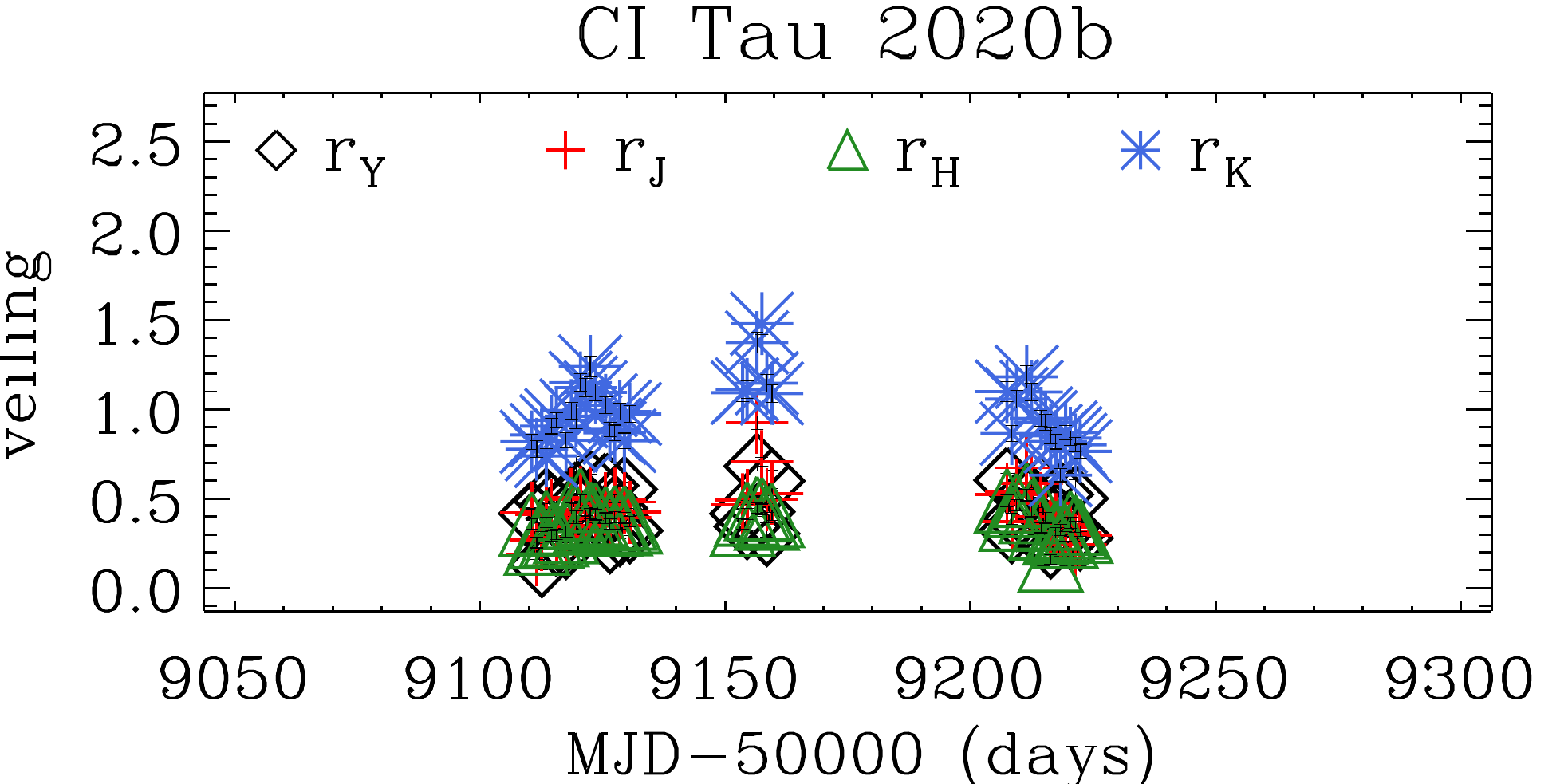}}
 {\includegraphics[width=4.4cm]{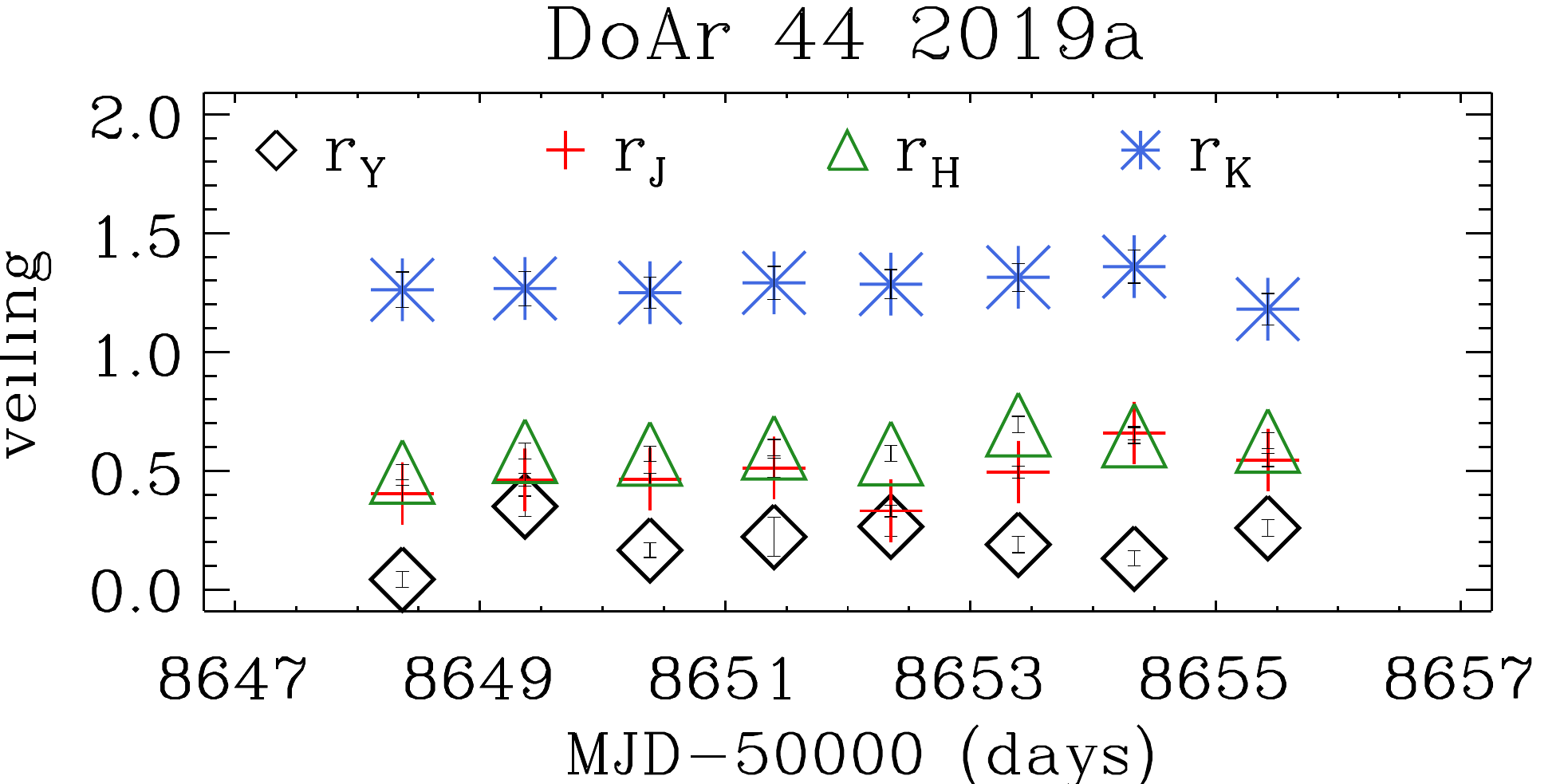}}
 {\includegraphics[width=4.4cm]{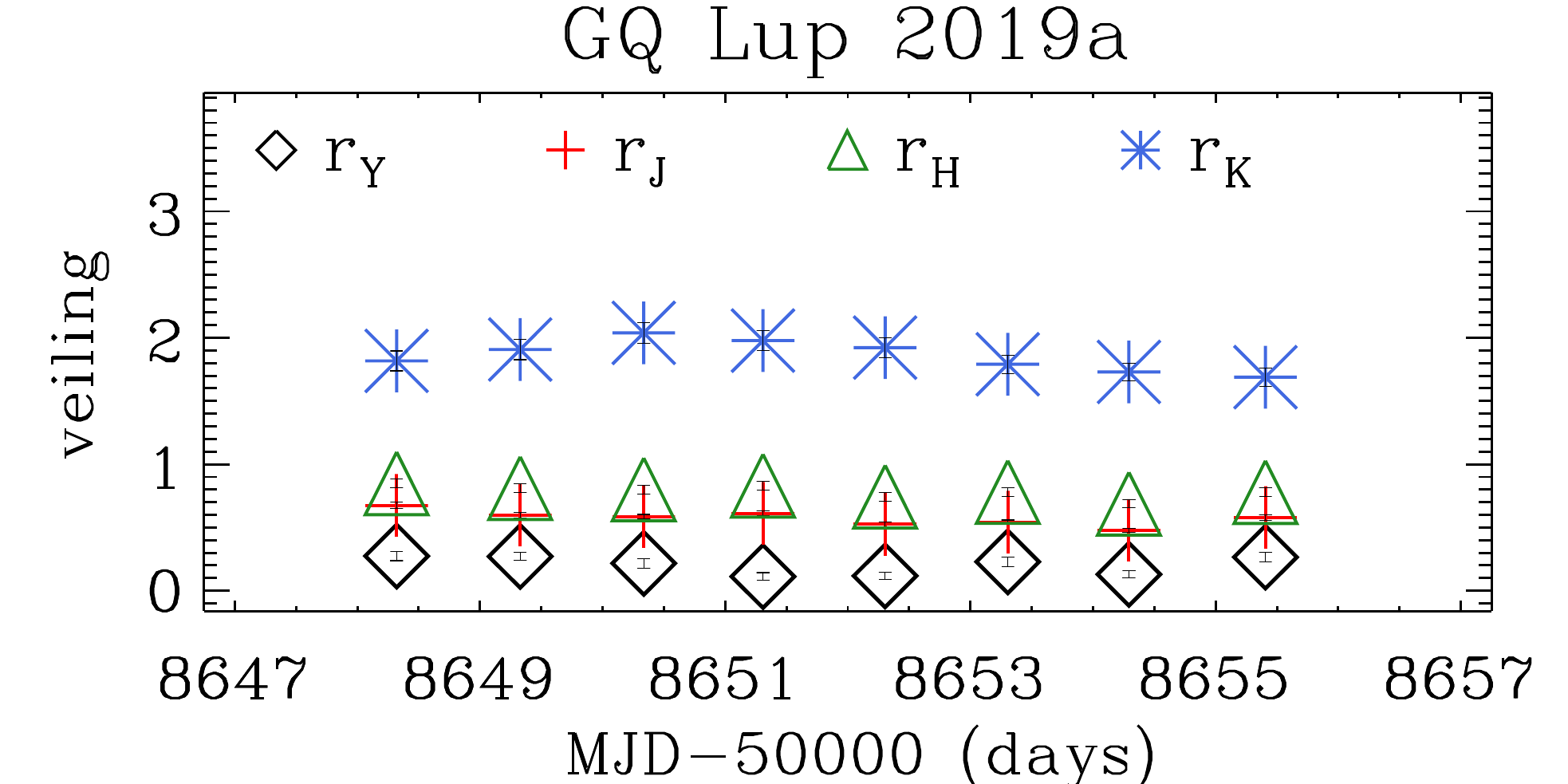}}
 {\includegraphics[width=4.4cm]{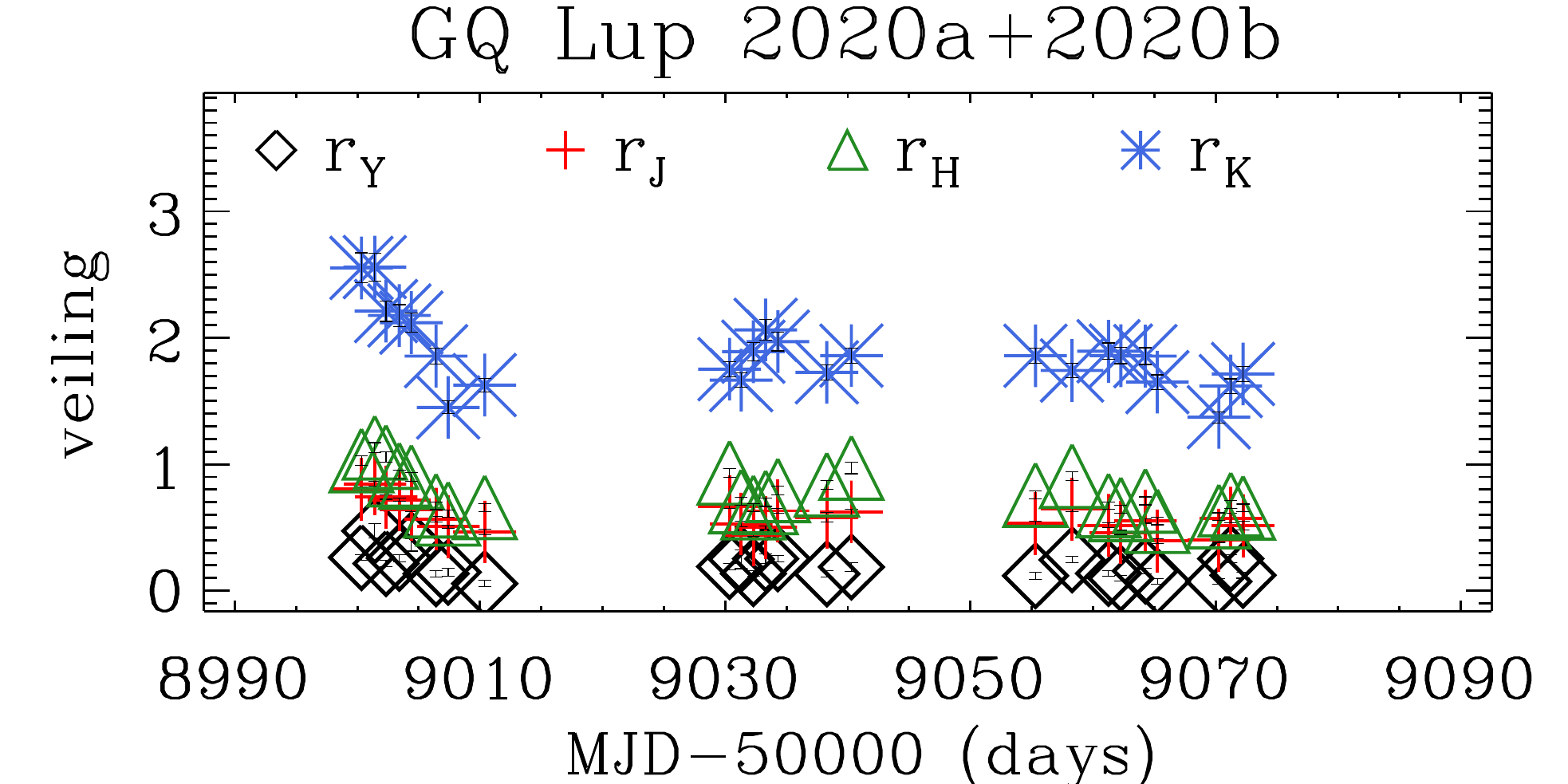}}
 {\includegraphics[width=4.4cm]{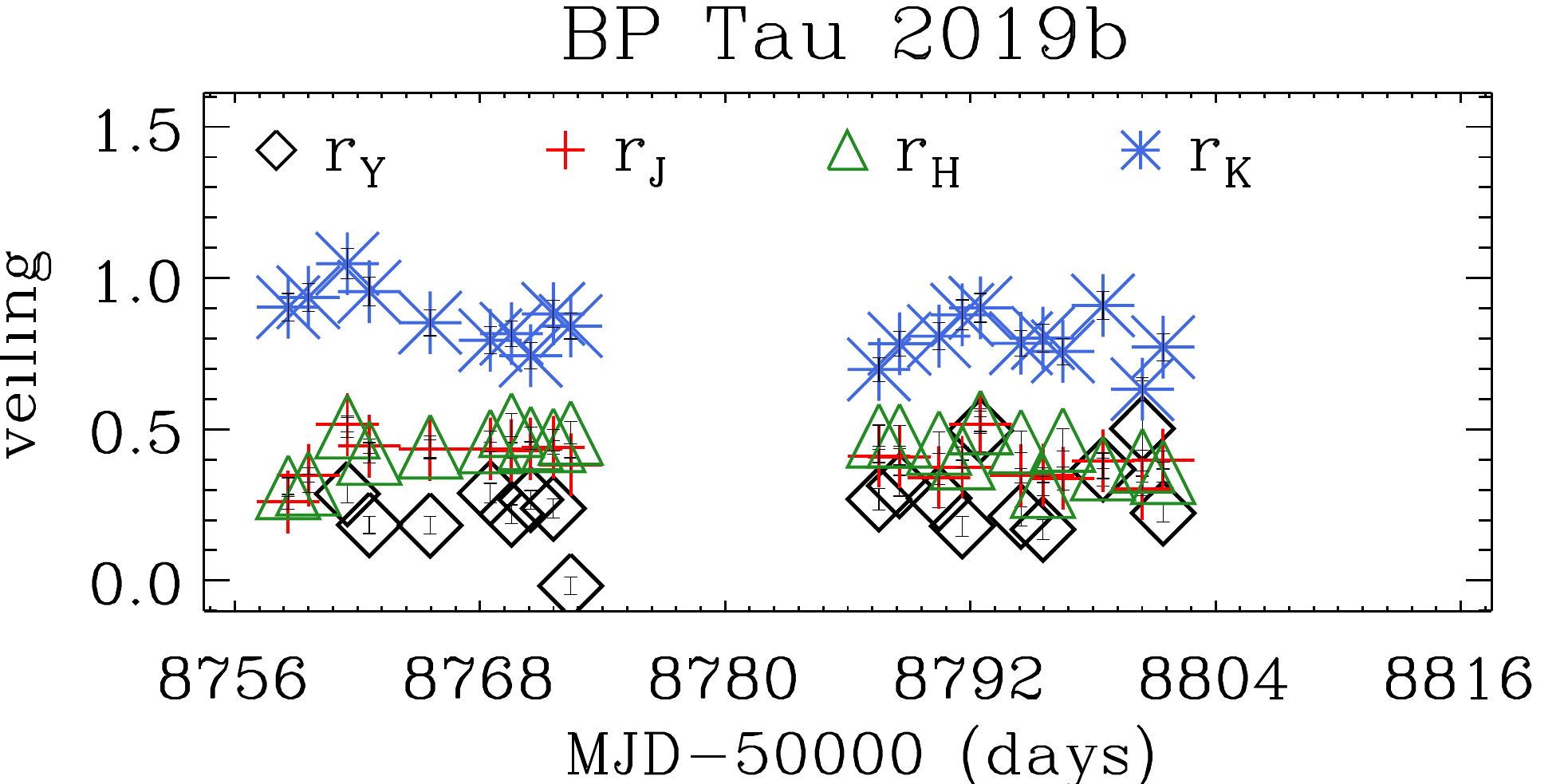}}
  {\includegraphics[width=4.4cm]{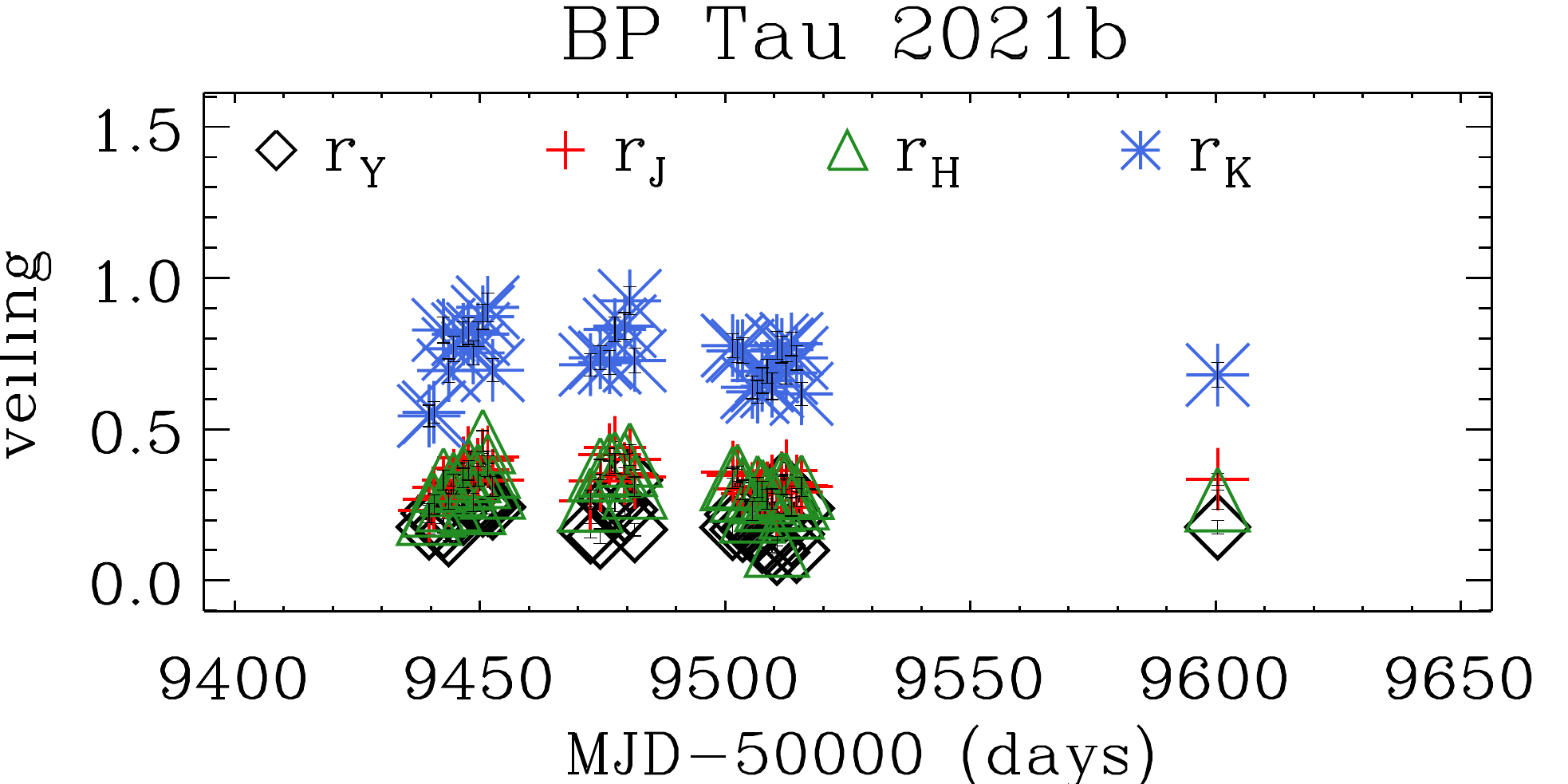}}
 {\includegraphics[width=4.4cm]{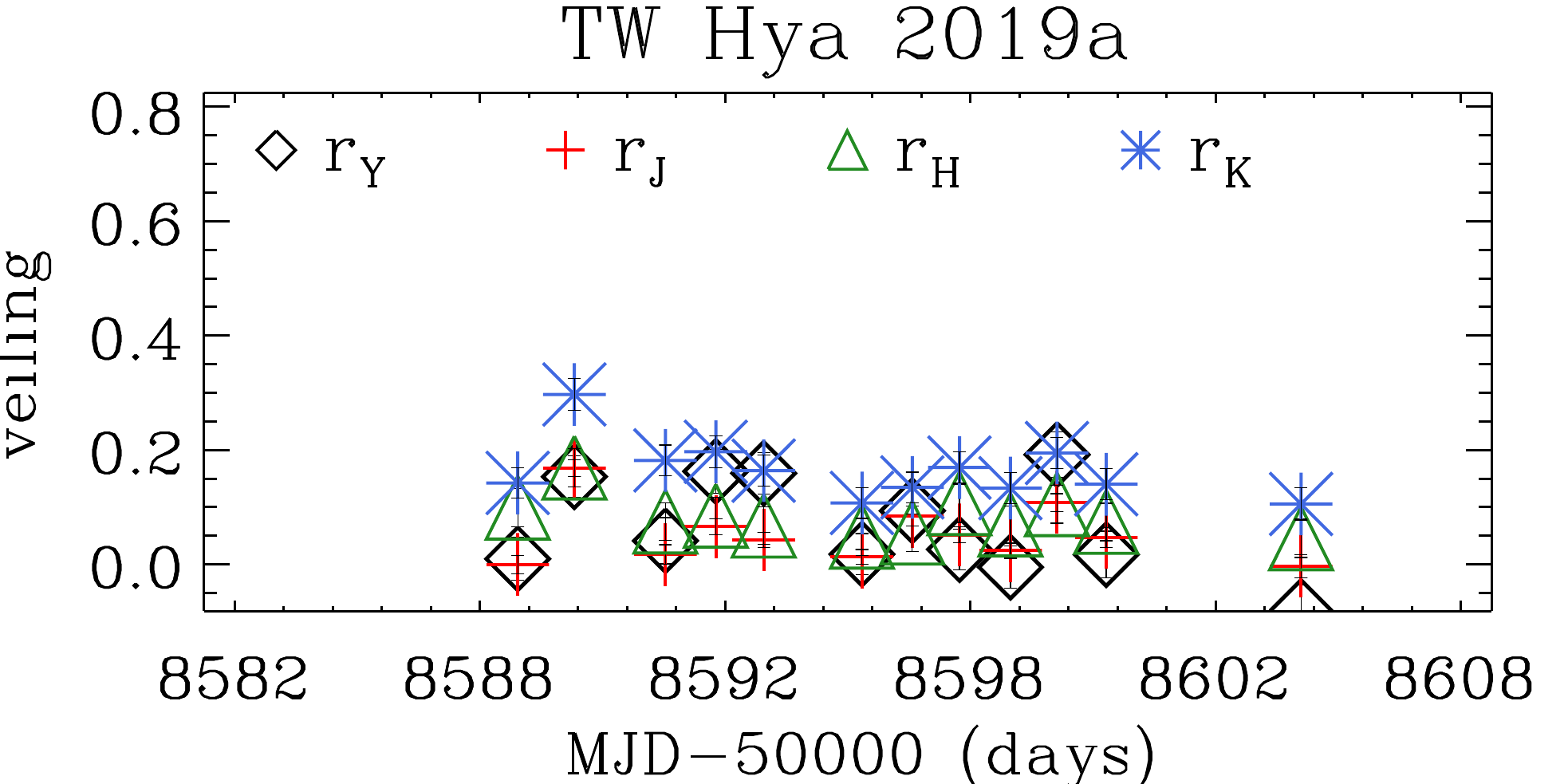}}
 {\includegraphics[width=4.4cm]{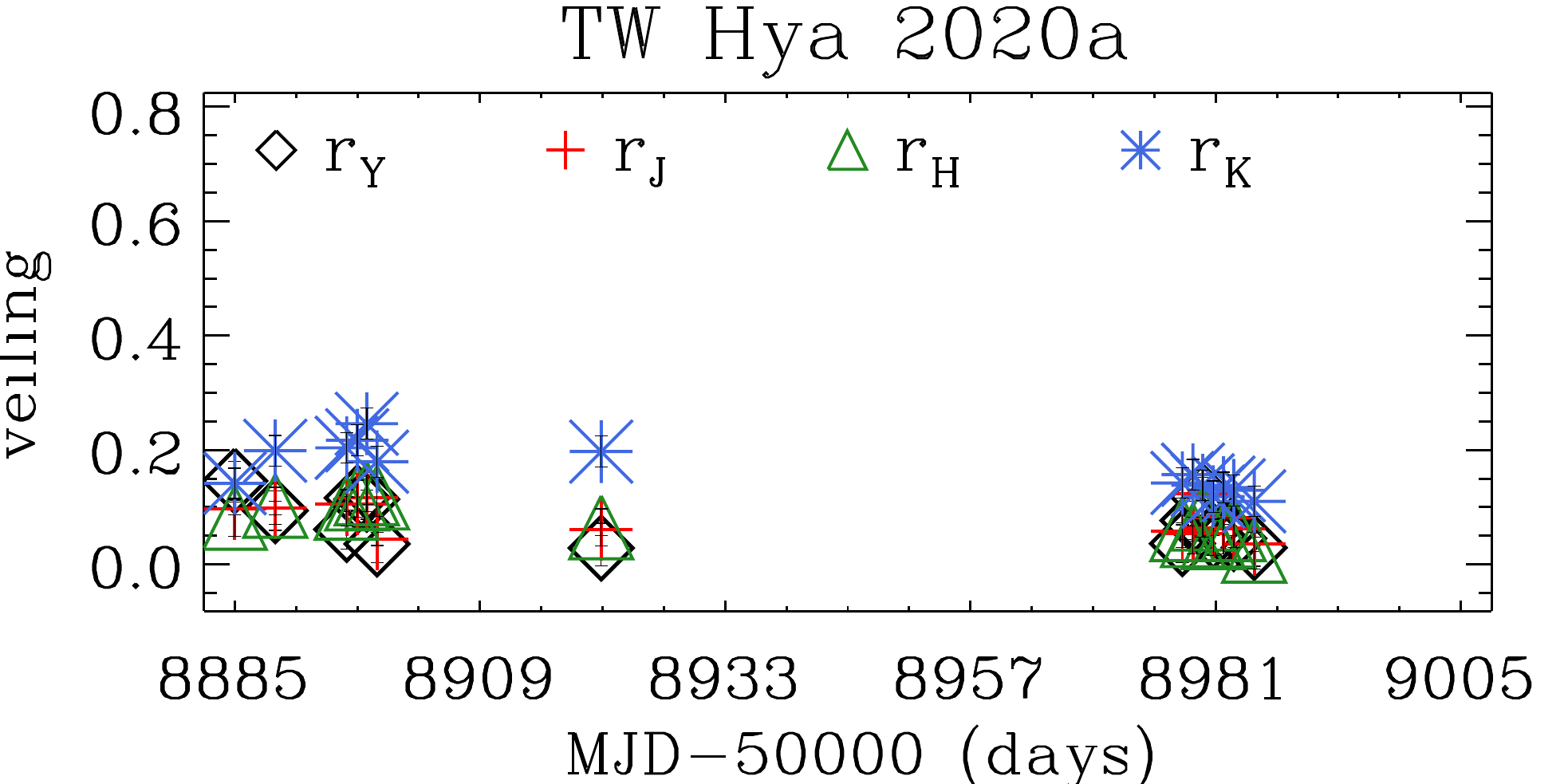}}
 {\includegraphics[width=4.4cm]{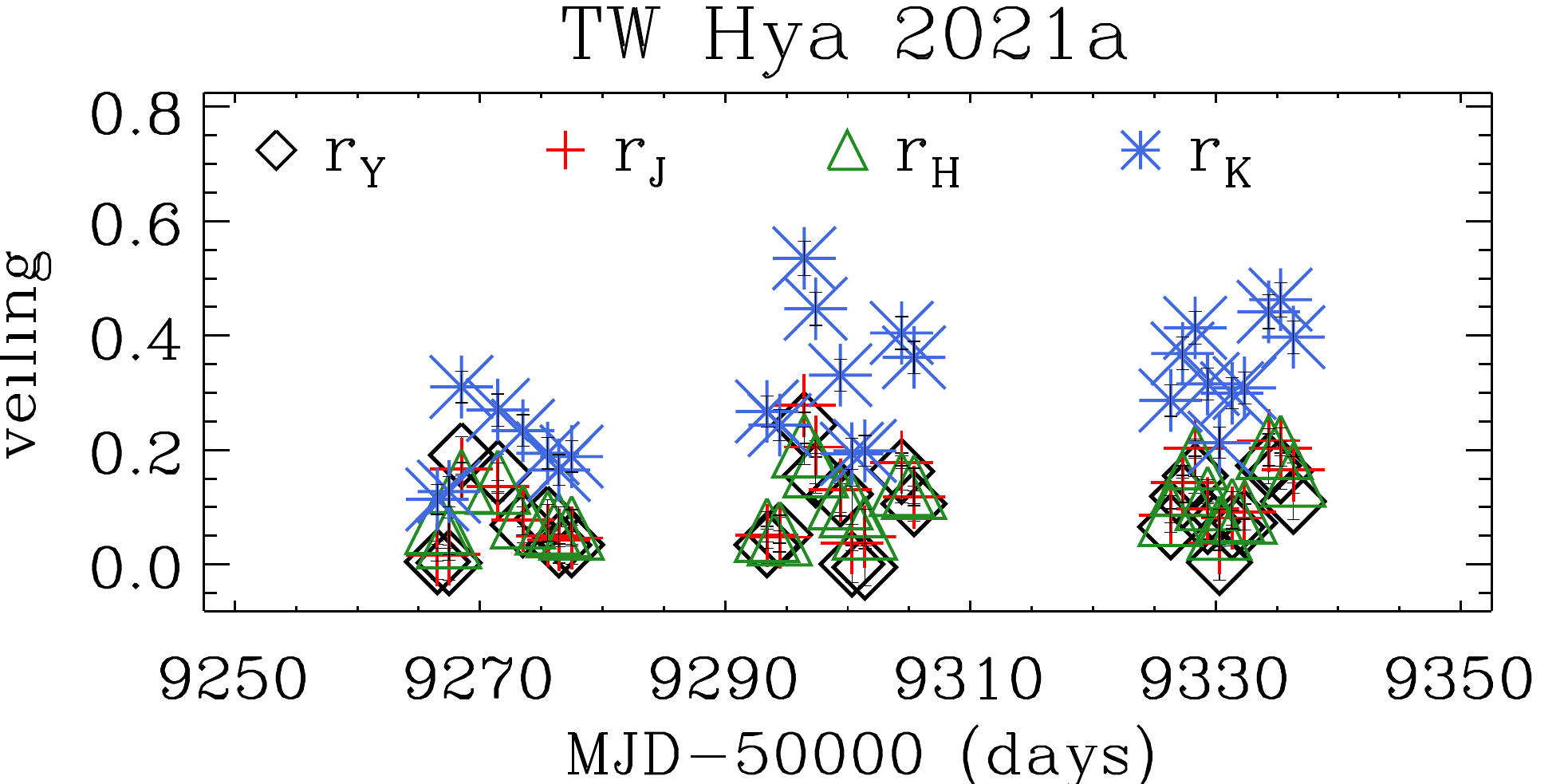}}
 {\includegraphics[width=4.4cm]{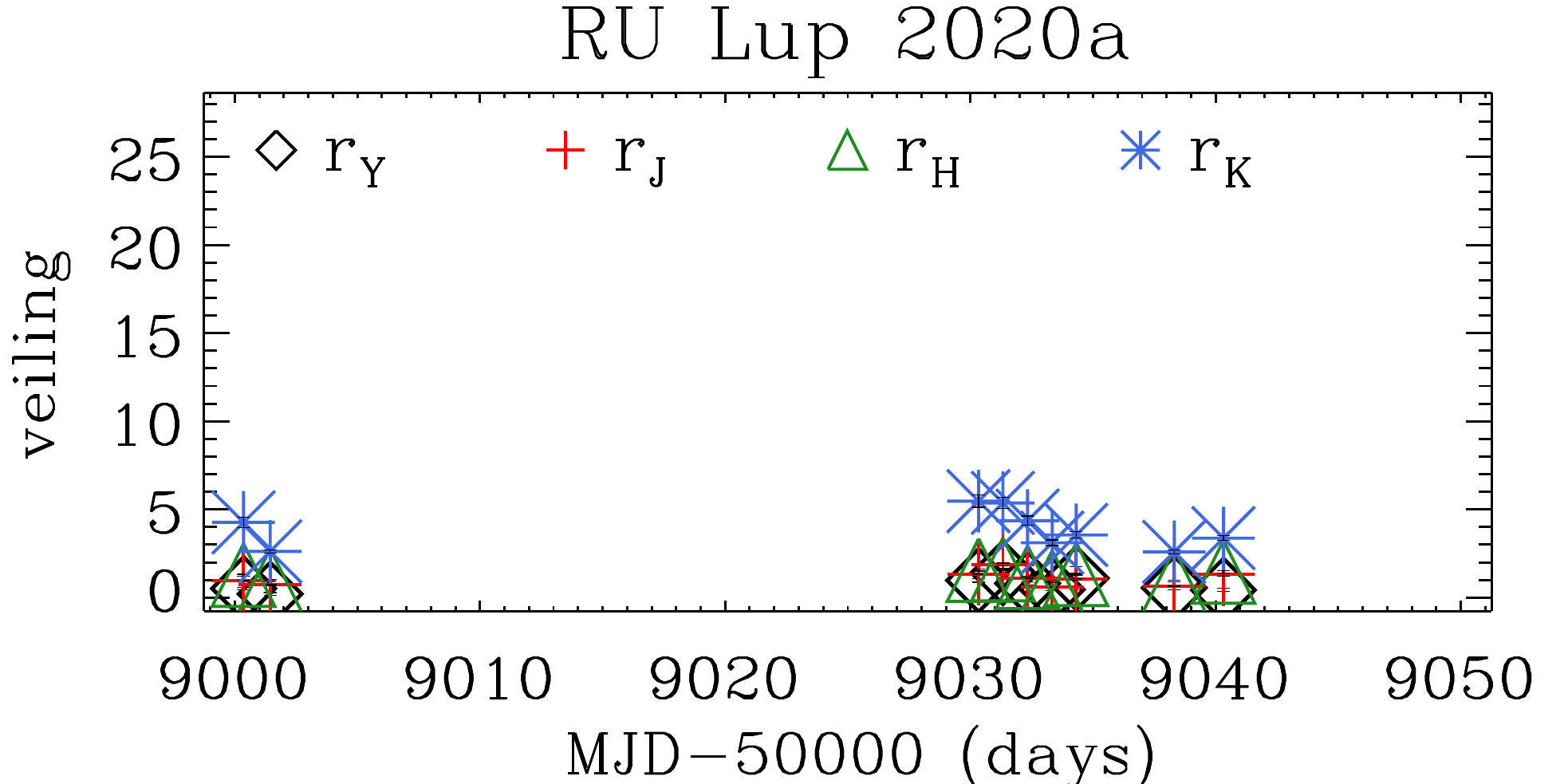}}
 {\includegraphics[width=4.4cm]{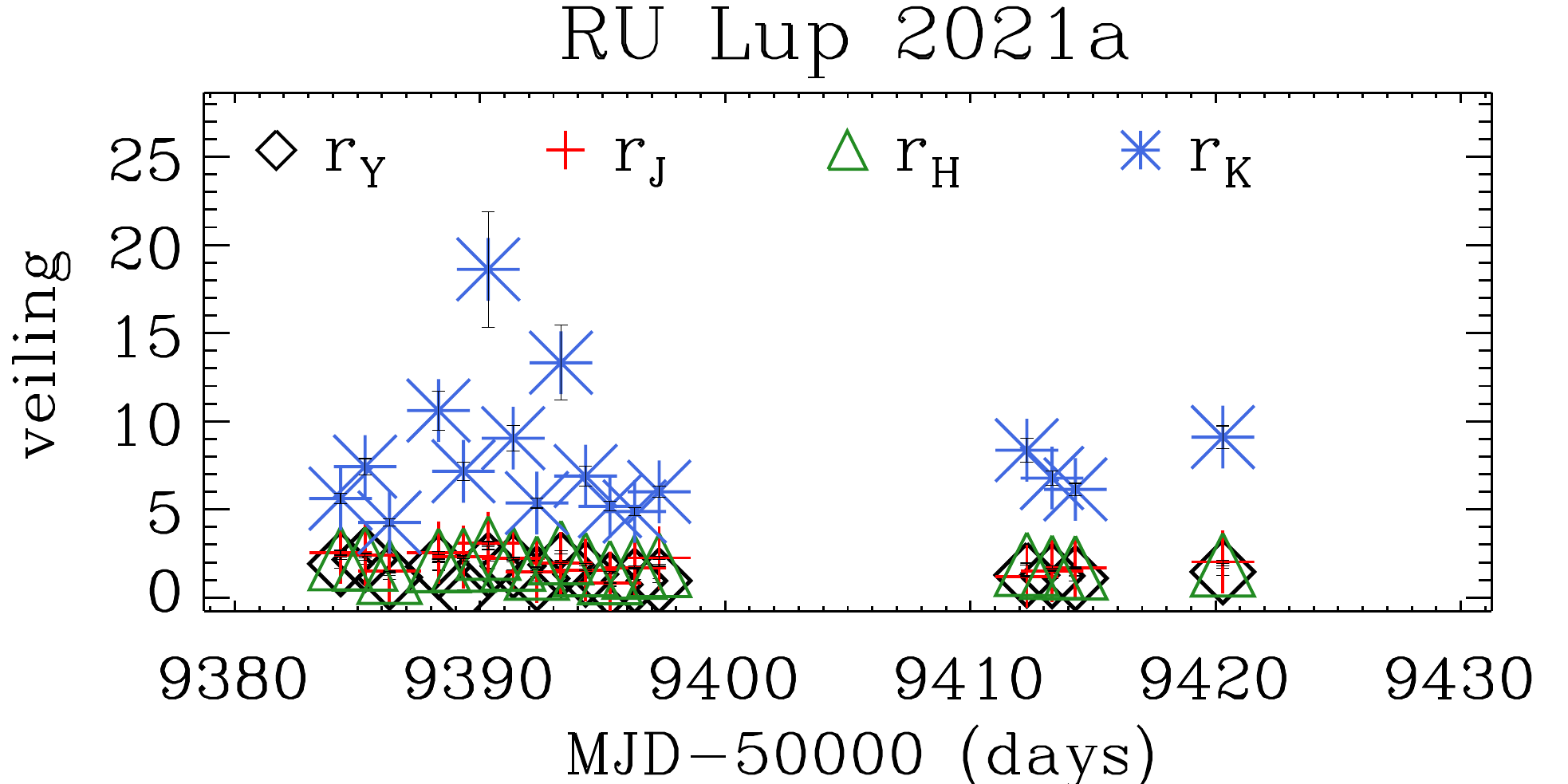}}
 {\includegraphics[width=4.4cm]{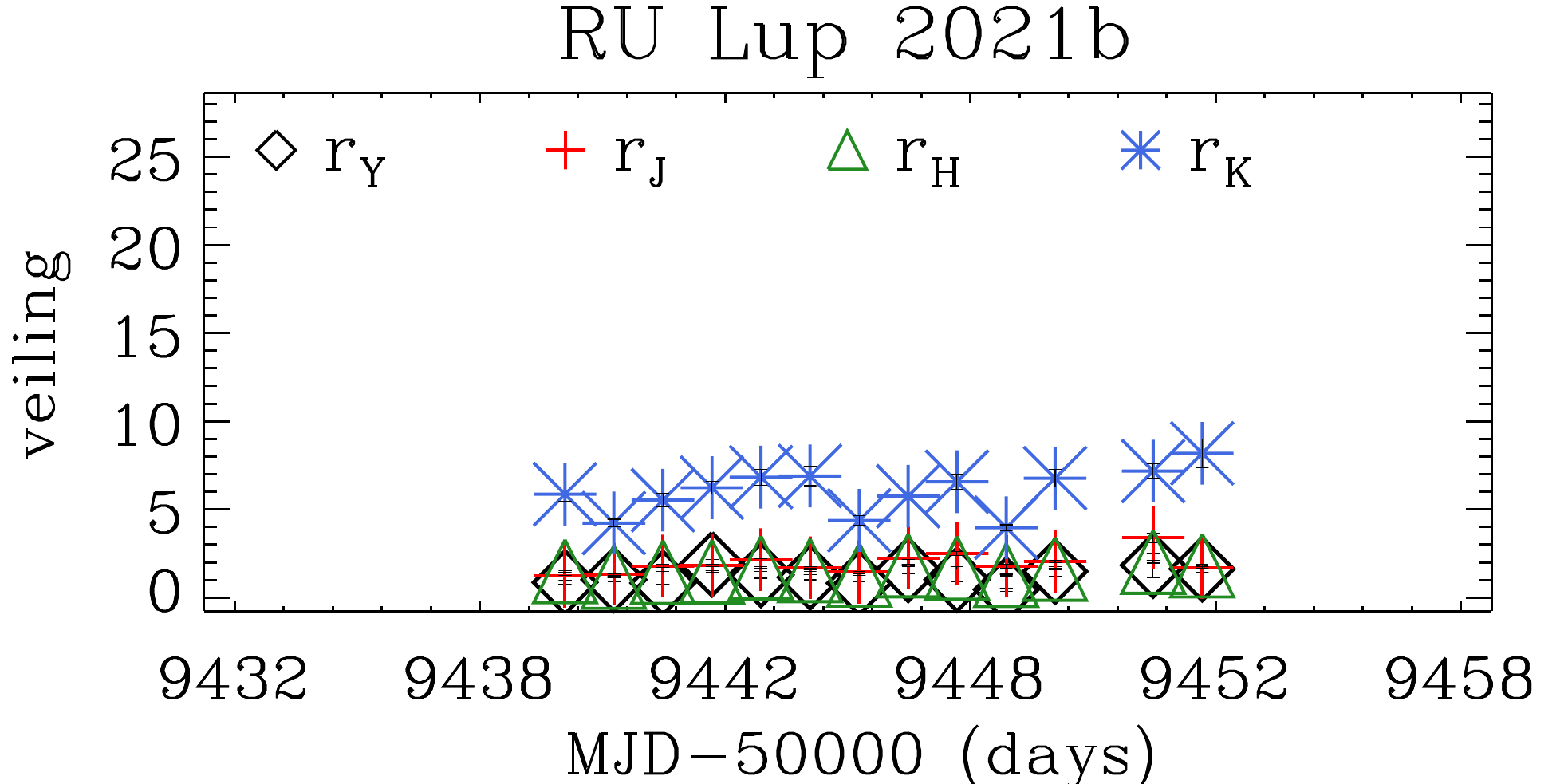}}
 {\includegraphics[width=4.4cm]{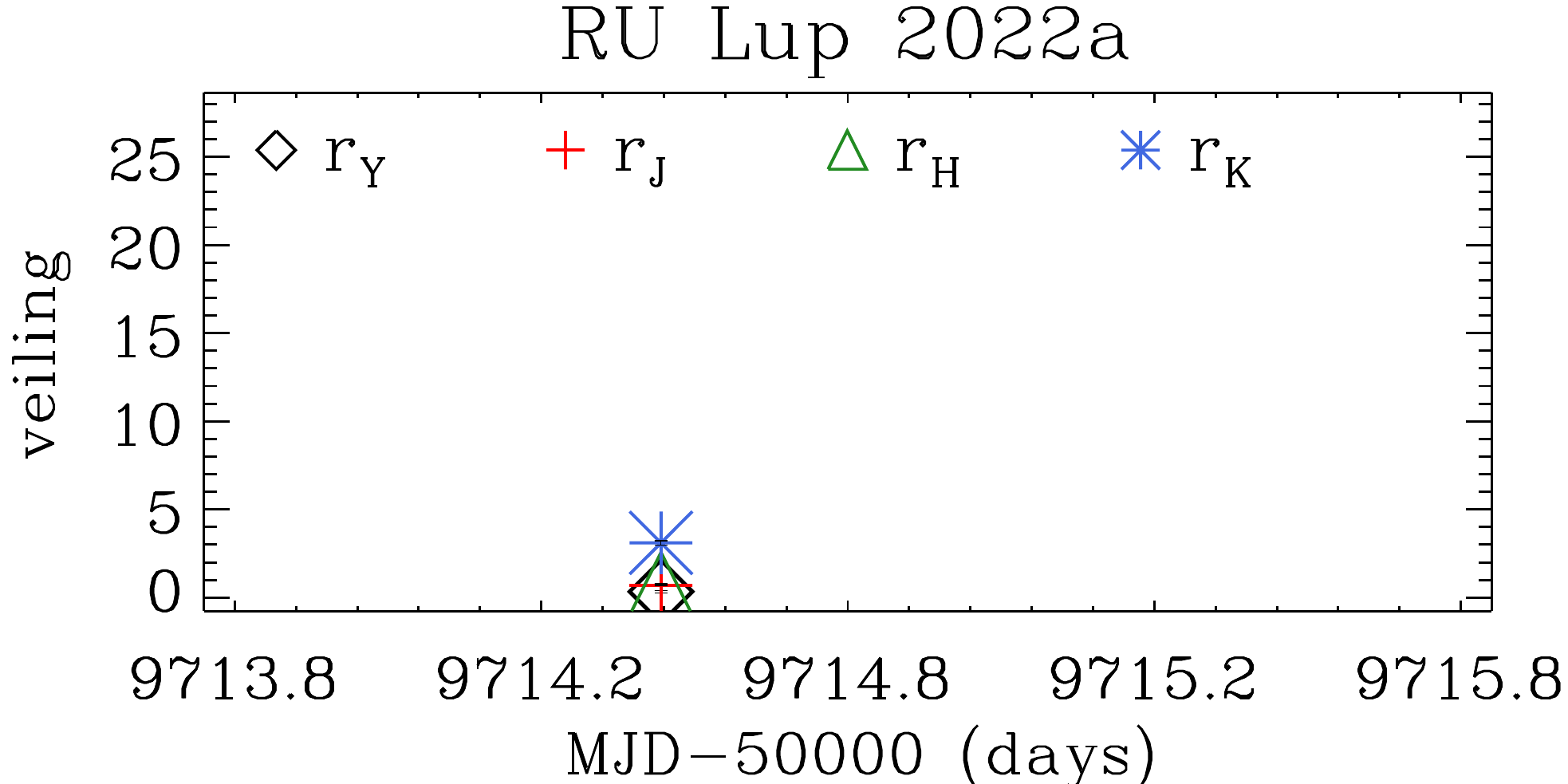}}
 {\includegraphics[width=4.4cm]{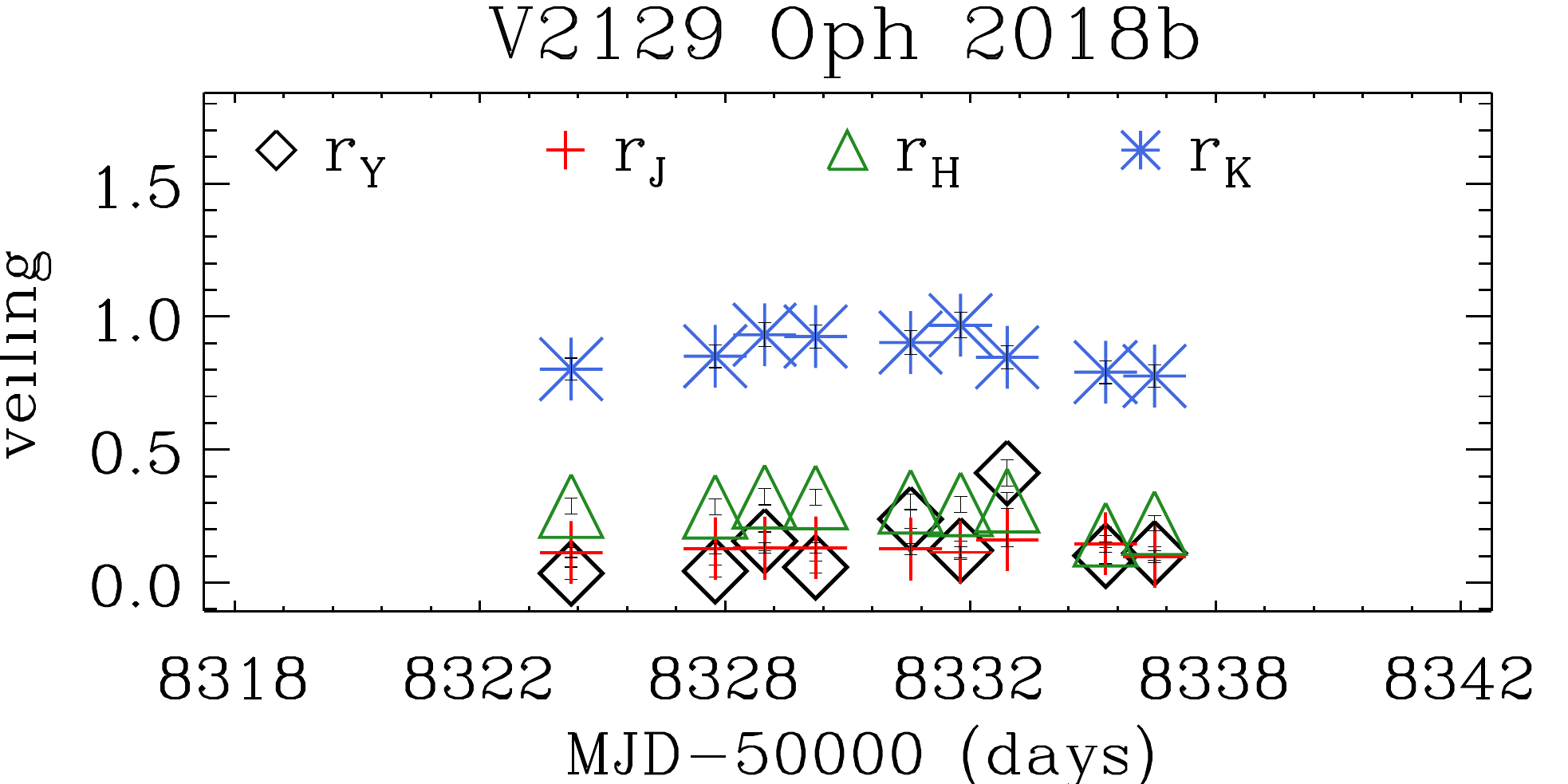}}
 {\includegraphics[width=4.4cm]{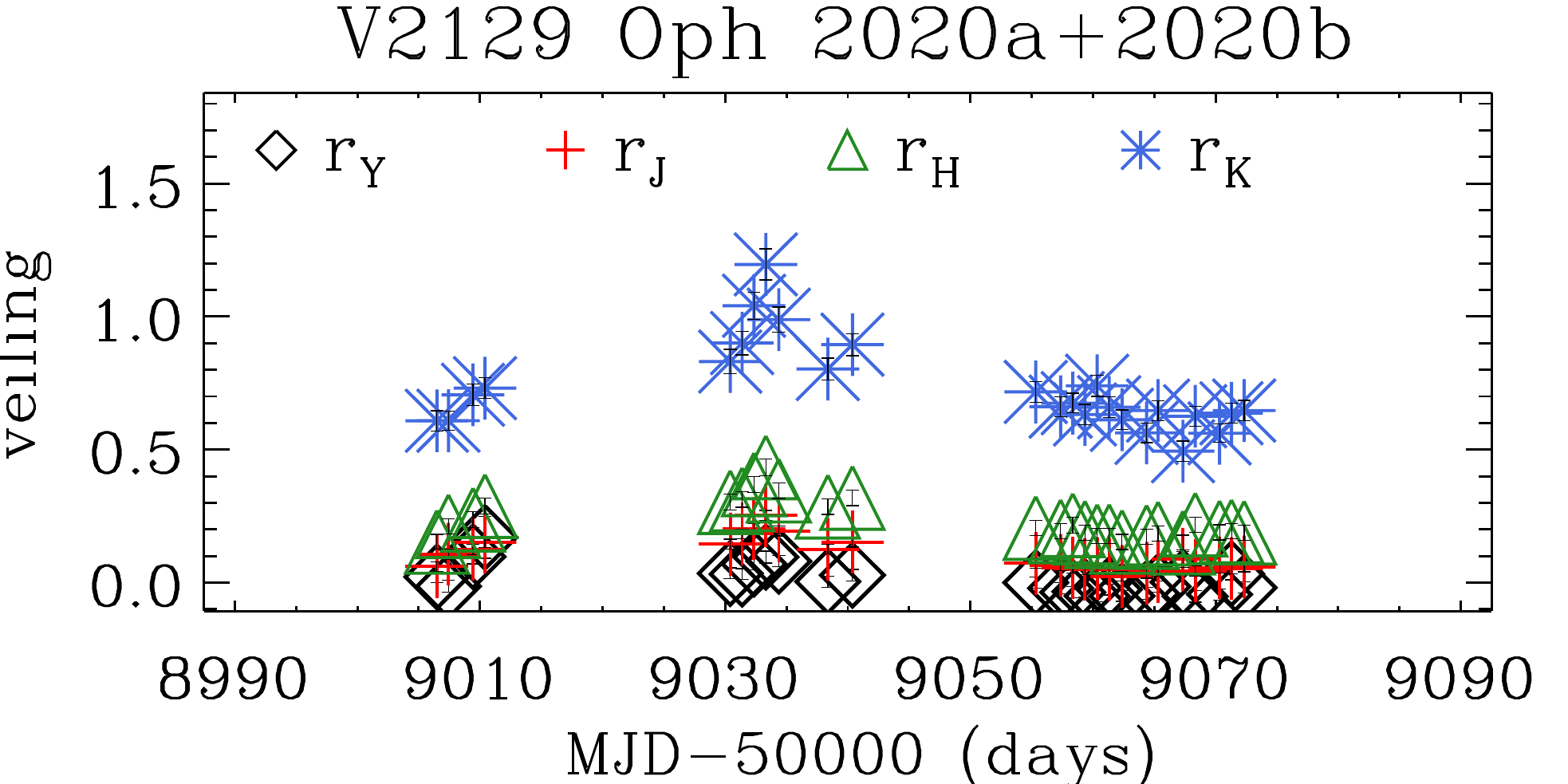}}
 {\includegraphics[width=4.4cm]{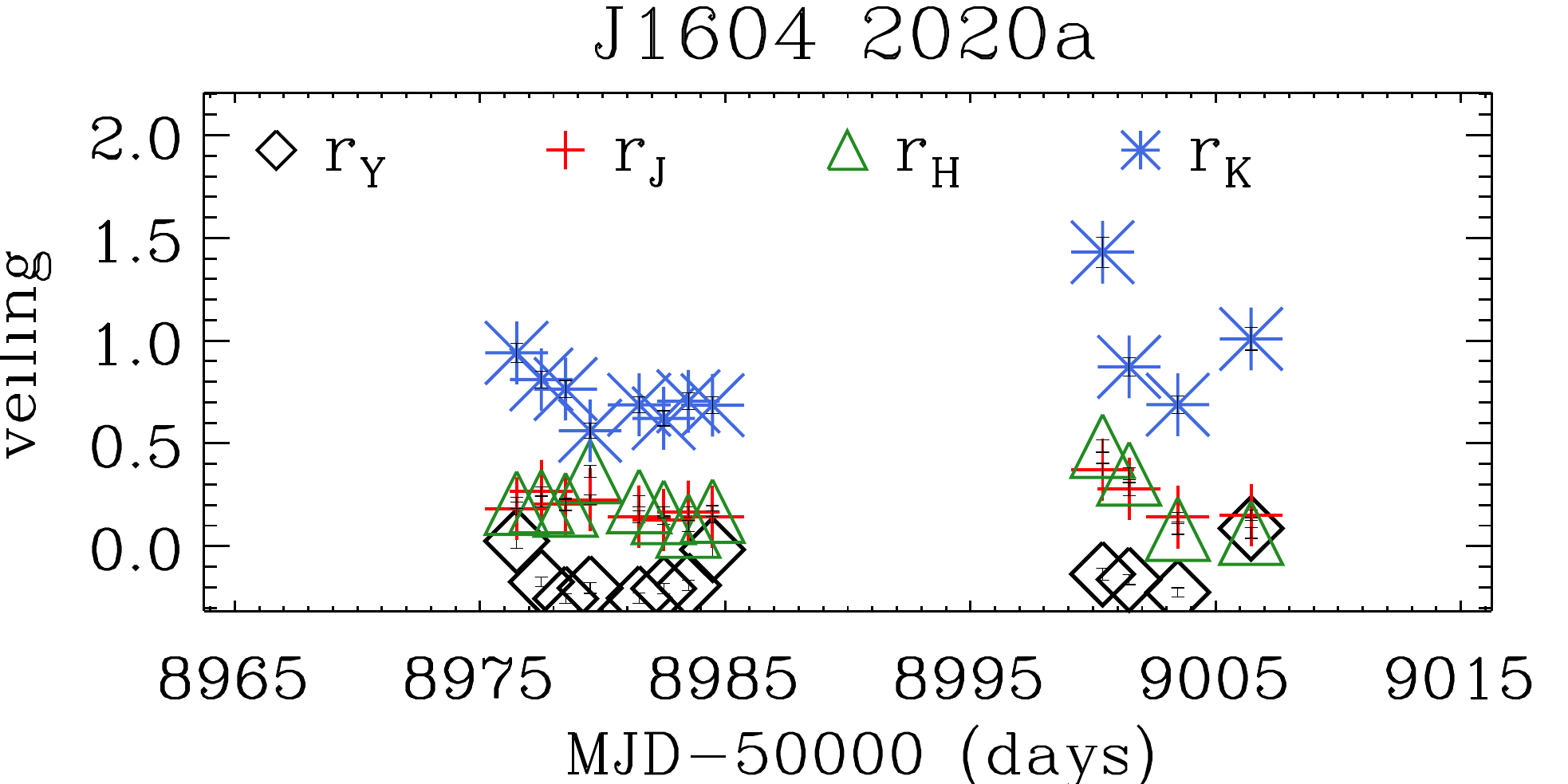}}
 {\includegraphics[width=4.4cm]{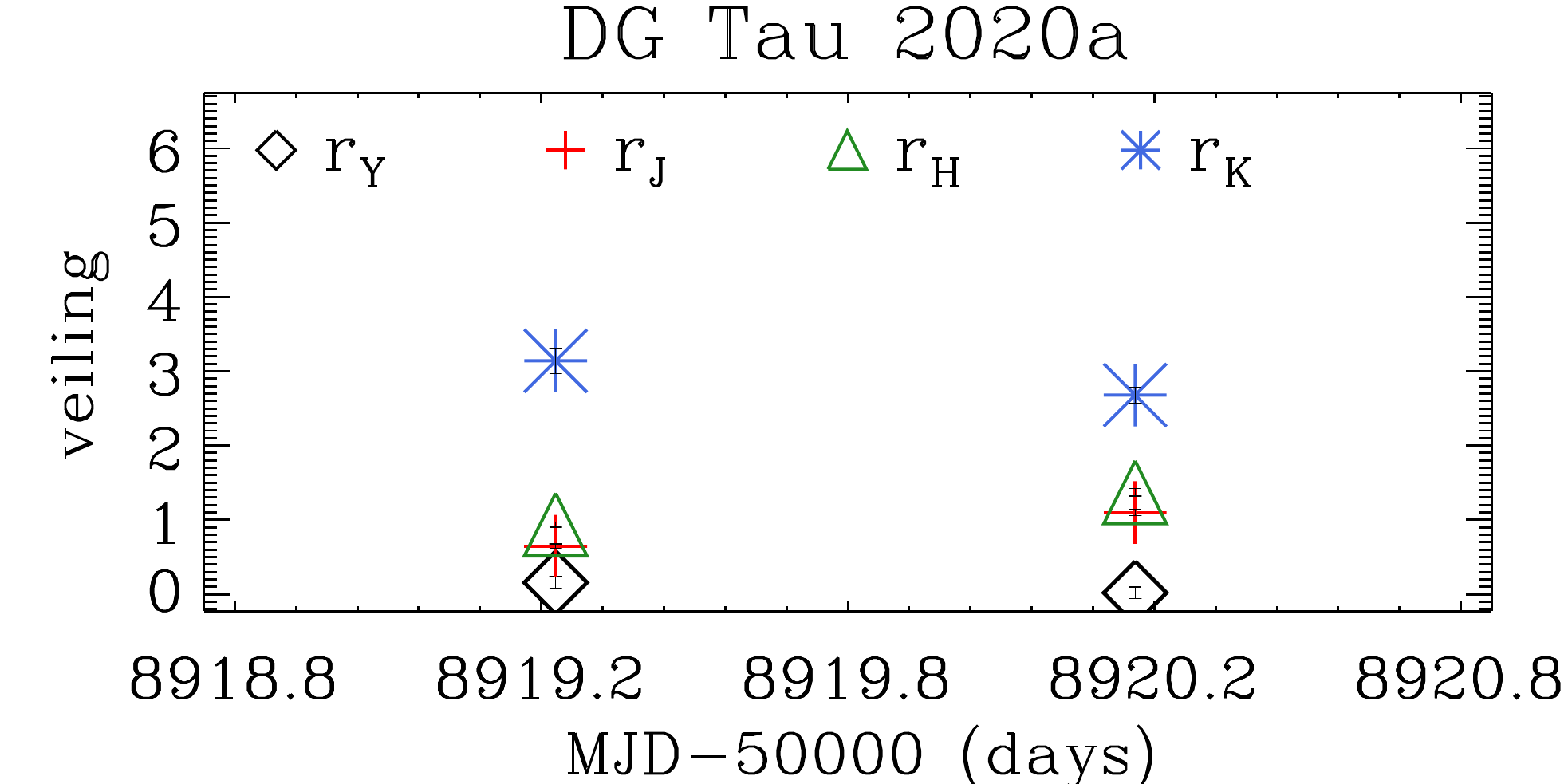}}
 {\includegraphics[width=4.4cm]{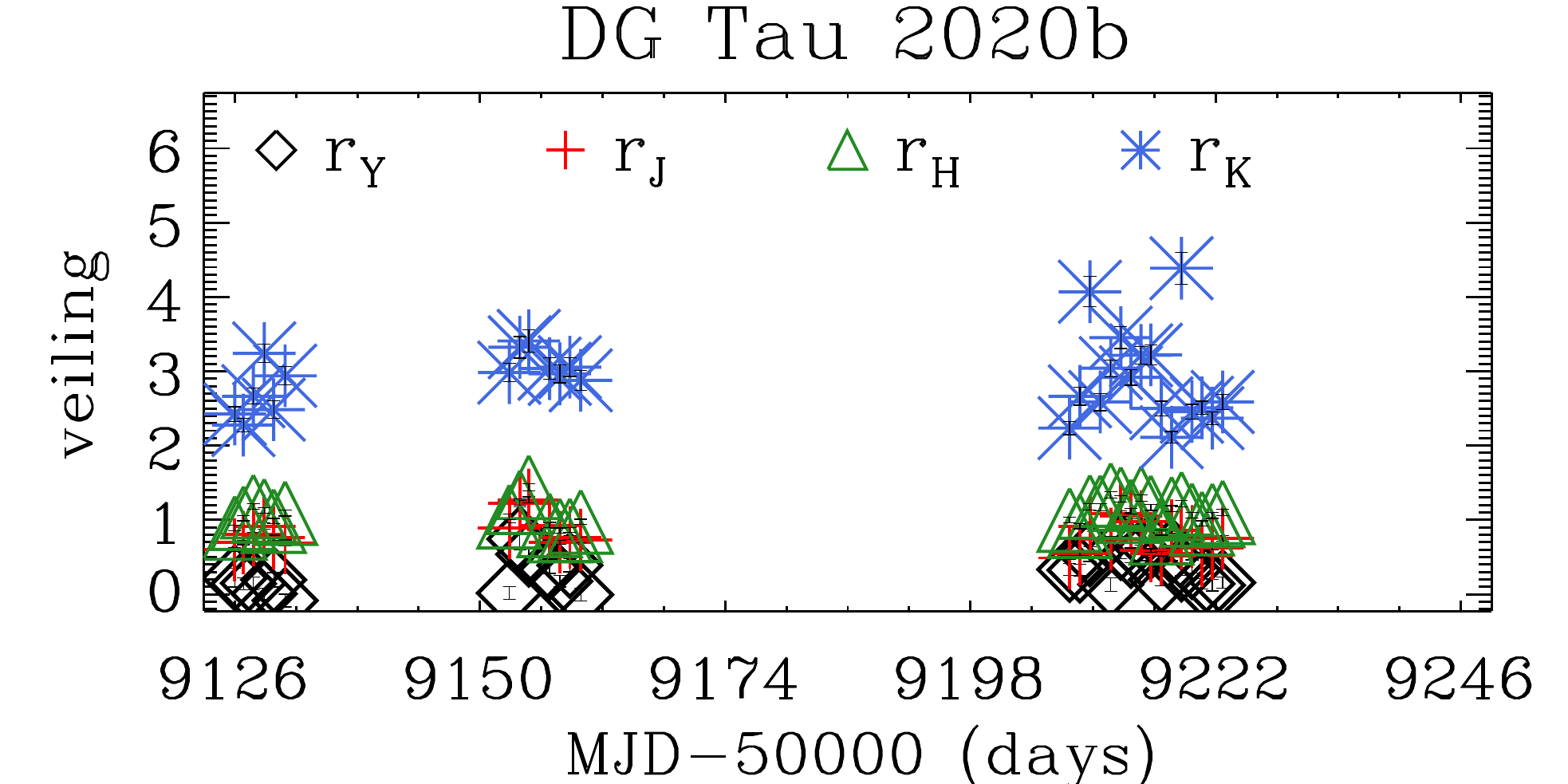}}
 {\includegraphics[width=4.4cm]{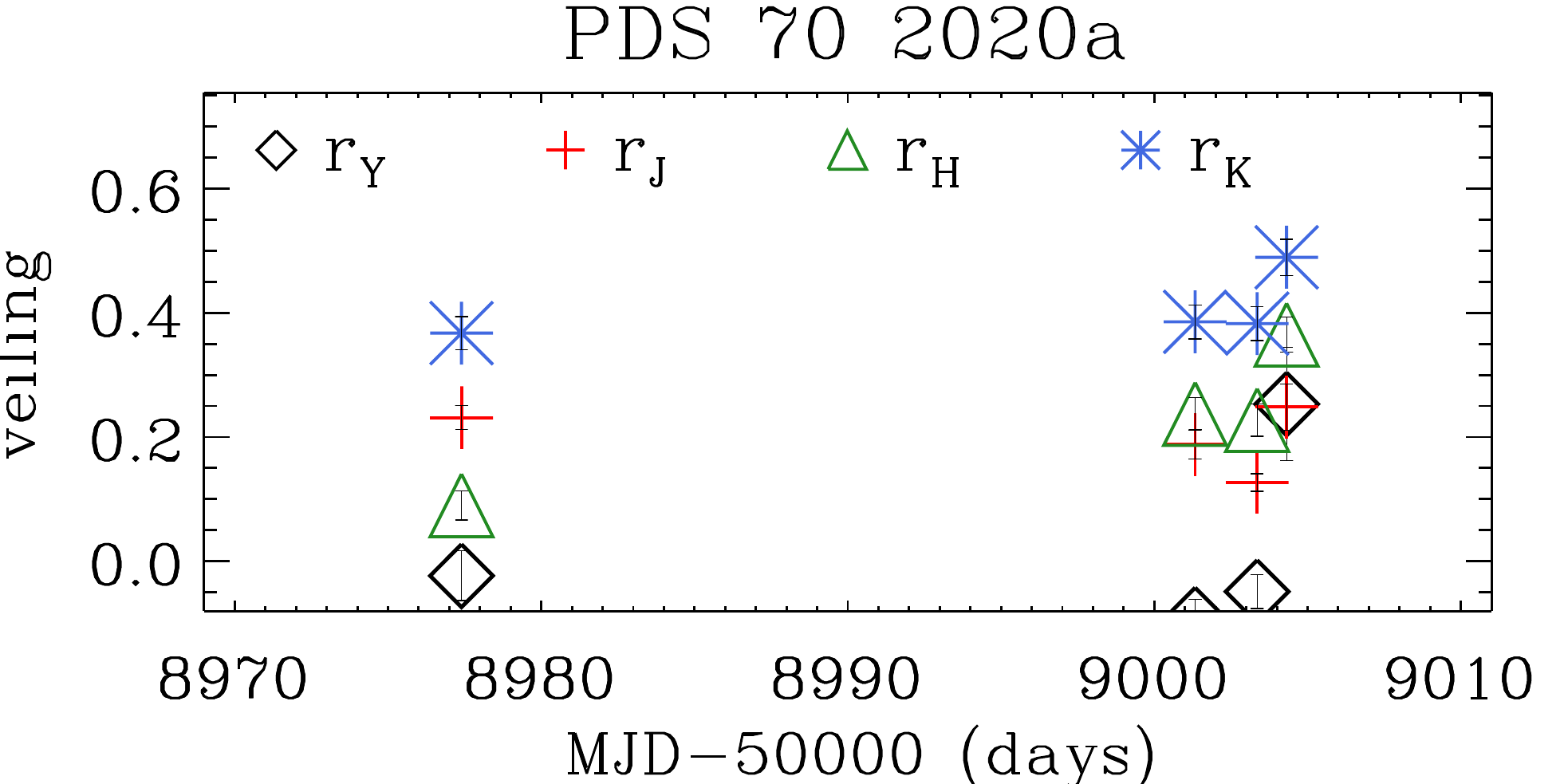}}
 {\includegraphics[width=4.4cm]{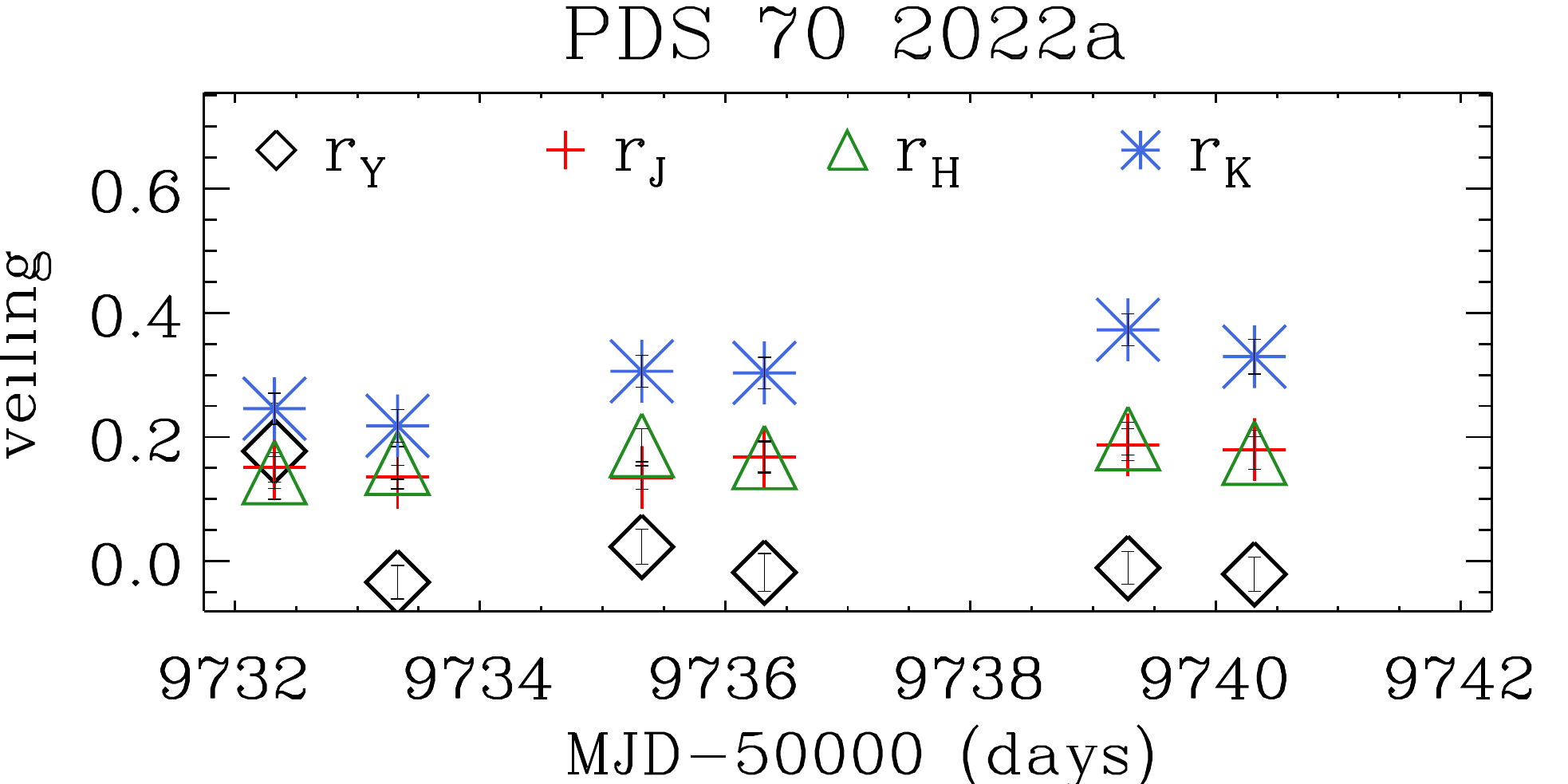}}
 {\includegraphics[width=4.4cm]{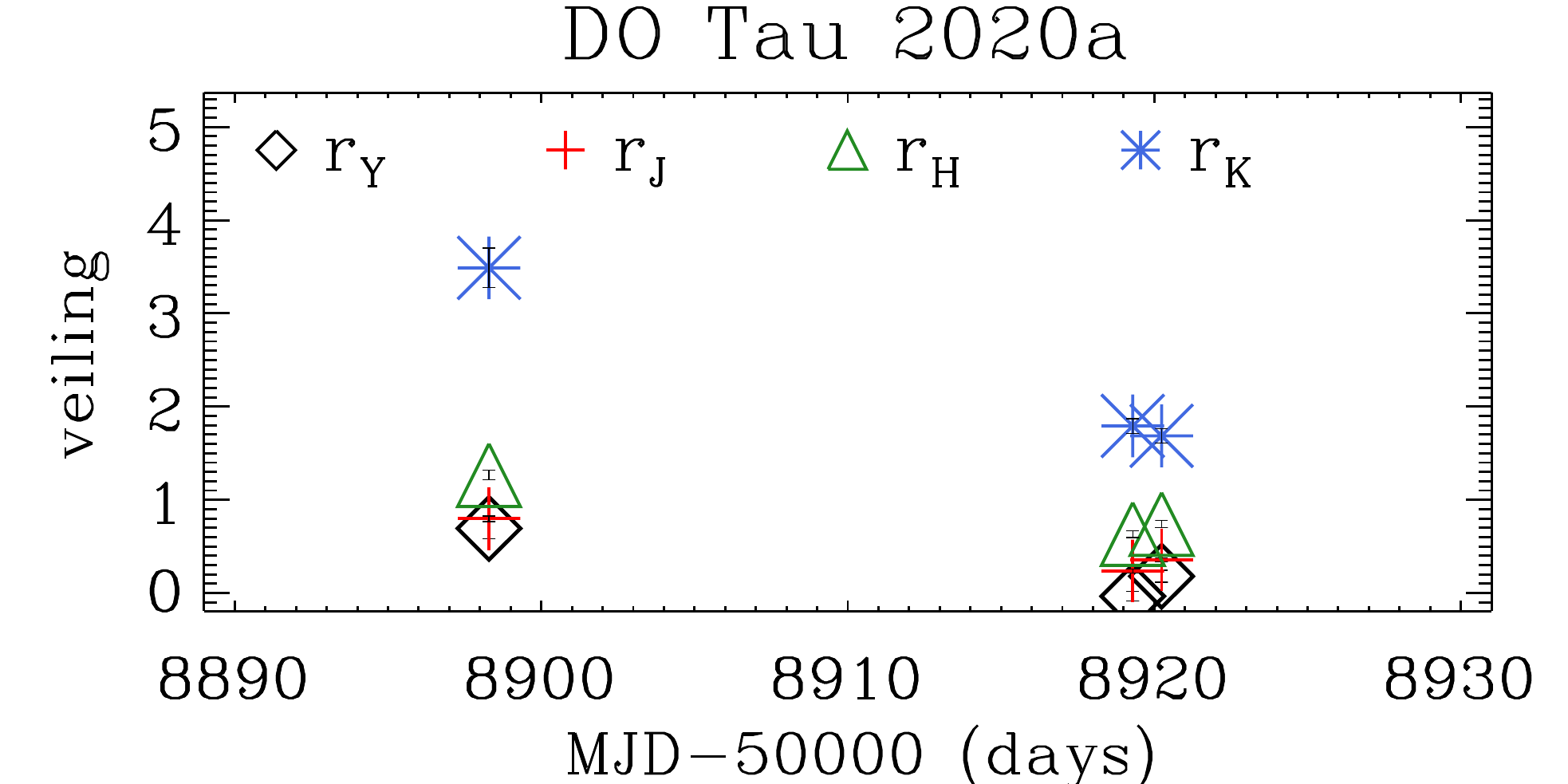}}
 {\includegraphics[width=4.4cm]{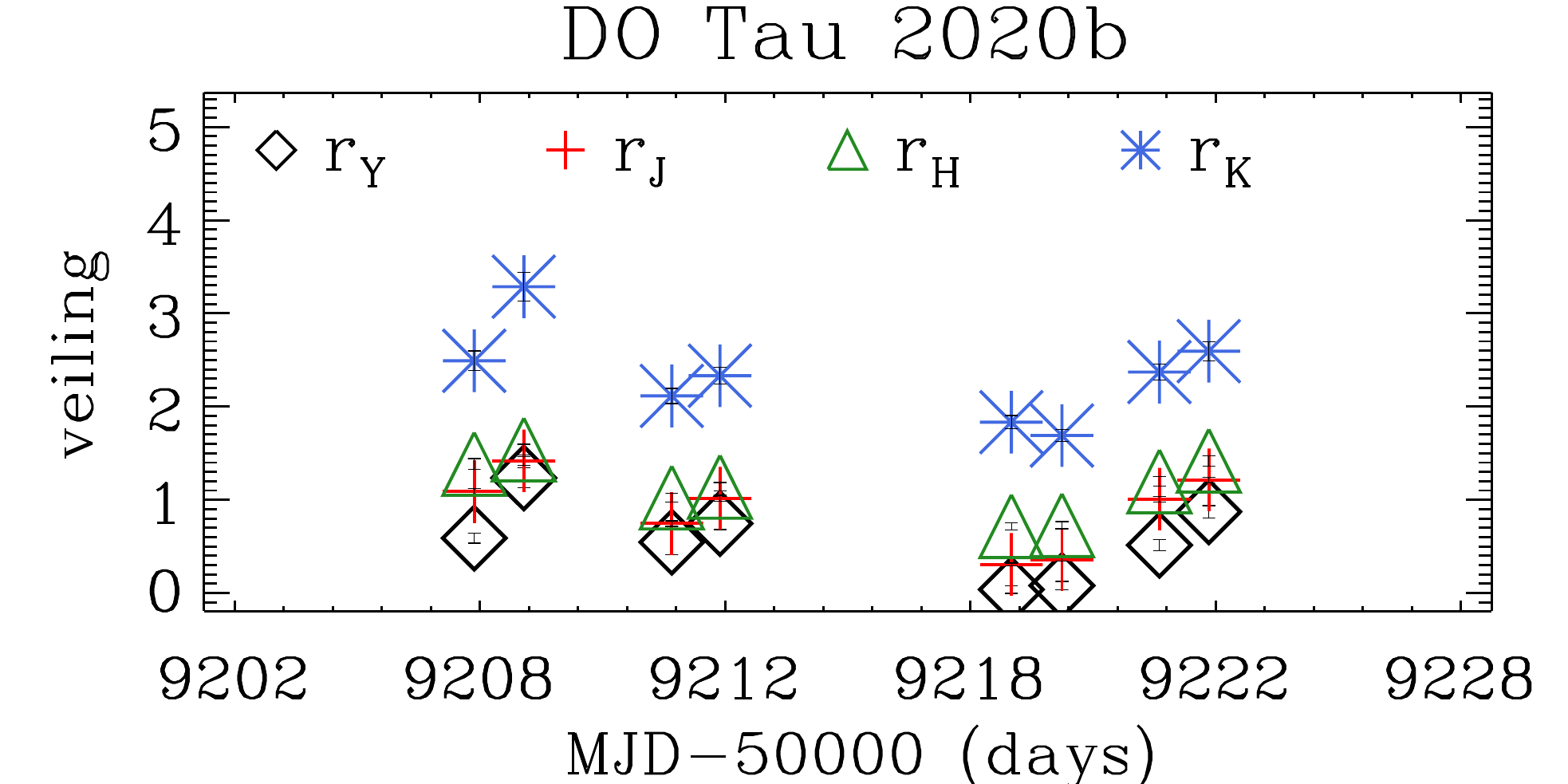}}
 {\includegraphics[width=4.4cm]{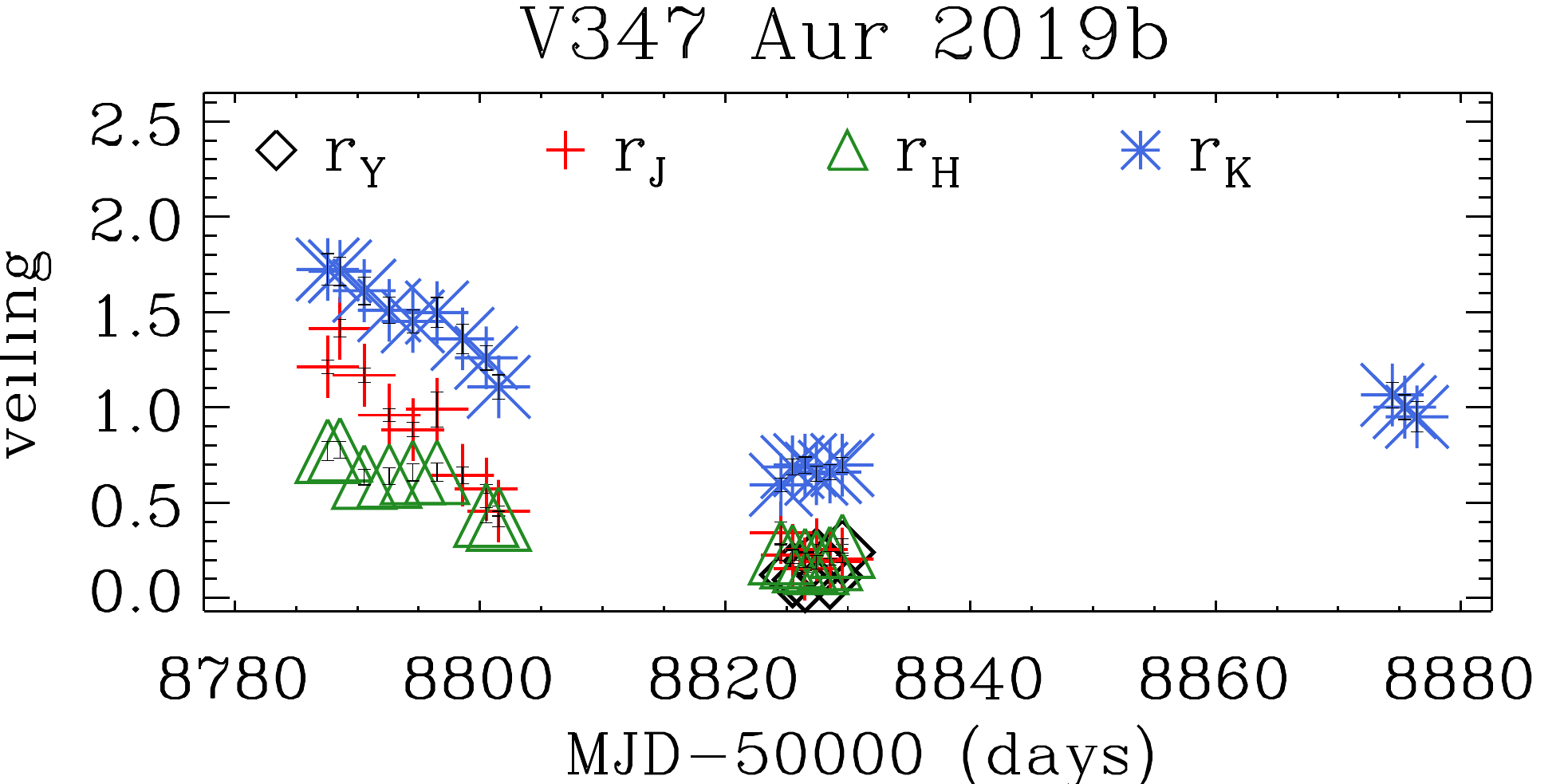}}
 {\includegraphics[width=4.4cm]{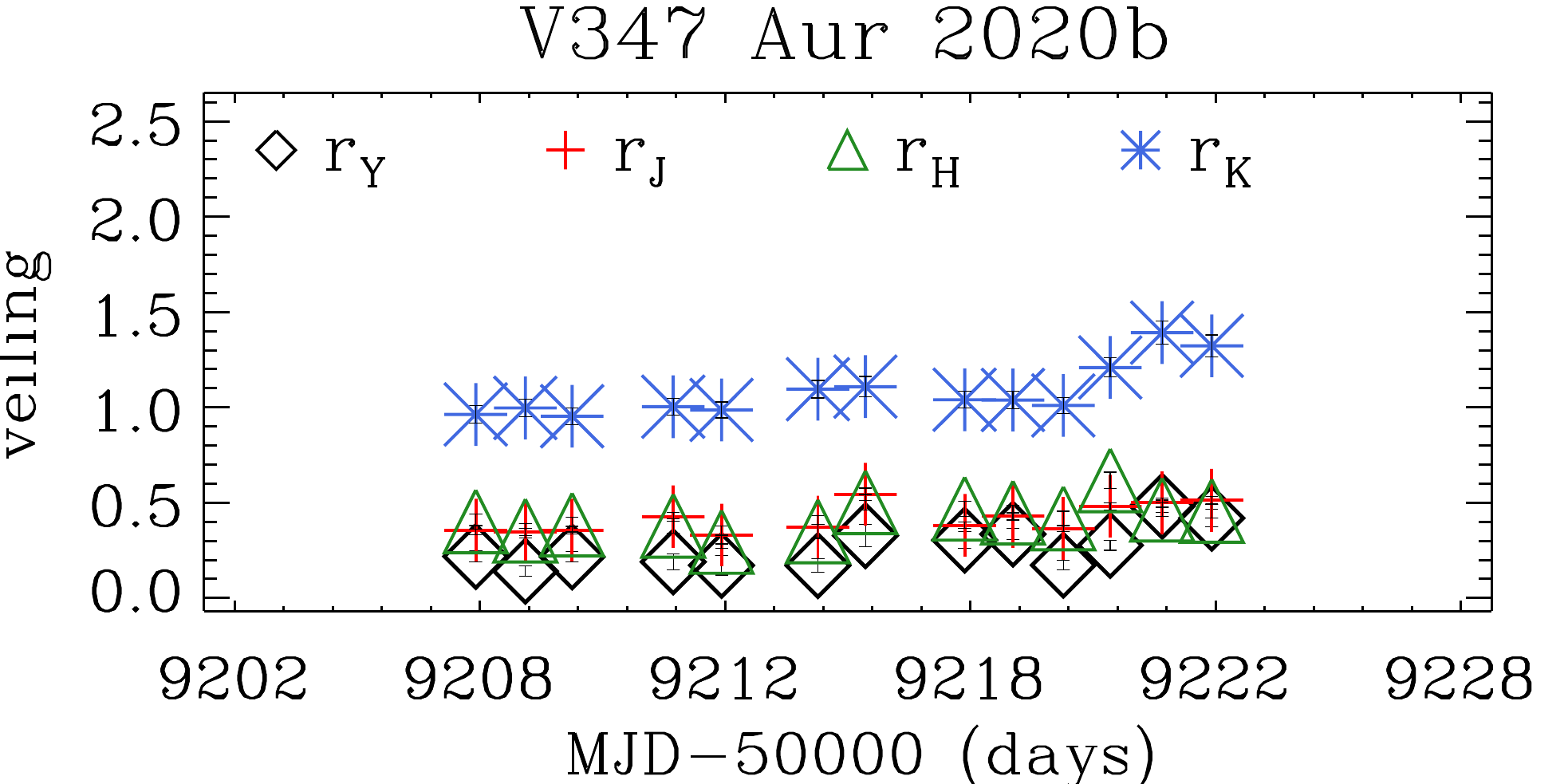}}
 {\includegraphics[width=4.4cm]{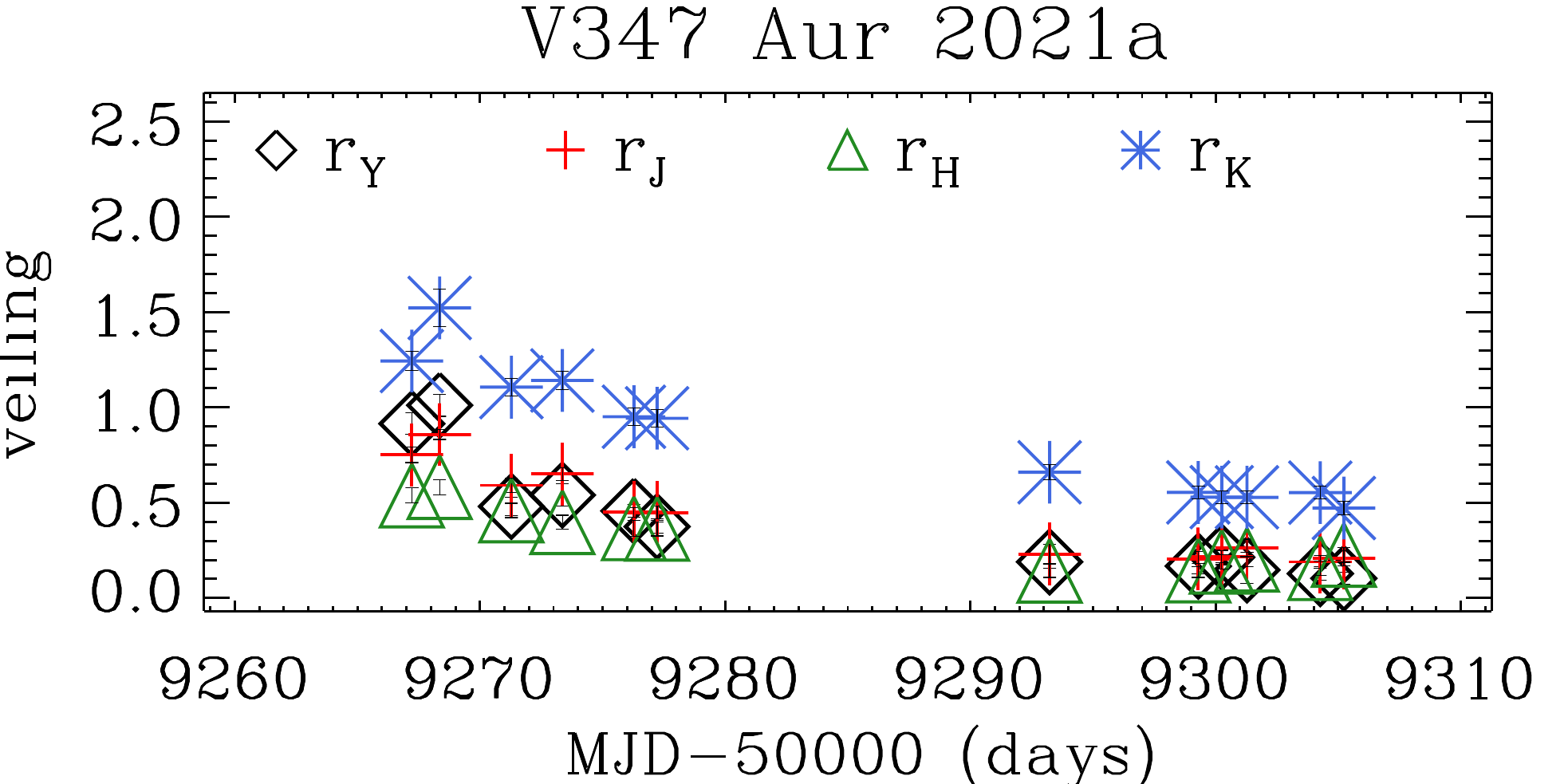}}
  {\includegraphics[width=4.4cm]{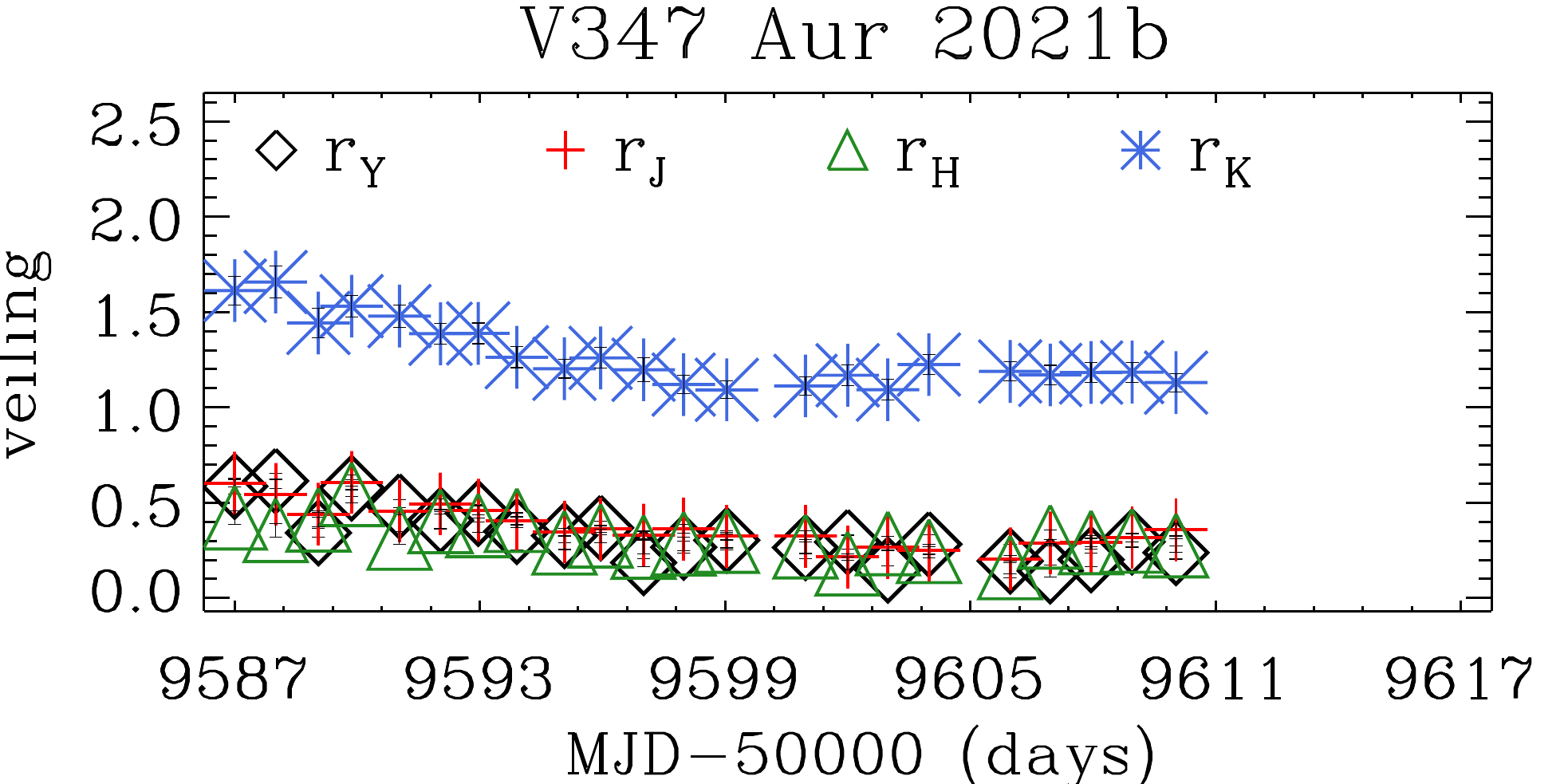}}
 {\includegraphics[width=4.4cm]{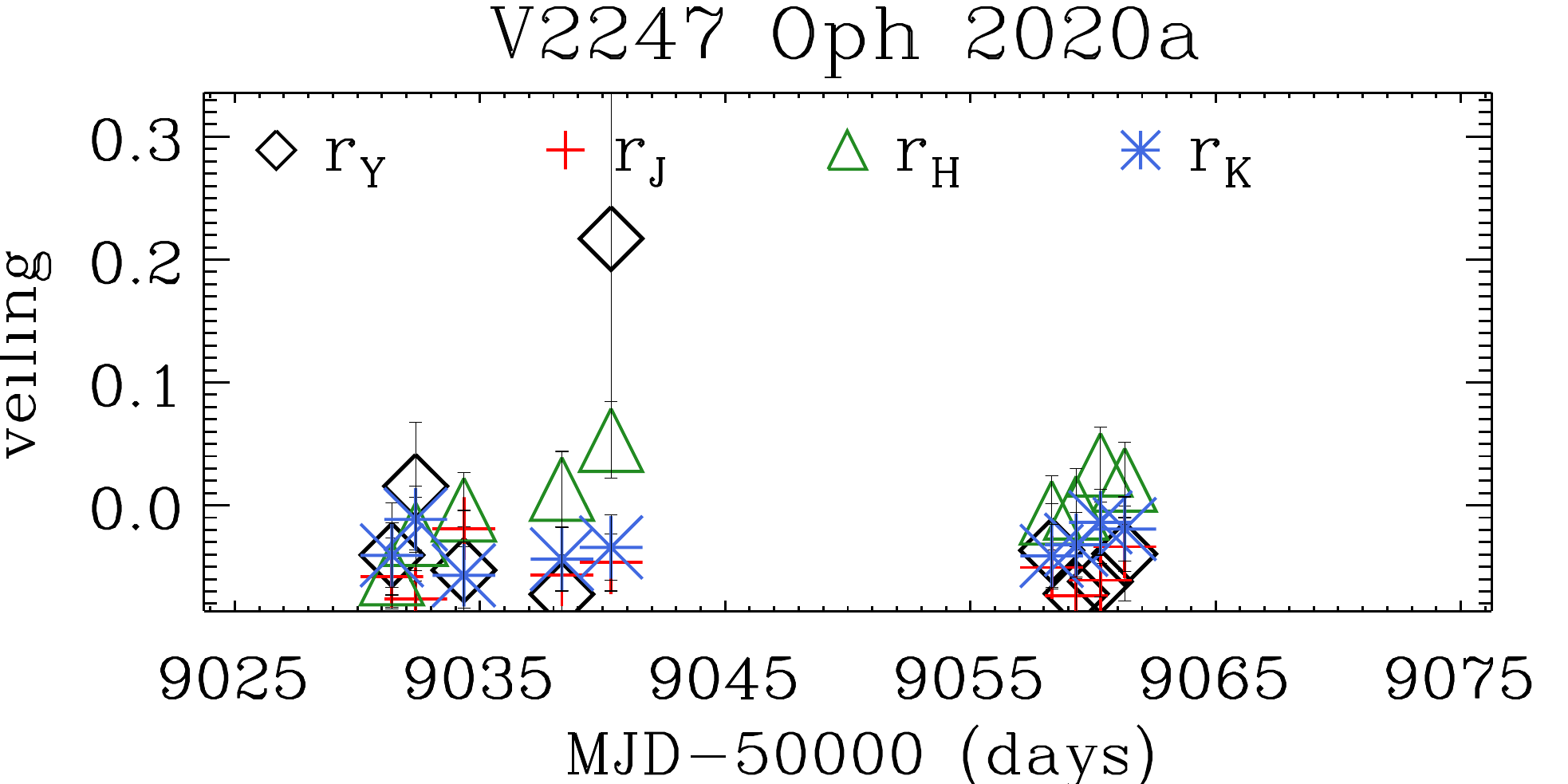}}
 {\includegraphics[width=4.4cm]{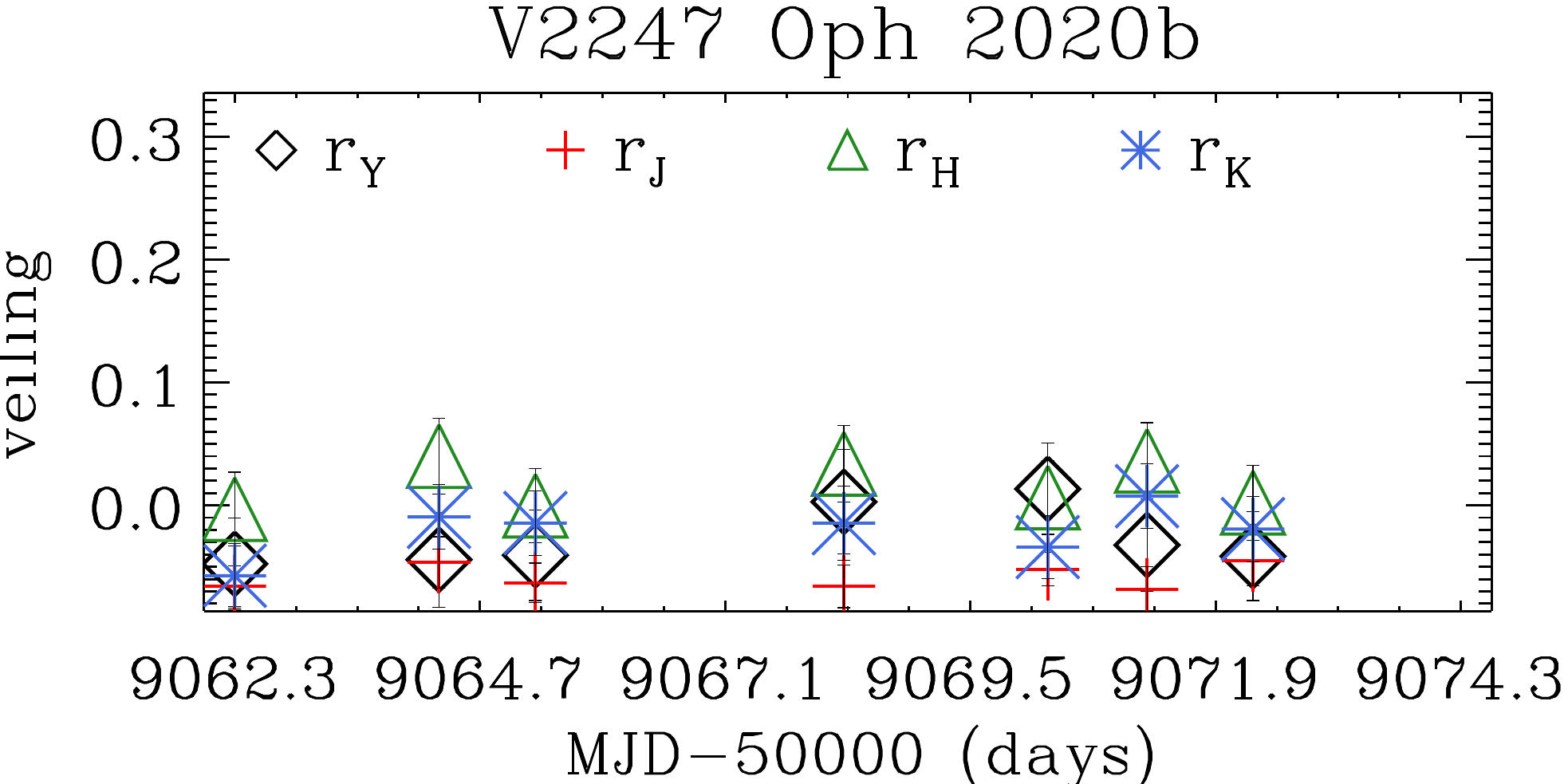}}
 {\includegraphics[width=4.4cm]{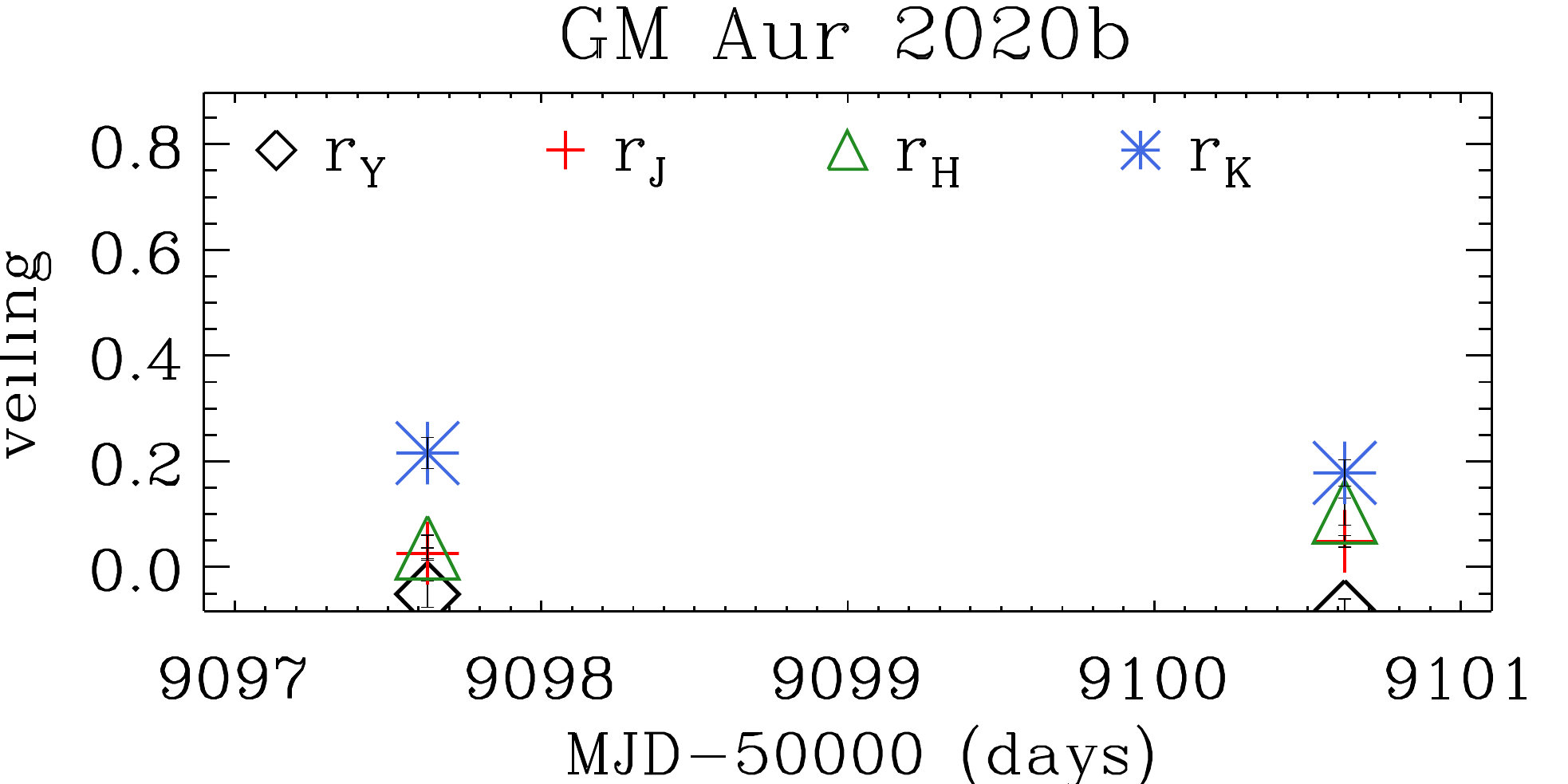}}
 {\includegraphics[width=4.4cm]{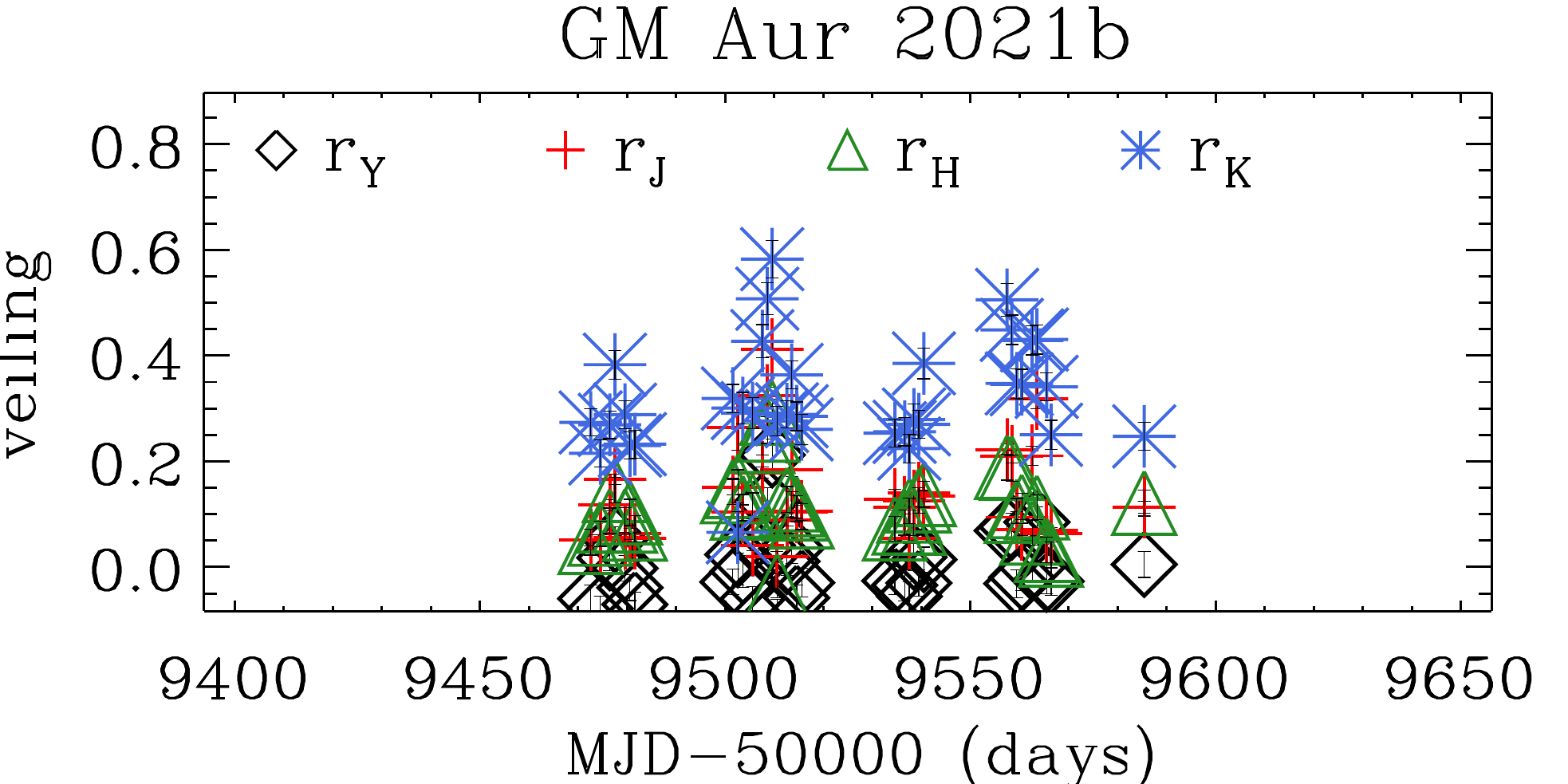}}
 \\
\caption{\label{fig:veil}Night-by-night veiling values measured in four different spectral regions. We show each observational period per target in an individual panel. A missing veiling point in a specific band means that the  spectral regions were subject to effects that prevented the veiling from being measured. In addition to the variability in terms of veiling, it is also not clearly periodic, at least on the timescale of stellar rotation. For more details, see text. } 
\end{figure*}

In this study, we have access to observations obtained on different nights and sometimes different observational periods for our sample of stars. This allowed us to analyze the night-to-night veiling variation and a possible long-term veiling variability on a timescale of two years. In Fig. \ref{fig:veil}, we show the veiling measured as a function of the observation dates. We used the modified Lomb-Scargle periodogram  \citep{1986ApJ...302..757H} to study a possible periodicity of the veiling variations.
We performed the periodogram analysis of the veiling in two ways: using all the observed nights and computing one periodogram per observational period. We searched for periods from 2 to 15 days or limited the search to the number of observed days, when the target was observed for less than 15 days. However, we did not find a significant periodic signal for most systems. We also phased the veiling using the known stellar rotation period of the targets, and again, most of the systems do not seem to vary in phase with the stellar rotation.

In Sect. \ref{sec:result}, we used the veiling error as a threshold of the veiling variability, using the RMS as a variability diagnostic, see Fig. 
\ref{fig:veilband}. Then, any point above this limit represents a true variability of the veiling; below this threshold, the variability is at the same level as the errors associated with the veiling.
 We can see that the veiling RMS presents values above the threshold of veiling variability for most of the targets and veiling bands. Only V2247 Oph shows most values below the threshold limit, while a few bands of TW Hya and DoAr 44 present most of the RMS values below the threshold. Then, for V2247 Oph, we cannot attribute a variability to the veiling, which is consistent with this target presenting a small veiling and weak level of accretion and inner disk emission diagnostics.  

We know the veiling is generally variable and we can trust the detection of veiling variability for most targets.  Figure \ref{fig:veil} shows that for all the veiling variable targets, the veiling varies on a timescale of at least one day and the veiling computed in the four spectral regions presents the same variability timescale. These results show that whatever region the IR veiling comes from, this region is dynamic and its flux changes on a timescale of days. 

\begin{figure}
\centering
{\includegraphics[width=9.cm]{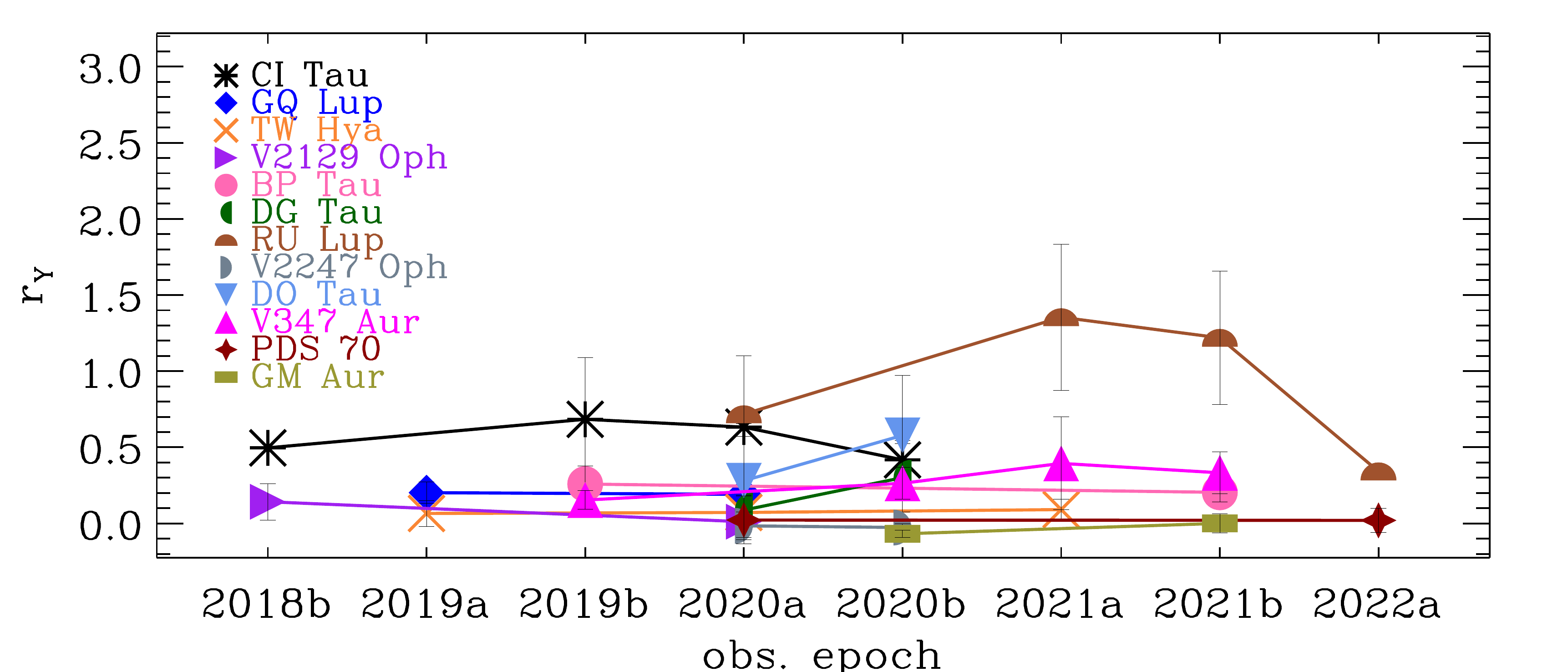}}\\
{\includegraphics[width=9.cm]{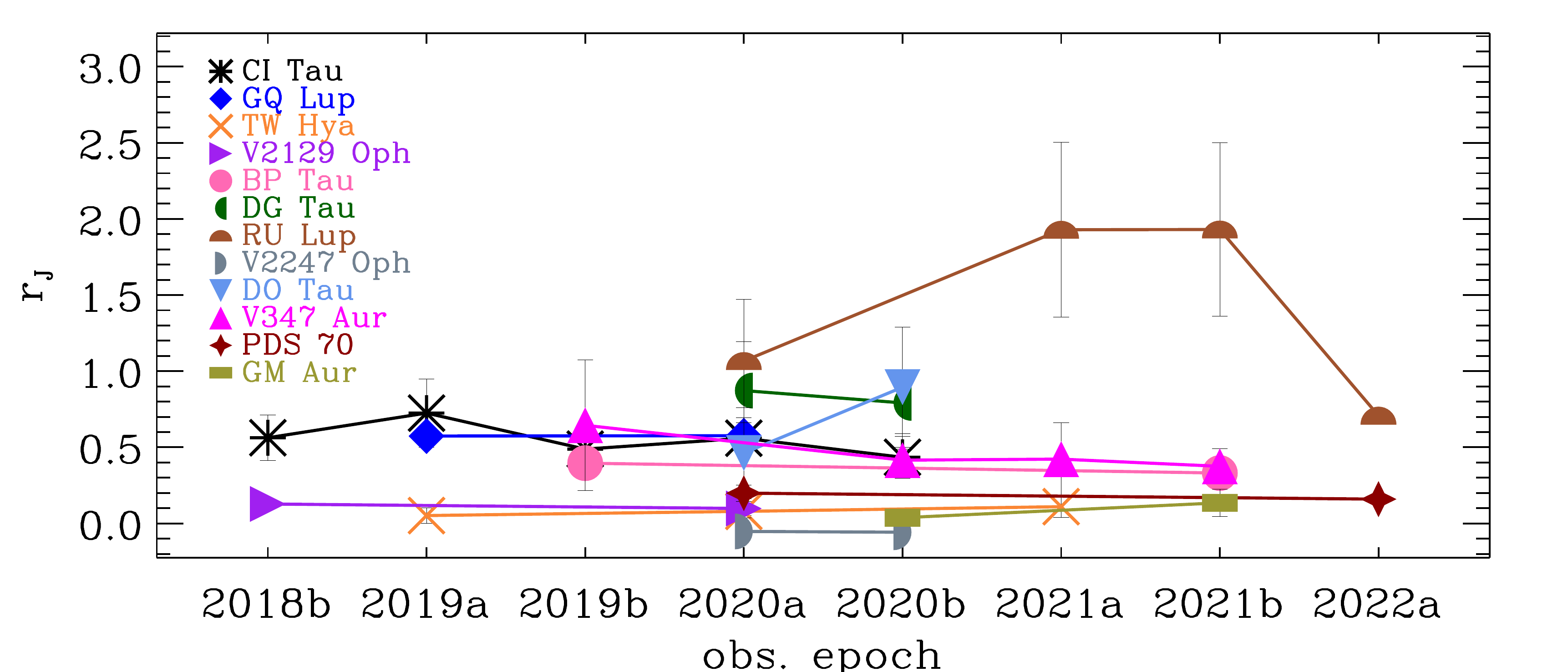}}\\
{\includegraphics[width=9.cm]{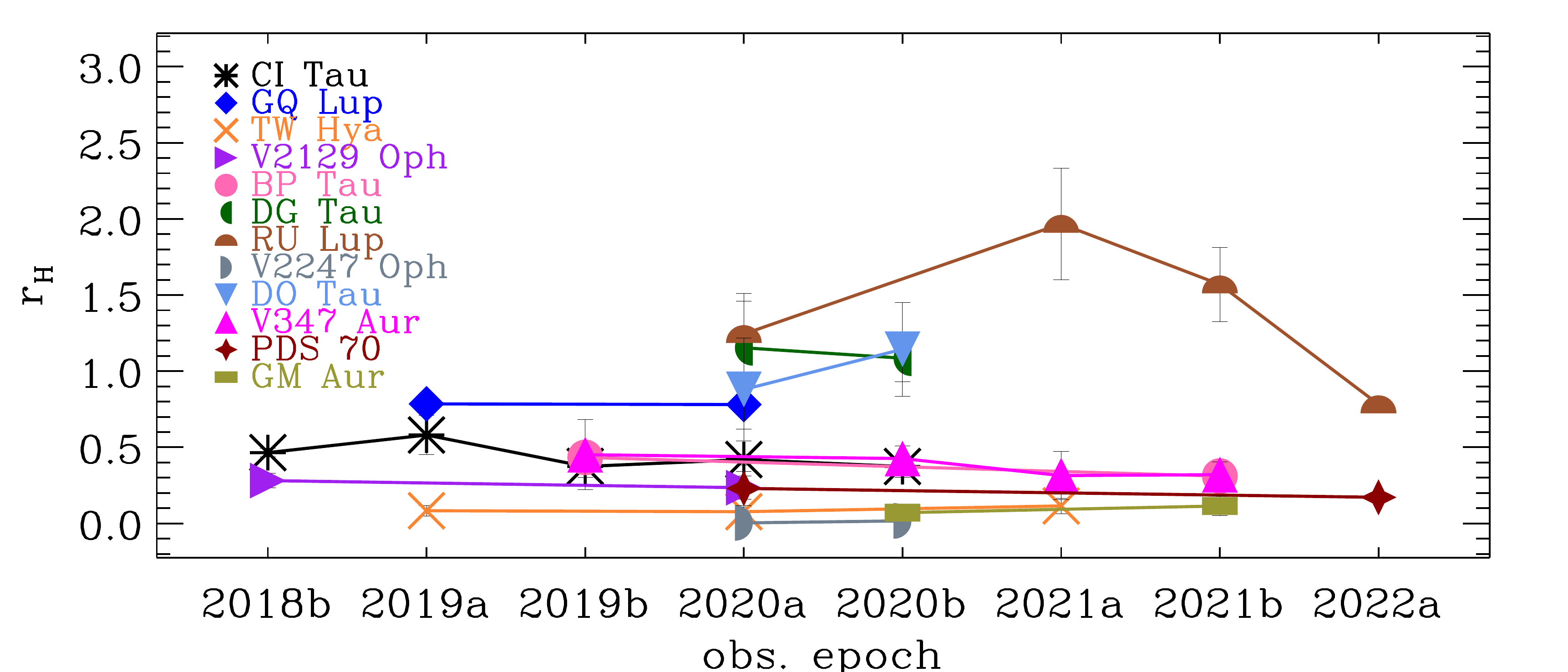}}\\
{\includegraphics[width=9.cm]{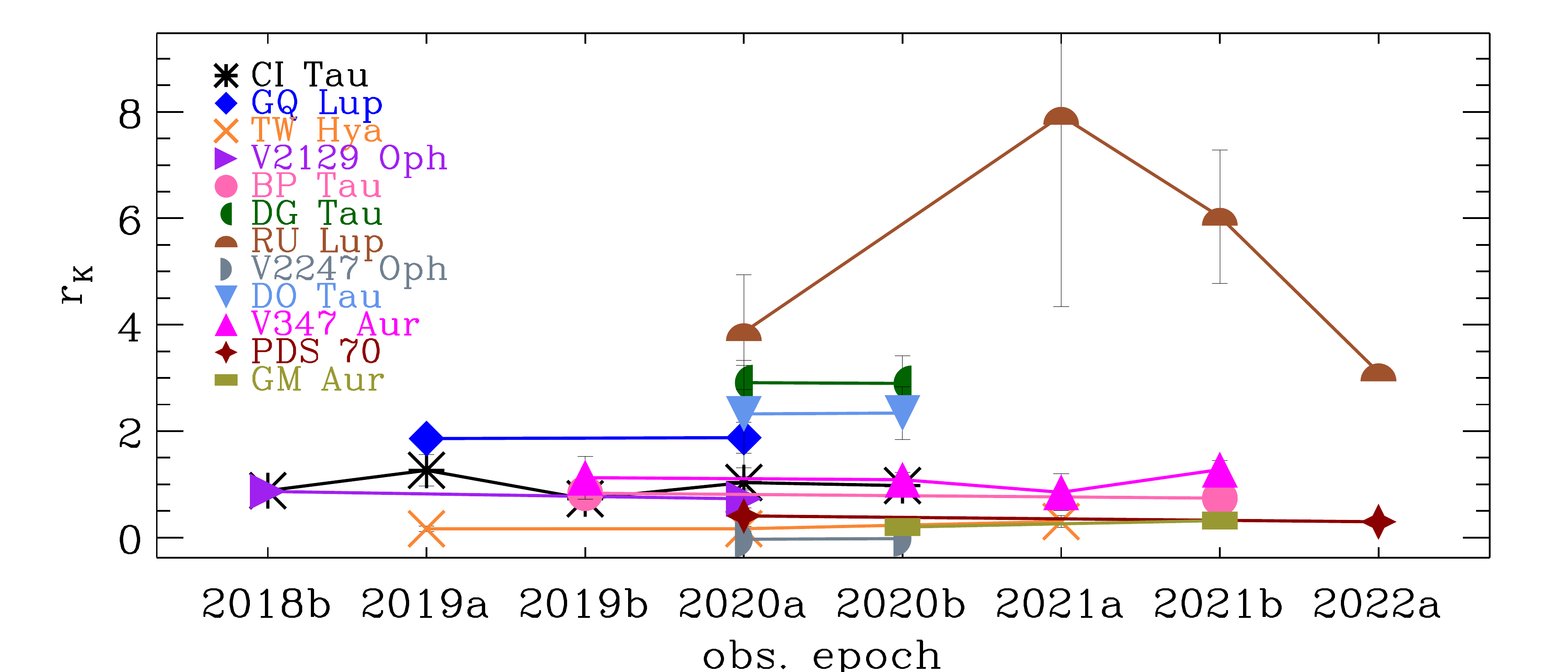}}
\caption{\label{fig:veil_epoch} Average of the NIR veiling computed at each observational period, with the error bar as the standard deviation of the computed mean veiling.  We merged the measured veiling in 2020a and 2020b of the stars V2129 Oph and GQ Lup, as they are  successive observations. The mean NIR veiling seems to be stable for most of the targets for a few months or years.} 
\end{figure}

We also investigated veiling variability on a timescale of months to a few years; a change in the veiling in that timescale can have a different origin from the day-scale veiling variability, discussed above. The latter can be associated with the dynamic of the system's rotation, while the possible long-term veiling variability reflects a change in the system's accretion and/or inner disk conditions. In Figure \ref{fig:veil_epoch}, we present the averaged veiling measured at each observational period. This plot displays nine systems that were observed in more than one observational season. Most targets do not present a significant difference in the veiling along the observational period. The $K$ band veiling of CI Tau and the $YJH$ veiling of DO Tau show a possible small change in this timescale.  Then, we conclude that the veiling variability on a timescale of months-to-years is on the same order of magnitude as the day-to-day variability.  RU Lup is the unique target with an evident change in the veiling along the observational periods in the four bands, but much more pronounced in the $K$ band, along with a high standard deviation. We associate this change in the veiling with an occasional high accretion episode that occurred in 2021a, and despite the veiling still being high in 2021b, it seems to start to diminish later on. In 2022a, it is even smaller, but we have only one observation to serve as the basis for this assumption. The circumstellar emission lines' equivalent widths corroborate with this assumption, as they increase in 2021a and start to decrease in the subsequent observation periods, similarly to the veiling. In contrast, the average veiling is stable, at least for most of the stars we analyzed, except for very high accreting systems, such as RU Lup, which can present episodic high veiling. Furthermore, in the same observational period, some targets show a few episodes of an increase in veiling, such as V347 Aur (Fig. \ref{fig:veil}); however,    the average veiling values are still sustained.

\section{Discussion}\label{sec:discussion}

The dependence of veiling on wavelength is ubiquitous from the UV to the NIR range.
The veiling in the optical domain decreases from the blue to the red part of the spectra, which is an effect of the decrease of the accretion spot continuum contribution \citep[e.g.,][]{1998ApJ...509..802C}; also, the veiling also does not vary for some wavelength ranges
\citep[e.g.,][]{1990ApJ...363..654B}. On the other hand, in the IR range, the veiling increases with wavelength, as seen in Figs. \ref{fig:veilband} and \ref{fig:veil}, which is in agreement with similar results in the literature \citep[e.g.,][]{2011ApJ...730...73F,2021A&A...652A..72A} 

\begin{figure*}
\centering
{\includegraphics[width=6cm]{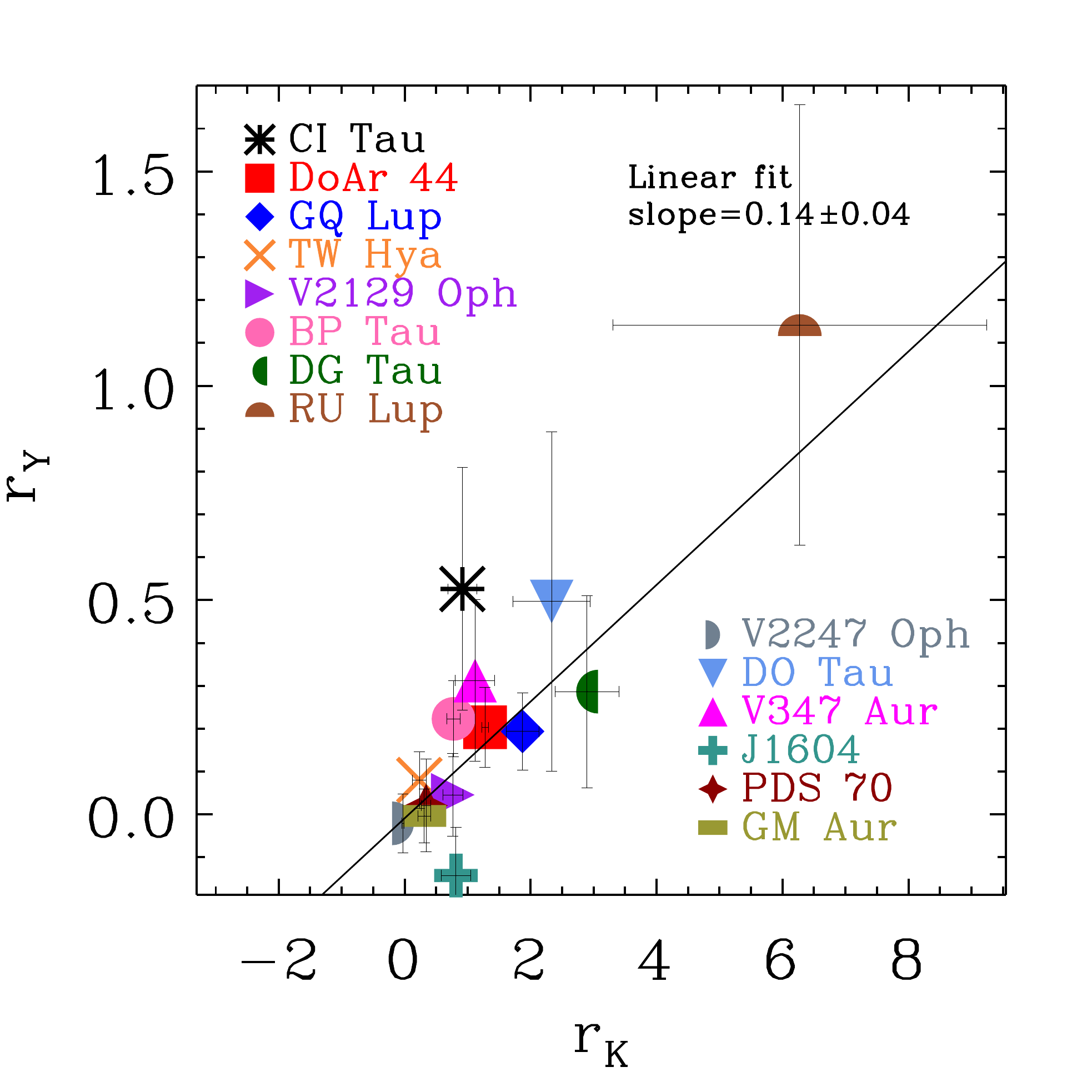}}
{\includegraphics[width=6cm]{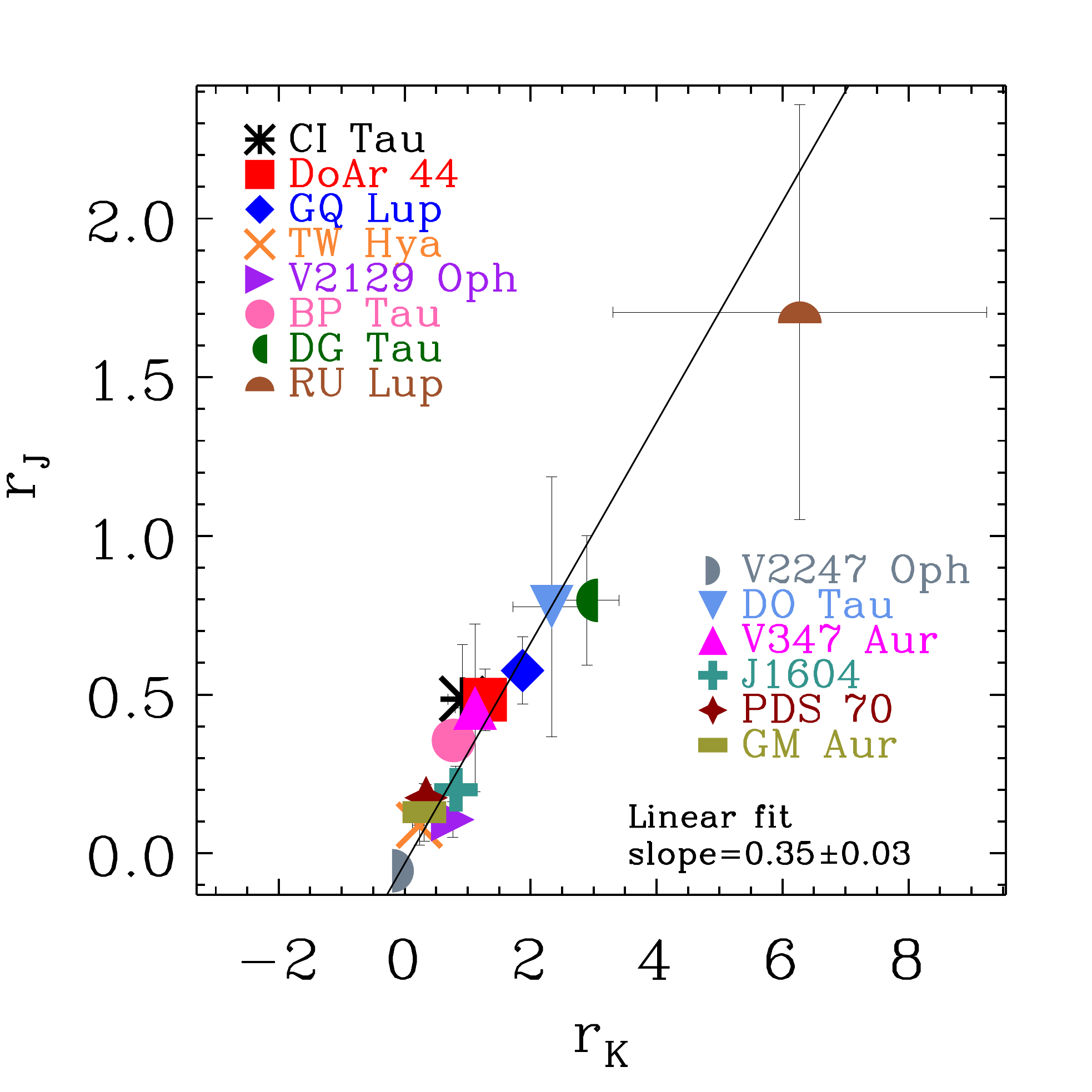}}
{\includegraphics[width=6cm]{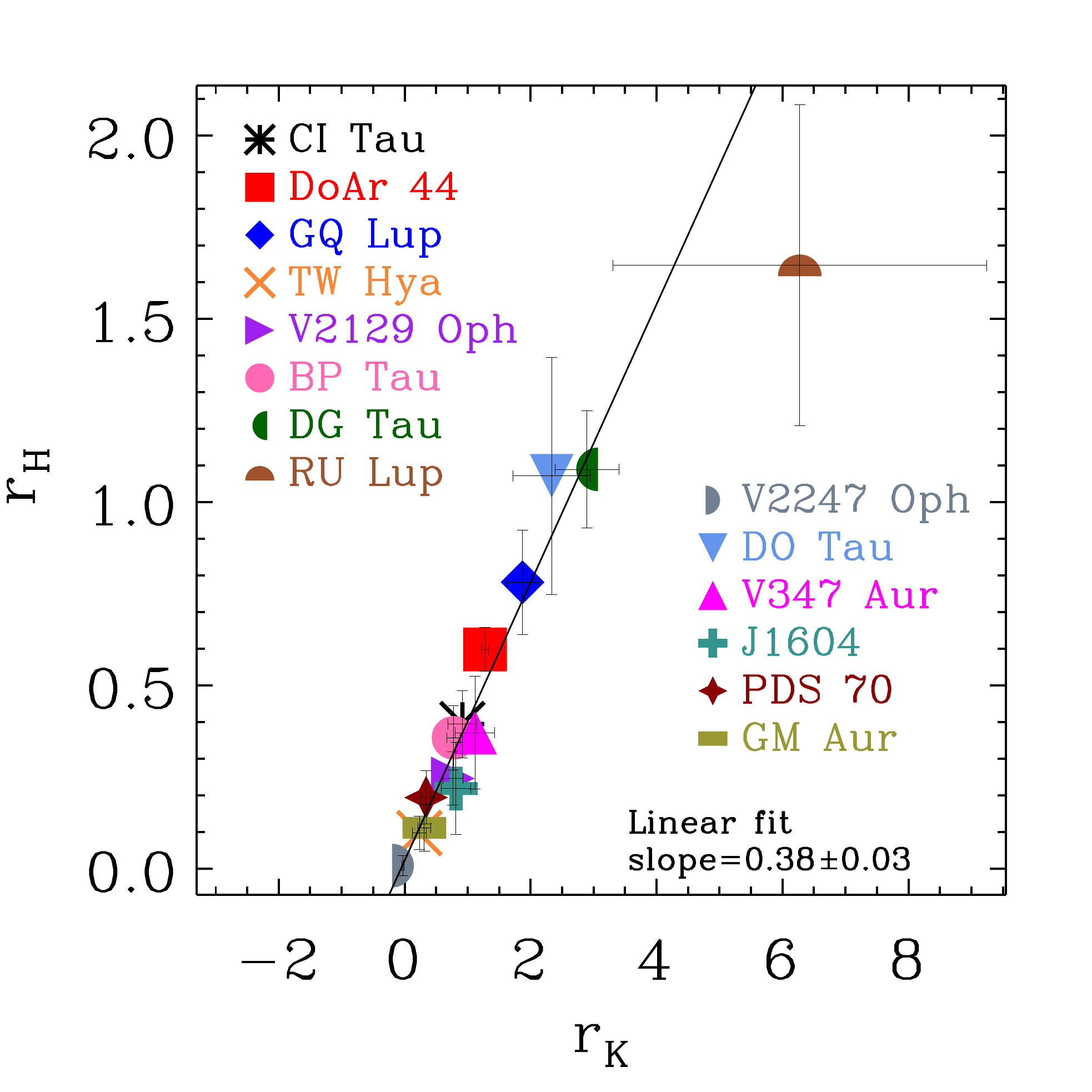}}
\caption{\label{fig:rYJHvsrK} $YJH$ veiling as a function of the $K$ band veiling. Each veiling value corresponds to the mean of all the observed nights, and the error bar is its standard deviation. The solid line is the linear fit to the data, and the fitted line's slope is written in each panel. The correlation between the $K$ band and the $YJH$ bands veilings increases from the $Y$ to the $H$ band.} 
\end{figure*}

 The average veiling value for the entire sample of CTTS is $\left<r_Y\right>=0.2\pm0.3$, $\left<r_J\right>=0.4\pm0.4$, $\left<r_H\right>=0.5\pm0.5$, and $\left<r_K\right>=1.4\pm1.6$. We note that in the $Y$ band, the average veiling is the lowest. Over these wavelengths from $Y$ to $K$, the veiling can have contributions from different sources. For example, we expect the veiling in the $K$ band to present more contributions from the inner disk than in the $Y$ band. In Fig. \ref{fig:rYJHvsrK}, we show the $YJH$ veiling as a function of the $K$ band veiling. We can see that the $J$ and $H$ veilings seem to increase as the $K$ band veiling increases; however, there is smaller correlation with the $Y$ band veiling. These results are supported by the analysis of the correlation of the veiling samples and the linear fit showed in the Fig. \ref{fig:rYJHvsrK}. We computed the linear correlation coefficient (r) between two samples, where r=1 represents a perfect correlation and when r=0 there is no correlation. The $Y$ and $K$ band veilings present a correlation coefficient of 0.87, while the coefficient of the $J$ and $H$ and the $K$ band are 0.98 and 0.96, respectively. Similar results were found by \cite{2005ApJ...635..422C}, where they compare the excess in the $J$ and $H$ bands with the $K$ band excess, showing that both presented a linear correlation with the $K$ band excess, and this was explained as due to the $JHK$ excess arising from the same region.

The NIR veiling should be the result of a combination of physical processes. \cite{2021A&A...652A..72A} computed the NIR veiling at several wavelengths for a sample of very high-mass accretion rate systems, including DG Tau (included in our sample). These authors fitted the veiling as a function of wavelength using a blackbody function and found temperatures compatible with the presence of dust in the inner disk. However, the temperature was too high ($>2000\,K$) for the dust to survive in a few of their cases. They also argued that the veiling should have a contribution from the hot gas inside the disk sublimation radius, and similar results were found by \cite{2011ApJ...730...73F}. To investigate this proposition, we looked at the CO bandhead at 2.3\mum\ of the CFHT/SPIRou data. This band, when in emission, is expected to form in the hot gas in the inner disk. Using the $K$ band veiling, we veiled the template and removed the photospheric lines of the CO bandhead to obtain the residual CO profiles. Most targets do not present clear signs of CO emission, showing that this band is strictly photospheric. However, {a few } residual profiles of V347 Aur, which is a Class I object, along with DO Tau and RU Lup, which  are strong accretors, present CO emission in some observations, indicating the presence of hot gas in the inner disk. In particular, RU Lup presents these hints of CO emission in the observational period when the veiling was high, and the system probably ensued a high episodic accretion. A further analysis of the CO bandhead is beyond the scope of this paper and it will be carried out in a dedicated paper exploring the significance of these CO emissions.       

In the previous section, we show that the NIR veiling, mainly $r_K$, presents a good correlation with the inner disk emission diagnostics obtained from the NIR photometric data and SED fit, demonstrating that the NIR veiling has an important contribution from the inner disk. However, we also see a correlation between veiling and accretion diagnostics. A high accretion-rate system presents larger veiling values in the IR. This shows that high-mass accretion rate systems should feature higher inner disk heating, thus higher temperatures and stronger inner disk emission excess as a consequence. \cite{2022AJ....163..114E} fit most of the continuum spectra from NUV to NIR of the accreting star CVSO 109A quite well,  using a combination of emission from the accretion shock (multiple funnel flow model) on the stellar surface and emission from the irradiated inner edge of the dusty disk. 
However, the inner disk and accretion shock model do not adequately reproduce the continuum excess in the $Y$ to $J$ band.

Indeed, while the veiling in the $J$ to $K$ band seems to point to a significant contribution on the part of the inner disk emission excess, the $Y$ band veiling origin is still unknown.  \cite{2013AstL...39..389D} predicted significant veiling in the $Y$ band from the accretion spot. They argued that the accretion spot (continuum emission and emission lines formed in the accretion shock) could account for optical and near IR veiling up to the $J$ band. Unfortunately, our $Y$ band veiling was not as well computed as for the other bands due to several issues in this spectral region and given the photospheric lines are not quite so prominent. Despite these obstacles, the Y band veiling agrees on a smaller scale with both accretion and inner disk diagnostics.

\begin{figure}
\centering
{\includegraphics[width=4.3cm]{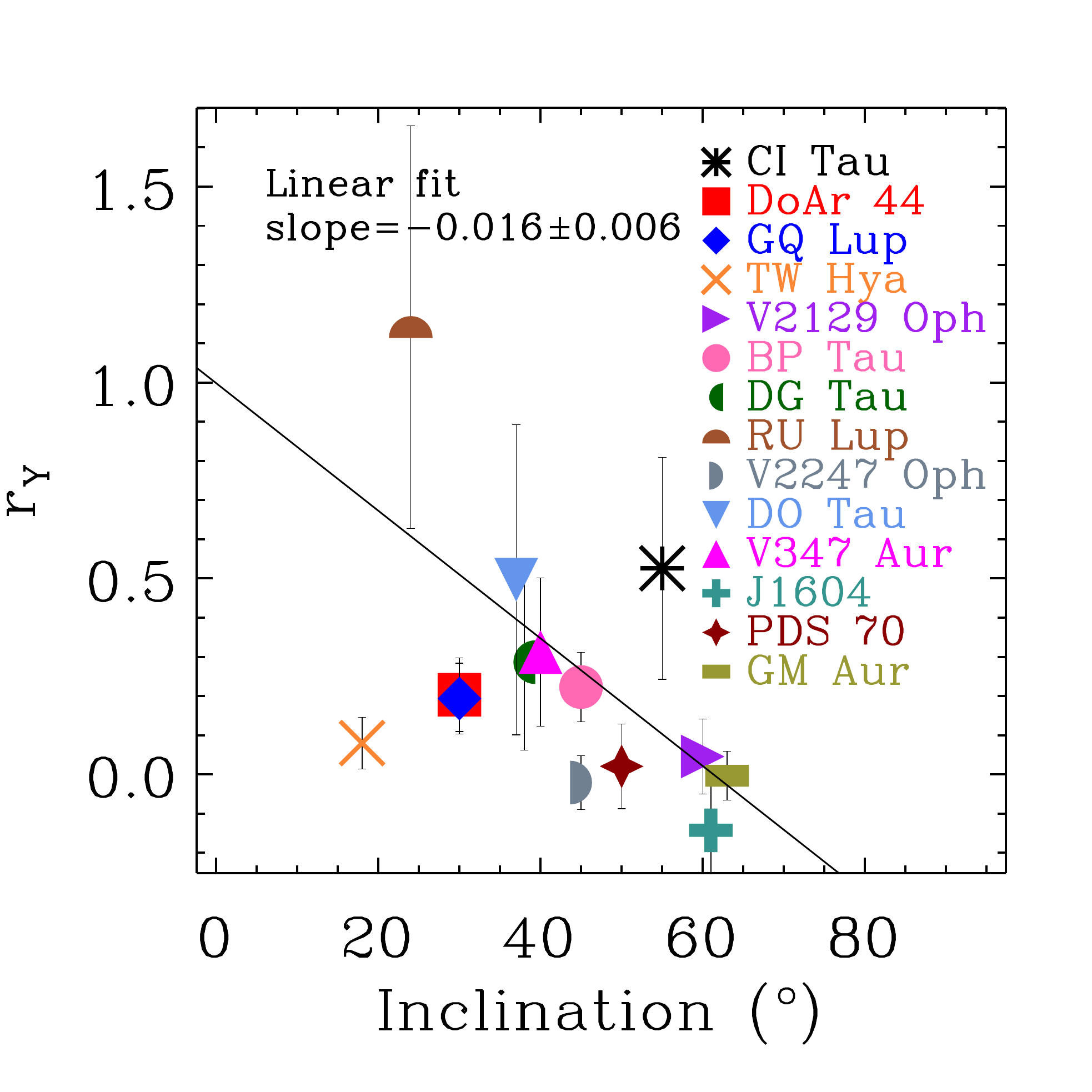}}
{\includegraphics[width=4.3cm]{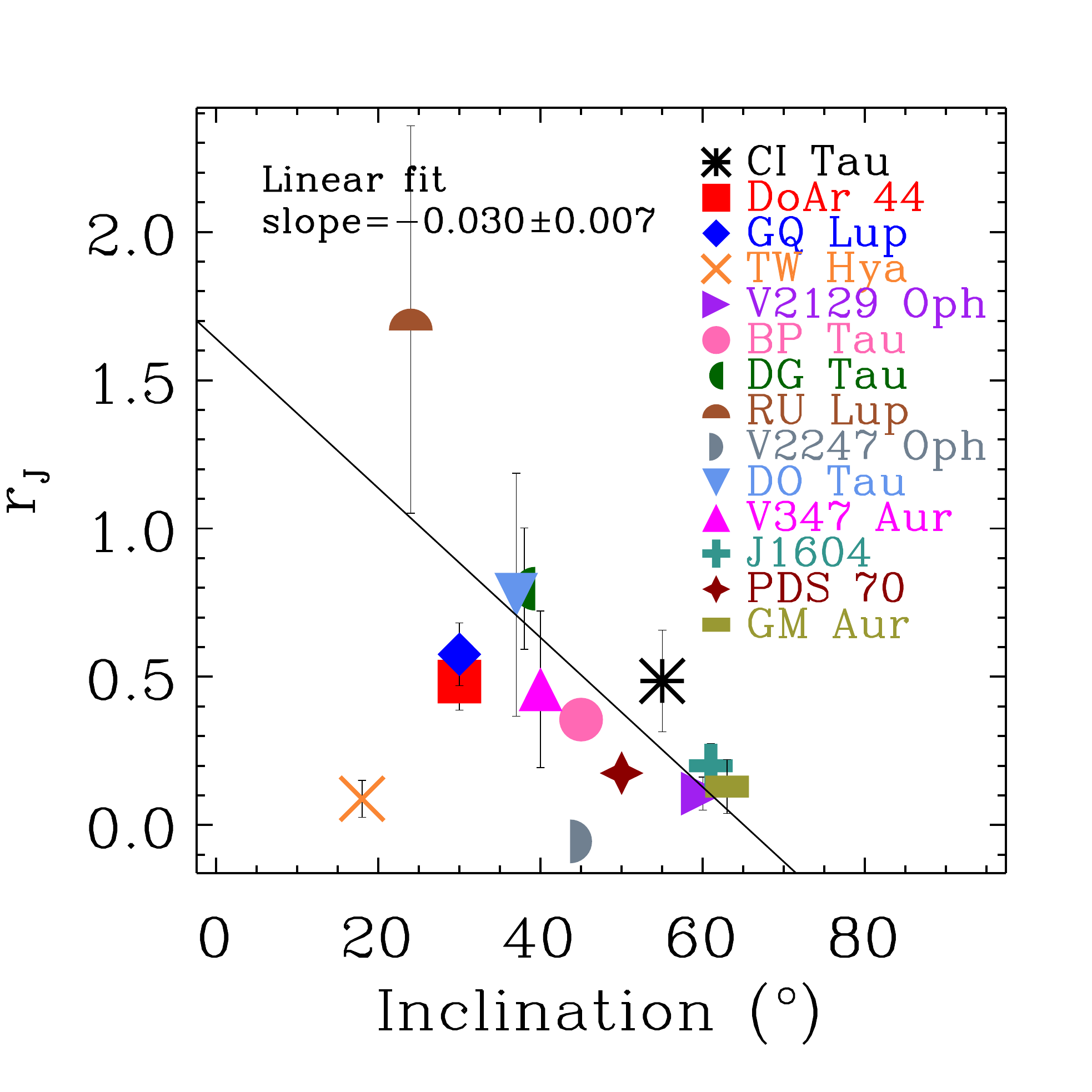}}\\
{\includegraphics[width=4.3cm]{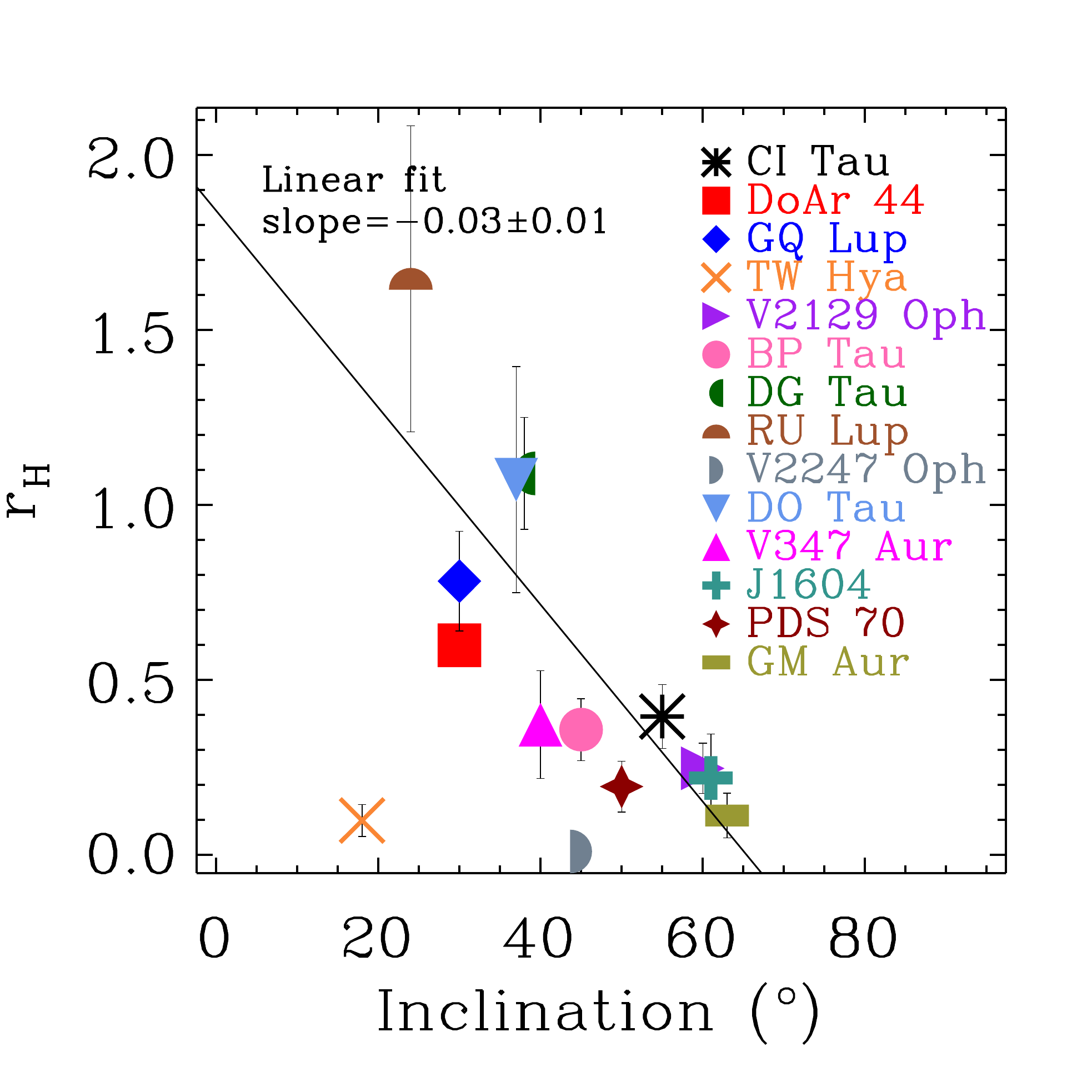}}
{\includegraphics[width=4.3cm]{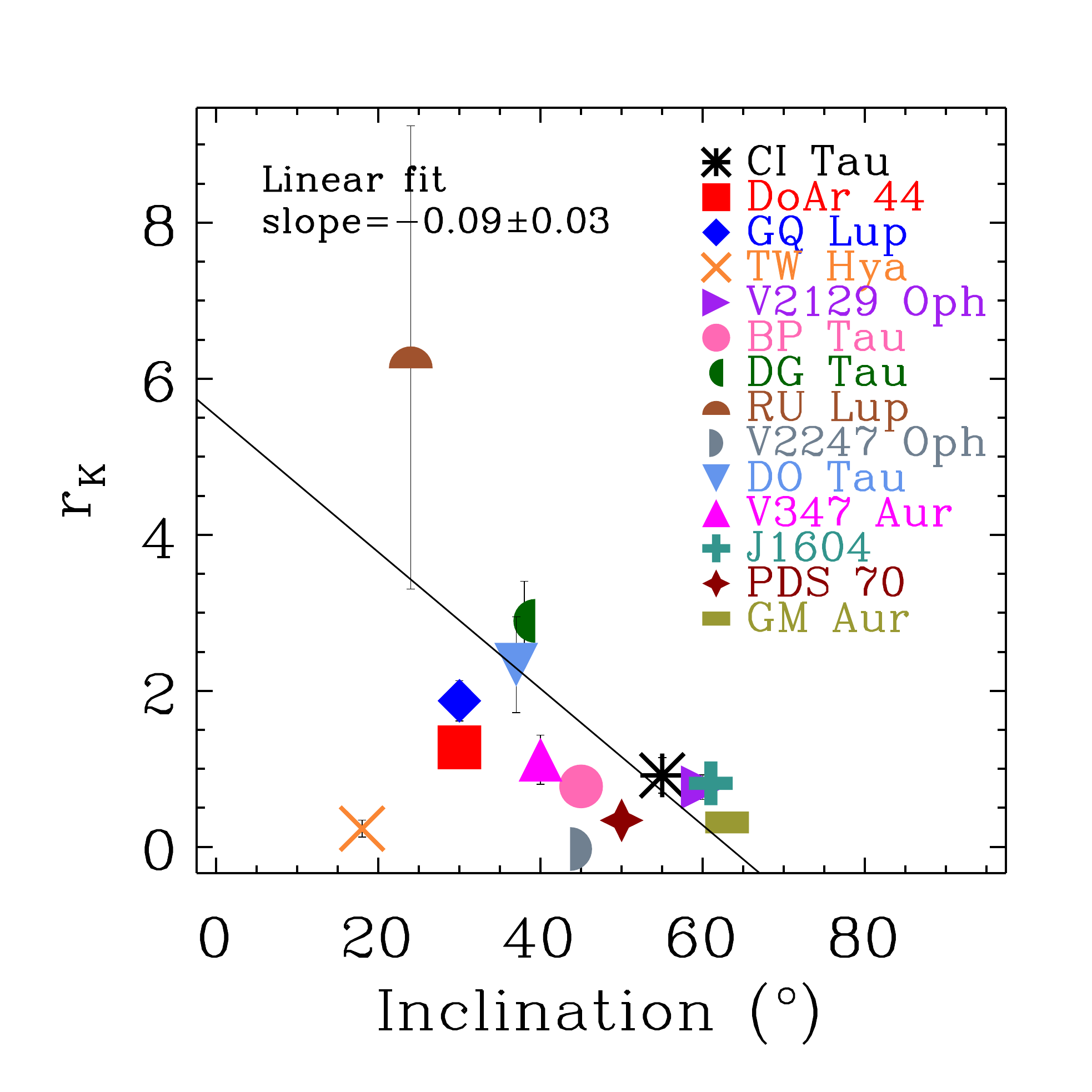}}
\caption{\label{fig:veilInc} Average NIR veiling as a function of the system's inclination with respect to our line of sight.  The solid line is the data's linear fit, and the fitted line's slope is given in each panel. We do not consider the discrepant targets (TW Hya and V2247 Oph) to fit the data. The NIR veilings tend to be anti-correlated with the system's inclination, see text.} 
\end{figure}

We checked if the inclination of the system has any impact on the measured veiling values. In Fig. \ref{fig:veilInc}, we show the NIR veiling as a function of the  inclination of the system with respect to our line of sight, listed in Table \ref{tab:obs}.  In addition, two discrepant systems (TW Hya and V2247 Oph), we can see an anti-correlation between veiling and the system's inclination. This anti-correlation is not pronounced, as we can see in the linear fit slope, due to the spread of points (the correlation coefficients between the inclination and the $YJHK$ veilings are -0.64, -0.76, -0.80, and -0.70, respectively), but the decreasing tendency of veiling with inclination is clear.  We also advise that the inclinations used for DG Tau and DO Tau are the outer disk inclination, while the disk and the stellar inclination are not necessarily the same \citep{2022A&A...658A.183B}.  If confirmed, this anti-correlation can be due to a geometric effect: the more the system is inclined, the less we see (from the inner disk edge) where the nIR veiling is supposed to arise. The two targets, TW Hya and V2247 Oph, which do not seem  to follow this tendency, do not have dust in the inner disk any longer and they are known to have gaps or holes in their inner disks \citep{2002ApJ...568.1008C,2008ApJ...684.1323P}. In that case, independently of the system's inclination, we would not expect to detect IR veiling, assuming that the IR veiling is due to dust emission in the inner disk.

\section{Conclusion}\label{sec:concl}

In this work, we analyze the NIR veiling computed using high-resolution data from CFHT/SPIRou of a sample of 14 low-mass young stars. We found the veiling to increase from the $Y$ to the $K$ band, as a result of the increase of the emission contribution from the inner disk as a function of wavelength. 

The veiling correlates with other photometric inner disk diagnostics, such as color excess and the slope of the spectral energy distribution, mainly in the $JHK$ band, providing further evidence that the NIR veiling arises from hot dust in the inner disk. We also found a linear correlation between veiling and the accretion properties of the system. This shows that accretion contributes to inner disk heating and, consequently, to the inner disk emission excess. This effect is enhanced in high-mass accretion rate systems that also present a denser inner disk and higher inner disk emission \citep[e.g.,][]{2022ApJ...928..134S}.

We analyzed the NIR veiling variability through the modified Lomb-Scargle periodogram and we did not find any significant periodic signal in the four bands in timescales typical of stellar rotation ($<15\,$days), which also suggests the veiling comes from the dust emission in the inner disk.  
However, we show that the veiling is variable for most targets on a timescale of at least one day.
Besides the night-by-night veiling variability, the mean NIR veiling per season appears to be mostly stable, for most targets and on timescales of several months to years. 

\begin{acknowledgements}
We thank the referee for the suggestions that helped to clarify this paper. We want to thank Claire Moutou, Sylvie Cabrit, Nicolas Grosso, and Konstantin Grankin for carefully reading the manuscript and giving suggestions to improve the paper.
This project has received funding from the European Research Council (ERC) under the European Union’s Horizon 2020 research and innovation programme (grant agreement No 742095; SPIDI: Star-Planets-Inner Disk-Interactions; http://www.spidi-eu.org and grant  agreement  No.  740651 NewWorlds). We acknowledge financial support from CNPq, CAPES and Fapemig. We acknowledge funding from the French National Research Agency (ANR) under contract number ANR-18-CE31-0019 (SPlaSH).
This research has made use of the SVO Filter Profile Service (http://svo2.cab.inta-csic.es/theory/fps/) supported from the Spanish MINECO through grant AYA2017-84089.
The authors wish to recognize and acknowledge the very significant cultural role and reverence that the summit of MaunaKea has always had within the indigenous Hawaiian community. We are most fortunate to have the opportunity to conduct observations from this mountain. 
We acknowledge with thanks the variable star observations from the AAVSO International Database contributed by observers worldwide and used in this research.

\end{acknowledgements}

\bibliographystyle{aa}   
\bibliography{ref}

\begin{thebibliography}{90}
\expandafter\ifx\csname natexlab\endcsname\relax\def\natexlab#1{#1}\fi

\bibitem[{{Alcal{\'a}} {et~al.}(2021){Alcal{\'a}}, {Gangi}, {Biazzo},
  {Antoniucci}, {Frasca}, {Giannini}, {Munari}, {Nisini}, {Harutyunyan},
  {Manara}, \& {Vitali}}]{2021A&A...652A..72A}
{Alcal{\'a}}, J.~M., {Gangi}, M., {Biazzo}, K., {et~al.} 2021, \aap, 652, A72

\bibitem[{{Alcal{\'a}} {et~al.}(2017){Alcal{\'a}}, {Manara}, {Natta}, {Frasca},
  {Testi}, {Nisini}, {Stelzer}, {Williams}, {Antoniucci}, {Biazzo}, {Covino},
  {Esposito}, {Getman}, \& {Rigliaco}}]{2017...600A..20A}
{Alcal{\'a}}, J.~M., {Manara}, C.~F., {Natta}, A., {et~al.} 2017, \aap, 600,
  A20

\bibitem[{{Alcal{\'a}} {et~al.}(2014){Alcal{\'a}}, {Natta}, {Manara}, {Spezzi},
  {Stelzer}, {Frasca}, {Biazzo}, {Covino}, {Randich}, {Rigliaco}, {Testi},
  {Comer{\'o}n}, {Cupani}, \& {D'Elia}}]{2014AA...561A...2A}
{Alcal{\'a}}, J.~M., {Natta}, A., {Manara}, C.~F., {et~al.} 2014, \aap, 561, A2

\bibitem[{{Alencar} \& {Batalha}(2002)}]{2002ApJ...571..378A}
{Alencar}, S. H.~P. \& {Batalha}, C. 2002, \apj, 571, 378

\bibitem[{{Alencar} {et~al.}(2012){Alencar}, {Bouvier}, {Walter}, {Dougados},
  {Donati}, {Kurosawa}, {Romanova}, {Bonfils}, {Lima}, {Massaro}, {Ibrahimov},
  \& {Poretti}}]{2012AA...541A.116A}
{Alencar}, S.~H.~P., {Bouvier}, J., {Walter}, F.~M., {et~al.} 2012, A\&A, 541,
  A116

\bibitem[{{Antoniucci} {et~al.}(2017){Antoniucci}, {Nisini}, {Biazzo},
  {Giannini}, {Lorenzetti}, {Sanna}, {Harutyunyan}, {Origlia}, \&
  {Oliva}}]{2017A&A...606A..48A}
{Antoniucci}, S., {Nisini}, B., {Biazzo}, K., {et~al.} 2017, \aap, 606, A48

\bibitem[{{Basri} \& {Batalha}(1990)}]{1990ApJ...363..654B}
{Basri}, G. \& {Batalha}, C. 1990, \apj, 363, 654

\bibitem[{{Bohn} {et~al.}(2022){Bohn}, {Benisty}, {Perraut}, {van der Marel},
  {W{\"o}lfer}, {van Dishoeck}, {Facchini}, {Manara}, {Teague}, {Francis},
  {Berger}, {Garcia-Lopez}, {Ginski}, {Henning}, {Kenworthy}, {Kraus},
  {M{\'e}nard}, {M{\'e}rand}, \& {P{\'e}rez}}]{2022A&A...658A.183B}
{Bohn}, A.~J., {Benisty}, M., {Perraut}, K., {et~al.} 2022, \aap, 658, A183

\bibitem[{{Bouvier} {et~al.}(2020){Bouvier}, {Alecian}, {Alencar}, {Sousa},
  {Donati}, {Perraut}, {Bayo}, {Rebull}, {Dougados}, {Duvert}, {Berger},
  {Benisty}, {Pouilly}, {Folsom}, {Moutou}, \& {SPIRou
  Consortium}}]{2020...643A..99B}
{Bouvier}, J., {Alecian}, E., {Alencar}, S.~H.~P., {et~al.} 2020, \aap, 643,
  A99

\bibitem[{{Calvet} {et~al.}(2002){Calvet}, {D'Alessio}, {Hartmann}, {Wilner},
  {Walsh}, \& {Sitko}}]{2002ApJ...568.1008C}
{Calvet}, N., {D'Alessio}, P., {Hartmann}, L., {et~al.} 2002, \apj, 568, 1008

\bibitem[{{Calvet} \& {Gullbring}(1998)}]{1998ApJ...509..802C}
{Calvet}, N. \& {Gullbring}, E. 1998, ApJ, 509, 802

\bibitem[{{Calvet} {et~al.}(1997){Calvet}, {Hartmann}, \&
  {Strom}}]{1997ApJ...481..912C}
{Calvet}, N., {Hartmann}, L., \& {Strom}, S.~E. 1997, \apj, 481, 912

\bibitem[{{Carpenter} {et~al.}(2001){Carpenter}, {Hillenbrand}, \&
  {Skrutskie}}]{2001AJ....121.3160C}
{Carpenter}, J.~M., {Hillenbrand}, L.~A., \& {Skrutskie}, M.~F. 2001, \aj, 121,
  3160

\bibitem[{{Chiang} \& {Goldreich}(1997)}]{1997ApJ...490..368C}
{Chiang}, E.~I. \& {Goldreich}, P. 1997, \apj, 490, 368

\bibitem[{{Chiang} \& {Goldreich}(1999)}]{1999ApJ...519..279C}
{Chiang}, E.~I. \& {Goldreich}, P. 1999, \apj, 519, 279

\bibitem[{{Cieza} {et~al.}(2005){Cieza}, {Kessler-Silacci}, {Jaffe}, {Harvey},
  \& {Evans}}]{2005ApJ...635..422C}
{Cieza}, L.~A., {Kessler-Silacci}, J.~E., {Jaffe}, D.~T., {Harvey}, P.~M., \&
  {Evans}, Neal~J., I. 2005, \apj, 635, 422

\bibitem[{{Cody} {et~al.}(2014){Cody}, {Stauffer}, {Baglin}, {Micela},
  {Rebull}, {Flaccomio}, {Morales-Calder{\'o}n}, {Aigrain}, {Bouvier},
  {Hillenbrand}, {Gutermuth}, {Song}, {Turner}, {Alencar}, {Zwintz},
  {Plavchan}, {Carpenter}, {Findeisen}, {Carey}, {Terebey}, {Hartmann},
  {Calvet}, {Teixeira}, {Vrba}, {Wolk}, {Covey}, {Poppenhaeger}, {G{\"u}nther},
  {Forbrich}, {Whitney}, {Affer}, {Herbst}, {Hora}, {Barrado}, {Holtzman},
  {Marchis}, {Wood}, {Medeiros Guimar{\~a}es}, {Lillo Box}, {Gillen},
  {McQuillan}, {Espaillat}, {Allen}, {D'Alessio}, \&
  {Favata}}]{2014AJ....147...82C}
{Cody}, A.~M., {Stauffer}, J., {Baglin}, A., {et~al.} 2014, \aj, 147, 82

\bibitem[{{Connelley} \& {Greene}(2010)}]{2010AJ....140.1214C}
{Connelley}, M.~S. \& {Greene}, T.~P. 2010, \aj, 140, 1214

\bibitem[{{Cook} {et~al.}(2022){Cook}, {Artigau}, {Doyon}, {Hobson},
  {Martioli}, {Bouchy}, {Moutou}, {Carmona}, {Usher}, {Fouqu{\'e}}, {Arnold},
  {Delfosse}, {Boisse}, {Cadieux}, {Vandal}, {Donati}, \&
  {Desli{\`e}res}}]{2022arXiv221101358C}
{Cook}, N.~J., {Artigau}, {\'E}., {Doyon}, R., {et~al.} 2022, arXiv e-prints,
  arXiv:2211.01358

\bibitem[{{Dahm} \& {Hillenbrand}(2020)}]{2020AJ....160..278D}
{Dahm}, S.~E. \& {Hillenbrand}, L.~A. 2020, \aj, 160, 278

\bibitem[{{Dahm} {et~al.}(2012){Dahm}, {Slesnick}, \&
  {White}}]{2012ApJ...745...56D}
{Dahm}, S.~E., {Slesnick}, C.~L., \& {White}, R.~J. 2012, \apj, 745, 56

\bibitem[{{Davies}(2019)}]{2019MNRAS.484.1926D}
{Davies}, C.~L. 2019, \mnras, 484, 1926

\bibitem[{{Dodin} \& {Lamzin}(2013)}]{2013AstL...39..389D}
{Dodin}, A.~V. \& {Lamzin}, S.~A. 2013, Astronomy Letters, 39, 389

\bibitem[{{Donati} {et~al.}(2007){Donati}, {Jardine}, {Gregory}, {Petit},
  {Bouvier}, {Dougados}, {M{\'e}nard}, {Cameron}, {Harries}, {Jeffers}, \&
  {Paletou}}]{2007MNRAS.380.1297D}
{Donati}, J., {Jardine}, M.~M., {Gregory}, S.~G., {et~al.} 2007, MNRAS, 380,
  1297

\bibitem[{{Donati} {et~al.}(2020{\natexlab{a}}){Donati}, {Bouvier}, {Alencar},
  {Moutou}, {Malo}, {Takami}, {M{\'e}nard}, {Dougados}, {Hussain}, \& {Matysse
  Collaboration}}]{2020MNRAS.491.5660D}
{Donati}, J.~F., {Bouvier}, J., {Alencar}, S.~H., {et~al.} 2020{\natexlab{a}},
  \mnras, 491, 5660

\bibitem[{{Donati} {et~al.}(2011){Donati}, {Gregory}, {Alencar}, {Bouvier},
  {Hussain}, {Skelly}, {Dougados}, {Jardine}, {M{\'e}nard}, {Romanova}, \&
  {Unruh}}]{2011MNRAS.417..472D}
{Donati}, J.~F., {Gregory}, S.~G., {Alencar}, S.~H.~P., {et~al.} 2011, \mnras,
  417, 472

\bibitem[{{Donati} {et~al.}(2012){Donati}, {Gregory}, {Alencar}, {Hussain},
  {Bouvier}, {Dougados}, {Jardine}, {M{\'e}nard}, \&
  {Romanova}}]{2012MNRAS.425.2948D}
{Donati}, J.~F., {Gregory}, S.~G., {Alencar}, S.~H.~P., {et~al.} 2012, \mnras,
  425, 2948

\bibitem[{{Donati} {et~al.}(2015){Donati}, {H{\'e}brard}, {Hussain}, {Moutou},
  {Malo}, {Grankin}, {Vidotto}, {Alencar}, {Gregory}, {Jardine}, {Herczeg},
  {Morin}, {Fares}, {M{\'e}nard}, {Bouvier}, {Delfosse}, {Doyon}, {Takami},
  {Figueira}, {Petit}, {Boisse}, \& {MaTYSSE
  Collaboration}}]{2015MNRAS.453.3706D}
{Donati}, J.~F., {H{\'e}brard}, E., {Hussain}, G.~A.~J., {et~al.} 2015, \mnras,
  453, 3706

\bibitem[{{Donati} {et~al.}(2008){Donati}, {Jardine}, {Gregory}, {Petit},
  {Paletou}, {Bouvier}, {Dougados}, {M{\'e}nard}, {Collier Cameron}, {Harries},
  {Hussain}, {Unruh}, {Morin}, {Marsden}, {Manset}, {Auri{\`e}re}, {Catala}, \&
  {Alecian}}]{2008MNRAS.386.1234D}
{Donati}, J.~F., {Jardine}, M.~M., {Gregory}, S.~G., {et~al.} 2008, \mnras,
  386, 1234

\bibitem[{{Donati} {et~al.}(2018){Donati}, {Kouach}, {Lacombe}, {Baratchart},
  {Doyon}, {Delfosse}, {Artigau}, {Moutou}, {H{\'e}brard}, {Bouchy}, {Bouvier},
  {Alencar}, {Saddlemyer}, {Par{\`e}s}, {Rabou}, {Micheau}, {Dolon}, {Barrick},
  {Hernandez}, {Wang}, {Reshetov}, {Striebig}, {Challita}, {Carmona},
  {Tibault}, {Martioli}, {Figueira}, {Boisse}, \& {Pepe}}]{2018haex.bookE.107D}
{Donati}, J.-F., {Kouach}, D., {Lacombe}, M., {et~al.} 2018, in Handbook of
  Exoplanets, ed. H.~J. {Deeg} \& J.~A. {Belmonte}, 107

\bibitem[{{Donati} {et~al.}(2020{\natexlab{b}}){Donati}, {Kouach}, {Moutou},
  {Doyon}, {Delfosse}, {Artigau}, {Baratchart}, {Lacombe}, {Barrick},
  {H{\'e}brard}, {Bouchy}, {Saddlemyer}, {Par{\`e}s}, {Rabou}, {Micheau},
  {Dolon}, {Reshetov}, {Challita}, {Carmona}, {Striebig}, {Thibault},
  {Martioli}, {Cook}, {Fouqu{\'e}}, {Vermeulen}, {Wang}, {Arnold}, {Pepe},
  {Boisse}, {Figueira}, {Bouvier}, {Ray}, {Feugeade}, {Morin}, {Alencar},
  {Hobson}, {Castilho}, {Udry}, {Santos}, {Hernandez}, {Benedict},
  {Vall{\'e}e}, {Gallou}, {Dupieux}, {Larrieu}, {Perruchot}, {Sottile},
  {Moreau}, {Usher}, {Baril}, {Wildi}, {Chazelas}, {Malo}, {Bonfils}, {Loop},
  {Kerley}, {Wevers}, {Dunn}, {Pazder}, {Macdonald}, {Dubois}, {Carri{\'e}},
  {Valentin}, {Henault}, {Yan}, \& {Steinmetz}}]{2020MNRAS.498.5684D}
{Donati}, J.~F., {Kouach}, D., {Moutou}, C., {et~al.} 2020{\natexlab{b}},
  \mnras, 498, 5684

\bibitem[{{Donati} {et~al.}(2010){Donati}, {Skelly}, {Bouvier}, {Jardine},
  {Gregory}, {Morin}, {Hussain}, {Dougados}, {M{\'e}nard}, \&
  {Unruh}}]{2010MNRAS.402.1426D}
{Donati}, J.~F., {Skelly}, M.~B., {Bouvier}, J., {et~al.} 2010, \mnras, 402,
  1426

\bibitem[{{Edwards} {et~al.}(2006){Edwards}, {Fischer}, {Hillenbrand}, \&
  {Kwan}}]{2006ApJ...646..319E}
{Edwards}, S., {Fischer}, W., {Hillenbrand}, L., \& {Kwan}, J. 2006, \apj, 646,
  319

\bibitem[{{Edwards} {et~al.}(2003){Edwards}, {Fischer}, {Kwan}, {Hillenbrand},
  \& {Dupree}}]{edwards03}
{Edwards}, S., {Fischer}, W., {Kwan}, J., {Hillenbrand}, L., \& {Dupree}, A.~K.
  2003, ApJL, 599, L41

\bibitem[{{Ercolano} {et~al.}(2009){Ercolano}, {Clarke}, \&
  {Robitaille}}]{2009MNRAS.394L.141E}
{Ercolano}, B., {Clarke}, C.~J., \& {Robitaille}, T.~P. 2009, \mnras, 394, L141

\bibitem[{{Espaillat} {et~al.}(2022){Espaillat}, {Herczeg}, {Thanathibodee},
  {Pittman}, {Calvet}, {Arulanantham}, {France}, {Serna}, {Hern{\'a}ndez},
  {K{\'o}sp{\'a}l}, {Walter}, {Frasca}, {Fischer}, {Johns-Krull}, {Schneider},
  {Robinson}, {Edwards}, {{\'A}brah{\'a}m}, {Fang}, {Erkal}, {Manara},
  {Alcal{\'a}}, {Alecian}, {Alexander}, {Alonso-Santiago}, {Antoniucci},
  {Ardila}, {Banzatti}, {Benisty}, {Bergin}, {Biazzo}, {Brice{\~n}o},
  {Campbell-White}, {Cleeves}, {Coffey}, {Eisl{\"o}ffel}, {Facchini}, {Fedele},
  {Fiorellino}, {Froebrich}, {Gangi}, {Giannini}, {Grankin}, {G{\"u}nther},
  {Guo}, {Hartmann}, {Hillenbrand}, {Hinton}, {Kastner}, {Koen}, {Mauc{\'o}},
  {Mendigut{\'\i}a}, {Nisini}, {Panwar}, {Principe}, {Robberto},
  {Sicilia-Aguilar}, {Valenti}, {Wendeborn}, {Williams}, {Xu}, \&
  {Yadav}}]{2022AJ....163..114E}
{Espaillat}, C.~C., {Herczeg}, G.~J., {Thanathibodee}, T., {et~al.} 2022, \aj,
  163, 114

\bibitem[{{Faesi} {et~al.}(2012){Faesi}, {Covey}, {Gutermuth},
  {Morales{\textendash}Calder{\'o}n}, {Stauffer}, {Plavchan}, {Rebull}, {Song},
  \& {Lloyd}}]{2012PASP..124.1137F}
{Faesi}, C.~M., {Covey}, K.~R., {Gutermuth}, R., {et~al.} 2012, \pasp, 124,
  1137

\bibitem[{{Fazio} {et~al.}(2004){Fazio}, {Hora}, {Allen}, {Ashby}, {Barmby},
  {Deutsch}, {Huang}, {Kleiner}, {Marengo}, {Megeath}, {Melnick}, {Pahre},
  {Patten}, {Polizotti}, {Smith}, {Taylor}, {Wang}, {Willner}, {Hoffmann},
  {Pipher}, {Forrest}, {McMurty}, {McCreight}, {McKelvey}, {McMurray}, {Koch},
  {Moseley}, {Arendt}, {Mentzell}, {Marx}, {Losch}, {Mayman}, {Eichhorn},
  {Krebs}, {Jhabvala}, {Gezari}, {Fixsen}, {Flores}, {Shakoorzadeh}, {Jungo},
  {Hakun}, {Workman}, {Karpati}, {Kichak}, {Whitley}, {Mann}, {Tollestrup},
  {Eisenhardt}, {Stern}, {Gorjian}, {Bhattacharya}, {Carey}, {Nelson},
  {Glaccum}, {Lacy}, {Lowrance}, {Laine}, {Reach}, {Stauffer}, {Surace},
  {Wilson}, {Wright}, {Hoffman}, {Domingo}, \& {Cohen}}]{2004ApJS..154...10F}
{Fazio}, G.~G., {Hora}, J.~L., {Allen}, L.~E., {et~al.} 2004, ApJS, 154, 10

\bibitem[{{Fischer} {et~al.}(2011){Fischer}, {Edwards}, {Hillenbrand}, \&
  {Kwan}}]{2011ApJ...730...73F}
{Fischer}, W., {Edwards}, S., {Hillenbrand}, L., \& {Kwan}, J. 2011, \apj, 730,
  73

\bibitem[{{Flores} {et~al.}(2019){Flores}, {Connelley}, {Reipurth}, \&
  {Boogert}}]{2019ApJ...882...75F}
{Flores}, C., {Connelley}, M.~S., {Reipurth}, B., \& {Boogert}, A. 2019, \apj,
  882, 75

\bibitem[{{Folha} \& {Emerson}(1999)}]{1999A&A...352..517F}
{Folha}, D.~F.~M. \& {Emerson}, J.~P. 1999, \aap, 352, 517

\bibitem[{{Frasca} {et~al.}(2017){Frasca}, {Biazzo}, {Alcal{\'a}}, {Manara},
  {Stelzer}, {Covino}, \& {Antoniucci}}]{2017AA...602A..33F}
{Frasca}, A., {Biazzo}, K., {Alcal{\'a}}, J.~M., {et~al.} 2017, \aap, 602, A33

\bibitem[{{Gaia Collaboration} {et~al.}(2018){Gaia Collaboration}, {Brown},
  {Vallenari}, {Prusti}, {de Bruijne}, {Babusiaux}, {Bailer-Jones}, {Biermann},
  {Evans}, \& {Eyer}}]{2018A&A...616A...1G}
{Gaia Collaboration}, {Brown}, A.~G.~A., {Vallenari}, A., {et~al.} 2018, A\&A,
  616, A1

\bibitem[{{Gaia Collaboration} {et~al.}(2021){Gaia Collaboration}, {Brown},
  {Vallenari}, {Prusti}, {de Bruijne}, {Babusiaux}, {Biermann}, {Creevey},
  {Evans}, {Eyer}, {Hutton}, {Jansen}, {Jordi}, {Klioner}, {Lammers},
  {Lindegren}, {Luri}, {Mignard}, {Panem}, {Pourbaix}, {Randich}, {Sartoretti},
  {Soubiran}, {Walton}, {Arenou}, {Bailer-Jones}, {Bastian}, {Cropper},
  {Drimmel}, {Katz}, {Lattanzi}, {van Leeuwen}, {Bakker}, {Cacciari},
  {Casta{\~n}eda}, {De Angeli}, {Ducourant}, {Fabricius}, {Fouesneau},
  {Fr{\'e}mat}, {Guerra}, {Guerrier}, {Guiraud}, {Jean-Antoine Piccolo},
  {Masana}, {Messineo}, {Mowlavi}, {Nicolas}, {Nienartowicz}, {Pailler},
  {Panuzzo}, {Riclet}, {Roux}, {Seabroke}, {Sordo}, {Tanga}, {Th{\'e}venin},
  {Gracia-Abril}, {Portell}, {Teyssier}, {Altmann}, {Andrae}, {Bellas-Velidis},
  {Benson}, {Berthier}, {Blomme}, {Brugaletta}, {Burgess}, {Busso}, {Carry},
  {Cellino}, {Cheek}, {Clementini}, {Damerdji}, {Davidson}, {Delchambre},
  {Dell'Oro}, {Fern{\'a}ndez-Hern{\'a}ndez}, {Galluccio}, {Garc{\'\i}a-Lario},
  {Garcia-Reinaldos}, {Gonz{\'a}lez-N{\'u}{\~n}ez}, {Gosset}, {Haigron},
  {Halbwachs}, {Hambly}, {Harrison}, {Hatzidimitriou}, {Heiter},
  {Hern{\'a}ndez}, {Hestroffer}, {Hodgkin}, {Holl}, {Jan{\ss}en}, {Jevardat de
  Fombelle}, {Jordan}, {Krone-Martins}, {Lanzafame}, {L{\"o}ffler}, {Lorca},
  {Manteiga}, {Marchal}, {Marrese}, {Moitinho}, {Mora}, {Muinonen}, {Osborne},
  {Pancino}, {Pauwels}, {Petit}, {Recio-Blanco}, {Richards}, {Riello},
  {Rimoldini}, {Robin}, {Roegiers}, {Rybizki}, {Sarro}, {Siopis}, {Smith},
  {Sozzetti}, {Ulla}, {Utrilla}, {van Leeuwen}, {van Reeven}, {Abbas}, {Abreu
  Aramburu}, {Accart}, {Aerts}, {Aguado}, {Ajaj}, {Altavilla}, {{\'A}lvarez},
  {{\'A}lvarez Cid-Fuentes}, {Alves}, {Anderson}, {Anglada Varela}, {Antoja},
  {Audard}, {Baines}, {Baker}, {Balaguer-N{\'u}{\~n}ez}, {Balbinot}, {Balog},
  {Barache}, {Barbato}, {Barros}, {Barstow}, {Bartolom{\'e}}, {Bassilana},
  {Bauchet}, {Baudesson-Stella}, {Becciani}, {Bellazzini}, {Bernet}, {Bertone},
  {Bianchi}, {Blanco-Cuaresma}, {Boch}, {Bombrun}, {Bossini}, {Bouquillon},
  {Bragaglia}, {Bramante}, {Breedt}, {Bressan}, {Brouillet}, {Bucciarelli},
  {Burlacu}, {Busonero}, {Butkevich}, {Buzzi}, {Caffau}, {Cancelliere},
  {C{\'a}novas}, {Cantat-Gaudin}, {Carballo}, {Carlucci}, {Carnerero},
  {Carrasco}, {Casamiquela}, {Castellani}, {Castro-Ginard}, {Castro Sampol},
  {Chaoul}, {Charlot}, {Chemin}, {Chiavassa}, {Cioni}, {Comoretto}, {Cooper},
  {Cornez}, {Cowell}, {Crifo}, {Crosta}, {Crowley}, {Dafonte}, {Dapergolas},
  {David}, {David}, {de Laverny}, {De Luise}, {De March}, {De Ridder}, {de
  Souza}, {de Teodoro}, {de Torres}, {del Peloso}, {del Pozo}, {Delbo},
  {Delgado}, {Delgado}, {Delisle}, {Di Matteo}, {Diakite}, {Diener},
  {Distefano}, {Dolding}, {Eappachen}, {Edvardsson}, {Enke}, {Esquej}, {Fabre},
  {Fabrizio}, {Faigler}, {Fedorets}, {Fernique}, {Fienga}, {Figueras},
  {Fouron}, {Fragkoudi}, {Fraile}, {Franke}, {Gai}, {Garabato},
  {Garcia-Gutierrez}, {Garc{\'\i}a-Torres}, {Garofalo}, {Gavras}, {Gerlach},
  {Geyer}, {Giacobbe}, {Gilmore}, {Girona}, {Giuffrida}, {Gomel}, {Gomez},
  {Gonzalez-Santamaria}, {Gonz{\'a}lez-Vidal}, {Granvik},
  {Guti{\'e}rrez-S{\'a}nchez}, {Guy}, {Hauser}, {Haywood}, {Helmi}, {Hidalgo},
  {Hilger}, {H{\l}adczuk}, {Hobbs}, {Holland}, {Huckle}, {Jasniewicz},
  {Jonker}, {Juaristi Campillo}, {Julbe}, {Karbevska}, {Kervella}, {Khanna},
  {Kochoska}, {Kontizas}, {Kordopatis}, {Korn}, {Kostrzewa-Rutkowska},
  {Kruszy{\'n}ska}, {Lambert}, {Lanza}, {Lasne}, {Le Campion}, {Le Fustec},
  {Lebreton}, {Lebzelter}, {Leccia}, {Leclerc}, {Lecoeur-Taibi}, {Liao},
  {Licata}, {Lindstr{\o}m}, {Lister}, {Livanou}, {Lobel}, {Madrero Pardo},
  {Managau}, {Mann}, {Marchant}, {Marconi}, {Marcos Santos}, {Marinoni},
  {Marocco}, {Marshall}, {Martin Polo}, {Mart{\'\i}n-Fleitas}, {Masip},
  {Massari}, {Mastrobuono-Battisti}, {Mazeh}, {McMillan}, {Messina},
  {Michalik}, {Millar}, {Mints}, {Molina}, {Molinaro}, {Moln{\'a}r},
  {Montegriffo}, {Mor}, {Morbidelli}, {Morel}, {Morris}, {Mulone}, {Munoz},
  {Muraveva}, {Murphy}, {Musella}, {Noval}, {Ord{\'e}novic}, {Orr{\`u}},
  {Osinde}, {Pagani}, {Pagano}, {Palaversa}, {Palicio}, {Panahi}, {Pawlak},
  {Pe{\~n}alosa Esteller}, {Penttil{\"a}}, {Piersimoni}, {Pineau}, {Plachy},
  {Plum}, {Poggio}, {Poretti}, {Poujoulet}, {Pr{\v{s}}a}, {Pulone}, {Racero},
  {Ragaini}, {Rainer}, {Raiteri}, {Rambaux}, {Ramos}, {Ramos-Lerate}, {Re
  Fiorentin}, {Regibo}, {Reyl{\'e}}, {Ripepi}, {Riva}, {Rixon}, {Robichon},
  {Robin}, {Roelens}, {Rohrbasser}, {Romero-G{\'o}mez}, {Rowell}, {Royer},
  {Rybicki}, {Sadowski}, {Sagrist{\`a} Sell{\'e}s}, {Sahlmann}, {Salgado},
  {Salguero}, {Samaras}, {Sanchez Gimenez}, {Sanna}, {Santove{\~n}a},
  {Sarasso}, {Schultheis}, {Sciacca}, {Segol}, {Segovia}, {S{\'e}gransan},
  {Semeux}, {Shahaf}, {Siddiqui}, {Siebert}, {Siltala}, {Slezak}, {Smart},
  {Solano}, {Solitro}, {Souami}, {Souchay}, {Spagna}, {Spoto}, {Steele},
  {Steidelm{\"u}ller}, {Stephenson}, {S{\"u}veges}, {Szabados}, {Szegedi-Elek},
  {Taris}, {Tauran}, {Taylor}, {Teixeira}, {Thuillot}, {Tonello}, {Torra},
  {Torra}, {Turon}, {Unger}, {Vaillant}, {van Dillen}, {Vanel}, {Vecchiato},
  {Viala}, {Vicente}, {Voutsinas}, {Weiler}, {Wevers}, {Wyrzykowski}, {Yoldas},
  {Yvard}, {Zhao}, {Zorec}, {Zucker}, {Zurbach}, \&
  {Zwitter}}]{2021A&A...649A...1G}
{Gaia Collaboration}, {Brown}, A.~G.~A., {Vallenari}, A., {et~al.} 2021, \aap,
  649, A1

\bibitem[{{Gullbring} {et~al.}(1998){Gullbring}, {Hartmann}, {Brice{\~n}o}, \&
  {Calvet}}]{1998ApJ...492..323G}
{Gullbring}, E., {Hartmann}, L., {Brice{\~n}o}, C., \& {Calvet}, N. 1998, APJ,
  492, 323

\bibitem[{{Gully-Santiago} {et~al.}(2017){Gully-Santiago}, {Herczeg},
  {Czekala}, {Somers}, {Grankin}, {Covey}, {Donati}, {Alencar}, {Hussain},
  {Shappee}, {Mace}, {Lee}, {Holoien}, {Jose}, \& {Liu}}]{2017ApJ...836..200G}
{Gully-Santiago}, M.~A., {Herczeg}, G.~J., {Czekala}, I., {et~al.} 2017, \apj,
  836, 200

\bibitem[{{Gunn} {et~al.}(1998){Gunn}, {Carr}, {Rockosi}, {Sekiguchi}, {Berry},
  {Elms}, {de Haas}, {Ivezi{\'c}}, {Knapp}, {Lupton}, {Pauls}, {Simcoe},
  {Hirsch}, {Sanford}, {Wang}, {York}, {Harris}, {Annis}, {Bartozek},
  {Boroski}, {Bakken}, {Haldeman}, {Kent}, {Holm}, {Holmgren}, {Petravick},
  {Prosapio}, {Rechenmacher}, {Doi}, {Fukugita}, {Shimasaku}, {Okada}, {Hull},
  {Siegmund}, {Mannery}, {Blouke}, {Heidtman}, {Schneider}, {Lucinio}, \&
  {Brinkman}}]{1998AJ....116.3040G}
{Gunn}, J.~E., {Carr}, M., {Rockosi}, C., {et~al.} 1998, AJ, 116, 3040

\bibitem[{{Hartigan} {et~al.}(1989){Hartigan}, {Hartmann}, {Kenyon}, {Hewett},
  \& {Stauffer}}]{1989ApJS...70..899H}
{Hartigan}, P., {Hartmann}, L., {Kenyon}, S., {Hewett}, R., \& {Stauffer}, J.
  1989, ApJS, 70, 899

\bibitem[{{Hartigan} {et~al.}(1991){Hartigan}, {Kenyon}, {Hartmann}, {Strom},
  {Edwards}, {Welty}, \& {Stauffer}}]{1991ApJ...382..617H}
{Hartigan}, P., {Kenyon}, S.~J., {Hartmann}, L., {et~al.} 1991, ApJ, 382, 617

\bibitem[{{Hartmann} \& {Kenyon}(1990)}]{1990ApJ...349..190H}
{Hartmann}, L.~W. \& {Kenyon}, S.~J. 1990, \apj, 349, 190

\bibitem[{{Herczeg} \& {Hillenbrand}(2014)}]{2014ApJ...786...97H}
{Herczeg}, G.~J. \& {Hillenbrand}, L.~A. 2014, \apj, 786, 97

\bibitem[{{Hillenbrand} {et~al.}(1998){Hillenbrand}, {Strom}, {Calvet},
  {Merrill}, {Gatley}, {Makidon}, {Meyer}, \&
  {Skrutskie}}]{1998AJ....116.1816H}
{Hillenbrand}, L.~A., {Strom}, S.~E., {Calvet}, N., {et~al.} 1998, \aj, 116,
  1816

\bibitem[{{Horne} \& {Baliunas}(1986)}]{1986ApJ...302..757H}
{Horne}, J.~H. \& {Baliunas}, S.~L. 1986, APJ, 302, 757

\bibitem[{{Ingleby} {et~al.}(2013){Ingleby}, {Calvet}, {Herczeg}, {Blaty},
  {Walter}, {Ardila}, {Alexander}, {Edwards}, {Espaillat}, {Gregory},
  {Hillenbrand}, \& {Brown}}]{2013ApJ...767..112I}
{Ingleby}, L., {Calvet}, N., {Herczeg}, G., {et~al.} 2013, \apj, 767, 112

\bibitem[{{Johns-Krull}(2007)}]{2007ApJ...664..975J}
{Johns-Krull}, C.~M. 2007, \apj, 664, 975

\bibitem[{{Johns-Krull} \& {Valenti}(2001)}]{2001ApJ...561.1060J}
{Johns-Krull}, C.~M. \& {Valenti}, J.~A. 2001, \apj, 561, 1060

\bibitem[{{Johns-Krull} {et~al.}(2000){Johns-Krull}, {Valenti}, \&
  {Linsky}}]{2000ApJ...539..815J}
{Johns-Krull}, C.~M., {Valenti}, J.~A., \& {Linsky}, J.~L. 2000, \apj, 539, 815

\bibitem[{{Kidder} {et~al.}(2021){Kidder}, {Mace}, {L{\'o}pez-Valdivia},
  {Sokal}, {Catlett}, {Guti{\'e}rrez}, {Tofflemire}, \&
  {Jaffe}}]{2021ApJ...922...27K}
{Kidder}, B., {Mace}, G., {L{\'o}pez-Valdivia}, R., {et~al.} 2021, \apj, 922,
  27

\bibitem[{{Kounkel} {et~al.}(2019){Kounkel}, {Covey}, {Moe}, {Kratter},
  {Su{\'a}rez}, {Stassun}, {Rom{\'a}n-Z{\'u}{\~n}iga}, {Hernandez}, {Kim},
  {Pe{\~n}a Ram{\'\i}rez}, {Roman-Lopes}, {Stringfellow}, {Jaehnig},
  {Borissova}, {Tofflemire}, {Krolikowski}, {Rizzuto}, {Kraus}, {Badenes},
  {Longa-Pe{\~n}a}, {G{\'o}mez Maqueo Chew}, {Barba}, {Nidever}, {Brown}, {De
  Lee}, {Pan}, {Bizyaev}, {Oravetz}, \& {Oravetz}}]{2019AJ....157..196K}
{Kounkel}, M., {Covey}, K., {Moe}, M., {et~al.} 2019, \aj, 157, 196

\bibitem[{{Kwan} {et~al.}(2007){Kwan}, {Edwards}, \&
  {Fischer}}]{2007ApJ...657..897K}
{Kwan}, J., {Edwards}, S., \& {Fischer}, W. 2007, \apj, 657, 897

\bibitem[{{Kwan} \& {Fischer}(2011)}]{2011MNRAS.411.2383K}
{Kwan}, J. \& {Fischer}, W. 2011, MNRAS, 411, 2383

\bibitem[{{Lada} {et~al.}(2006){Lada}, {Muench}, {Luhman}, {Allen}, {Hartmann},
  {Megeath}, {Myers}, {Fazio}, {Wood}, {Muzerolle}, {Rieke}, {Siegler}, \&
  {Young}}]{Lada2006}
{Lada}, C.~J., {Muench}, A.~A., {Luhman}, K.~L., {et~al.} 2006, ApJ, 131, 1574

\bibitem[{{McClure} {et~al.}(2013){McClure}, {Calvet}, {Espaillat}, {Hartmann},
  {Hern{\'a}ndez}, {Ingleby}, {Luhman}, {D'Alessio}, \&
  {Sargent}}]{2013ApJ...769...73M}
{McClure}, M.~K., {Calvet}, N., {Espaillat}, C., {et~al.} 2013, \apj, 769, 73

\bibitem[{{M{\"u}ller} {et~al.}(2018){M{\"u}ller}, {Keppler}, {Henning},
  {Samland}, {Chauvin}, {Beust}, {Maire}, {Molaverdikhani}, {van Boekel},
  {Benisty}, {Boccaletti}, {Bonnefoy}, {Cantalloube}, {Charnay}, {Baudino},
  {Gennaro}, {Long}, {Cheetham}, {Desidera}, {Feldt}, {Fusco}, {Girard},
  {Gratton}, {Hagelberg}, {Janson}, {Lagrange}, {Langlois}, {Lazzoni}, {Ligi},
  {M{\'e}nard}, {Mesa}, {Meyer}, {Molli{\`e}re}, {Mordasini}, {Moulin},
  {Pavlov}, {Pawellek}, {Quanz}, {Ramos}, {Rouan}, {Sissa}, {Stadler}, {Vigan},
  {Wahhaj}, {Weber}, \& {Zurlo}}]{2018AA...617L...2M}
{M{\"u}ller}, A., {Keppler}, M., {Henning}, T., {et~al.} 2018, \aap, 617, L2

\bibitem[{{Muzerolle} {et~al.}(2010){Muzerolle}, {Allen}, {Megeath},
  {Hern{\'a}ndez}, \& {Gutermuth}}]{2010ApJ...708.1107M}
{Muzerolle}, J., {Allen}, L.~E., {Megeath}, S.~T., {Hern{\'a}ndez}, J., \&
  {Gutermuth}, R.~A. 2010, ApJ, 708, 1107

\bibitem[{{Muzerolle} {et~al.}(1998){Muzerolle}, {Hartmann}, \&
  {Calvet}}]{1998AJ....116..455M}
{Muzerolle}, J., {Hartmann}, L., \& {Calvet}, N. 1998, AJ, 116, 455

\bibitem[{{Nguyen} {et~al.}(2012){Nguyen}, {Brandeker}, {van Kerkwijk}, \&
  {Jayawardhana}}]{2012ApJ...745..119N}
{Nguyen}, D.~C., {Brandeker}, A., {van Kerkwijk}, M.~H., \& {Jayawardhana}, R.
  2012, \apj, 745, 119

\bibitem[{{Pecaut} \& {Mamajek}(2013)}]{2013ApJS..208....9P}
{Pecaut}, M.~J. \& {Mamajek}, E.~E. 2013, \apjs, 208, 9

\bibitem[{{Pecaut} \& {Mamajek}(2016)}]{2016MNRAS.461..794P}
{Pecaut}, M.~J. \& {Mamajek}, E.~E. 2016, \mnras, 461, 794

\bibitem[{{Pontoppidan} {et~al.}(2008){Pontoppidan}, {Blake}, {van Dishoeck},
  {Smette}, {Ireland}, \& {Brown}}]{2008ApJ...684.1323P}
{Pontoppidan}, K.~M., {Blake}, G.~A., {van Dishoeck}, E.~F., {et~al.} 2008,
  \apj, 684, 1323

\bibitem[{{Preibisch} \& {Zinnecker}(1999)}]{1999AJ....117.2381P}
{Preibisch}, T. \& {Zinnecker}, H. 1999, \aj, 117, 2381

\bibitem[{{Rebull}(2001)}]{2001AJ....121.1676R}
{Rebull}, L.~M. 2001, \aj, 121, 1676

\bibitem[{{Rebull} {et~al.}(2002){Rebull}, {Makidon}, {Strom}, {Hillenbrand},
  {Birmingham}, {Patten}, {Jones}, {Yagi}, \& {Adams}}]{2002AJ....123.1528R}
{Rebull}, L.~M., {Makidon}, R.~B., {Strom}, S.~E., {et~al.} 2002, \aj, 123,
  1528

\bibitem[{{Rei} {et~al.}(2018){Rei}, {Petrov}, \&
  {Gameiro}}]{2018A&A...610A..40R}
{Rei}, A.~C.~S., {Petrov}, P.~P., \& {Gameiro}, J.~F. 2018, A\&A, 610, A40

\bibitem[{{Ricci} {et~al.}(2010){Ricci}, {Testi}, {Natta}, {Neri}, {Cabrit}, \&
  {Herczeg}}]{2010A&A...512A..15R}
{Ricci}, L., {Testi}, L., {Natta}, A., {et~al.} 2010, \aap, 512, A15

\bibitem[{{Rieke} {et~al.}(2004){Rieke}, {Young}, {Engelbracht}, {Kelly},
  {Low}, {Haller}, {Beeman}, {Gordon}, {Stansberry}, {Misselt}, {Cadien},
  {Morrison}, {Rivlis}, {Latter}, {Noriega-Crespo}, {Padgett}, {Stapelfeldt},
  {Hines}, {Egami}, {Muzerolle}, {Alonso-Herrero}, {Blaylock}, {Dole}, {Hinz},
  {Le Floc'h}, {Papovich}, {P{\'e}rez-Gonz{\'a}lez}, {Smith}, {Su}, {Bennett},
  {Frayer}, {Henderson}, {Lu}, {Masci}, {Pesenson}, {Rebull}, {Rho}, {Keene},
  {Stolovy}, {Wachter}, {Wheaton}, {Werner}, \&
  {Richards}}]{2004ApJS..154...25R}
{Rieke}, G.~H., {Young}, E.~T., {Engelbracht}, C.~W., {et~al.} 2004, ApJS, 154,
  25

\bibitem[{{Rodrigo} \& {Solano}(2020)}]{2020sea..confE.182R}
{Rodrigo}, C. \& {Solano}, E. 2020, in XIV.0 Scientific Meeting (virtual) of
  the Spanish Astronomical Society, 182

\bibitem[{{Rodrigo} {et~al.}(2012){Rodrigo}, {Solano}, \&
  {Bayo}}]{2012ivoa.rept.1015R}
{Rodrigo}, C., {Solano}, E., \& {Bayo}, A. 2012, {SVO Filter Profile Service
  Version 1.0}, IVOA Working Draft 15 October 2012

\bibitem[{{Sicilia-Aguilar} {et~al.}(2015){Sicilia-Aguilar}, {Fang},
  {Roccatagliata}, {Collier Cameron}, {K{\'o}sp{\'a}l}, {Henning},
  {{\'A}brah{\'a}m}, \& {Sipos}}]{2015A&A...580A..82S}
{Sicilia-Aguilar}, A., {Fang}, M., {Roccatagliata}, V., {et~al.} 2015, \aap,
  580, A82

\bibitem[{{Sicilia-Aguilar} {et~al.}(2020){Sicilia-Aguilar}, {Manara}, {de
  Boer}, {Benisty}, {Pinilla}, \& {Bouvier}}]{2020AA...633A..37S}
{Sicilia-Aguilar}, A., {Manara}, C.~F., {de Boer}, J., {et~al.} 2020, \aap,
  633, A37

\bibitem[{{Simon} {et~al.}(2016){Simon}, {Pascucci}, {Edwards}, {Feng},
  {Gorti}, {Hollenbach}, {Rigliaco}, \& {Keane}}]{2016ApJ...831..169S}
{Simon}, M.~N., {Pascucci}, I., {Edwards}, S., {et~al.} 2016, \apj, 831, 169

\bibitem[{{Sousa} {et~al.}(2021){Sousa}, {Bouvier}, {Alencar}, {Donati},
  {Alecian}, {Roquette}, {Perraut}, {Dougados}, {Carmona}, {Covino}, {Fugazza},
  {Molinari}, {Moutou}, {Santerne}, {Grankin}, {Artigau}, {Delfosse},
  {Hebrard}, \& {SPIRou Consortium}}]{2021A&A...649A..68S}
{Sousa}, A.~P., {Bouvier}, J., {Alencar}, S.~H.~P., {et~al.} 2021, \aap, 649,
  A68

\bibitem[{{Stempels} {et~al.}(2007){Stempels}, {Gahm}, \&
  {Petrov}}]{2007A&A...461..253S}
{Stempels}, H.~C., {Gahm}, G.~F., \& {Petrov}, P.~P. 2007, \aap, 461, 253

\bibitem[{{Sullivan} \& {Kraus}(2022)}]{2022ApJ...928..134S}
{Sullivan}, K. \& {Kraus}, A.~L. 2022, \apj, 928, 134

\bibitem[{{Teixeira} {et~al.}(2012){Teixeira}, {Lada}, {Marengo}, \&
  {Lada}}]{2012A&A...540A..83T}
{Teixeira}, P.~S., {Lada}, C.~J., {Marengo}, M., \& {Lada}, E.~A. 2012, A\&A,
  540, A83

\bibitem[{{Thanathibodee} {et~al.}(2020){Thanathibodee}, {Molina}, {Calvet},
  {Serna}, {Bae}, {Reynolds}, {Hern{\'a}ndez}, {Muzerolle}, \&
  {Hern{\'a}ndez}}]{2020ApJ...892...81T}
{Thanathibodee}, T., {Molina}, B., {Calvet}, N., {et~al.} 2020, \apj, 892, 81

\bibitem[{{Torres} {et~al.}(2006){Torres}, {Quast}, {da Silva}, {de La Reza},
  {Melo}, \& {Sterzik}}]{2006...460..695T}
{Torres}, C.~A.~O., {Quast}, G.~R., {da Silva}, L., {et~al.} 2006, \aap, 460,
  695

\bibitem[{{Valenti} {et~al.}(1993){Valenti}, {Basri}, \&
  {Johns}}]{1993AJ....106.2024V}
{Valenti}, J.~A., {Basri}, G., \& {Johns}, C.~M. 1993, \aj, 106, 2024

\bibitem[{{White} \& {Basri}(2003)}]{2003ApJ...582.1109W}
{White}, R.~J. \& {Basri}, G. 2003, ApJ, 582, 1109

\bibitem[{{Wright} {et~al.}(2010){Wright}, {Eisenhardt}, {Mainzer}, {Ressler},
  {Cutri}, {Jarrett}, {Kirkpatrick}, {Padgett}, {McMillan}, {Skrutskie},
  {Stanford}, {Cohen}, {Walker}, {Mather}, {Leisawitz}, {Gautier}, {McLean},
  {Benford}, {Lonsdale}, {Blain}, {Mendez}, {Irace}, {Duval}, {Liu}, {Royer},
  {Heinrichsen}, {Howard}, {Shannon}, {Kendall}, {Walsh}, {Larsen}, {Cardon},
  {Schick}, {Schwalm}, {Abid}, {Fabinsky}, {Naes}, \&
  {Tsai}}]{2010AJ....140.1868W}
{Wright}, E.~L., {Eisenhardt}, P.~R.~M., {Mainzer}, A.~K., {et~al.} 2010, AJ,
  140, 1868

\end{thebibliography}

\end{document}